\documentclass[lettersize,onecolumn]{IEEEtran}
\IEEEoverridecommandlockouts
\usepackage[utf8]{inputenc}
\usepackage{stackengine}
\usepackage{graphicx}
\usepackage{amssymb}
\usepackage{amsmath}
\usepackage{mathrsfs}
\usepackage{ulem}
\usepackage{epsf,epsfig}
\usepackage{cite}
\usepackage{color,xcolor}
\usepackage{dsfont}
\usepackage{bbm}
\usepackage{amsthm}
\usepackage{physics}
\usepackage{tcolorbox}
\usepackage{wrapfig}
\usepackage{url}
\usepackage{graphicx}
\usepackage{stmaryrd}
\usepackage{hyperref}
\usepackage{eurosym}
\usepackage{enumitem}
\usepackage{nopageno}
\usepackage{fancyhdr}
\usepackage{mathtools}
\usepackage{tabularx}
\usepackage{array} 
\usepackage{multirow}

\def\complex{\mathbb{C}}

\def\integers{\mathbb{Z}}

\def\Expectation{\mathbb{E}}

\newcommand{\msout}[1]{\text{\sout{\ensuremath{#1}}}}
\definecolor{darkgreen}{rgb}{0,0.5,0}
\definecolor{darkmagenta}{rgb}{.5,0,.5}
\definecolor{darkblue}{rgb}{0,0,0.5}

\def\CalC{\mathcal{C}}
\def\CalD{\mathcal{D}}

\def\CalF{\mathcal{F}}

\def\CalH{\mathcal{H}}

\def\CalM{\mathcal{M}}
\def\CalN{\mathcal{N}}

\def\CalP{\mathcal{P}}

\def\CalT{\mathcal{T}}
\def\CalU{\mathcal{U}}
\def\CalV{\mathcal{V}}

\def\CalX{\mathcal{X}}

\def\ulineD{\underline{D}}

\def\ulineM{\underline{M}}

\def\ulineR{\underline{R}}

\def\ulineU{\underline{U}}
\def\ulineV{\underline{V}}
\def\ulineW{\underline{W}}
\def\ulineX{\underline{X}}
\def\ulineY{\underline{Y}}
\def\ulineZ{\underline{Z}}

\def\ulinea{\underline{a}}

\def\ulined{\underline{d}}
\def\ulinee{\underline{e}}

\def\ulinem{\underline{m}}

\def\ulineu{\underline{u}}
\def\ulinev{\underline{v}}

\def\ulinex{\underline{x}}

\def\ulinez{\underline{z}}

\def\ulinemu{\underline{\mu}}
\def\ulinetau{\underline{\tau}}
\def\CalC{\mathcal{C}}
\def\CalD{\mathcal{D}}

\def\CalF{\mathcal{F}}

\def\CalH{\mathcal{H}}

\def\CalM{\mathcal{M}}
\def\CalN{\mathcal{N}}

\def\CalP{\mathcal{P}}

\def\CalT{\mathcal{T}}
\def\CalU{\mathcal{U}}
\def\CalV{\mathcal{V}}

\def\CalX{\mathcal{X}}

\def\ulineCalM{\underline{\mathcal{M}}}

\def\ulineCalU{\underline{\mathcal{U}}}
\def\ulineCalV{\underline{\mathcal{V}}}

\def\ulineCalX{\underline{\mathcal{X}}}

\def\parsec{\par\noindent}
\def\med{\medskip\parsec}

\def\define{\mathrel{\ensurestackMath{\stackon[1pt]{=}{\scriptstyle\Delta}}}}

\def\olineB{\overline{B}}

\def\olineG{\overline{G}}

\def\olineZ{\overline{Z}}

\def\ScrC{\mathscr{C}}
\def\ScrD{\mathscr{D}}

\def\ScrQ{\mathscr{Q}}
\def\ScrR{\mathscr{R}}

\def\sfb{\mathsf{b}}

\def\sfs{\mathsf{s}}

\def\tildea{\tilde{a}}

\def\tildez{\tilde{z}}

\newtheorem{Notation}{Notation}
\newtheorem{remark}{Remark}
\newtheorem{theorem}{Theorem}

\newtheorem{definition}{Definition}
\newtheorem{lemma}{Lemma}
\newtheorem{example}{Example}
\newtheorem{proposition}{Proposition}
\newtheorem{fact}{Fact}

\def\span{\mbox{span}}

\newcommand{\colBlue}[1]{\textcolor{blue}{#1}}
\newcommand{\colRed}[1]{\textcolor{red}{#1}}

\newcommand{\colGreen}[1]{\textcolor{darkgreen}{#1}}
\usepackage{float}
\usepackage{accents}

\usepackage{caption}

\mathchardef\mhyphen="2D
\def\olinekappa{\overline{\kappa}}
\def\3To1BC{$3-$to$-1$}

\def\dbrackthree{\llbracket 3 \rrbracket}

\def\SemiPrivateRVSet{\CalU}

\def\Prime{\upsilon}

\allowdisplaybreaks
\begin{document}

\sloppy
\title{\huge Simultaneous Decoding of Classical Coset Codes over $3-$User
Quantum Interference Channel : New Achievable Rate Regions}

\author{
\IEEEauthorblockN{Fatma Gouiaa and Arun Padakandla\\
\vspace{-0.15in}
}
}
\maketitle

\thispagestyle{empty}

\begin{abstract}
\noindent{We undertake a Shannon theoretic study of the problem of communicating bit streams over a 3-user classical-quantum interference channel ($3-$CQIC) and focus on characterizing inner bounds. We design a new coding strategy based on (i) coset codes possessing algebraic closure properties and (ii) decoding POVMs to decode bi-variate interference efficiently. Needing to perform simultaneous decoding, we enhance Sen's powerful technique of tilting, smoothing, and augmentation - originally designed only for IID codes - to decode into `functions of codebooks'. Developing analysis techniques to combine all of these elements, we derive a new inner bound to the capacity region of a $3-$CQIC. The derived inner bound subsumes all currently known inner bounds and is analytically proven to be strictly larger for identified examples including non-commutative `additive' and `non-additive' ones.}
\end{abstract}

\vspace{-0.1in}
\section{Introduction}
\label{Sec:Introduction}

\begin{figure}
\vspace{-0.15in}
\begin{center}
\includegraphics[width=3.6in]{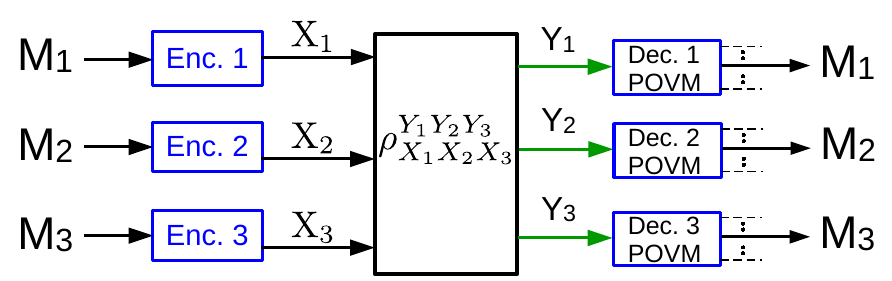}
\vspace{0.1in}
\caption{Three bit streams have to be communicated from three distributed Txs to three distributed Rxs over a $3-$CQIC.}
    \label{Fig:3CQIC}
    \end{center}
   \vspace{-0.15in}
\end{figure}

{Consider a $3-$user classical-quantum interference channel ($3-$CQIC) shared amidst three distributed transmitter-receiver (Tx-Rx) pairs depicted in Fig.~\ref{Fig:3CQIC}. The three Txs wish to communicate their bit streams reliably to their corresponding Rxs. We undertake a Shannon-theoretic study, and the problem of interest is to characterize the capacity region of the $3-$CQIC. Our specific focus is on designing coding strategies, analyzing their performance, and deriving inner bounds to the capacity of the $3-$CQIC. Let us begin by discussing the current known best coding strategy and the corresponding largest known inner bound.}

{The best known coding strategy and the largest known inner bound to the capacity {region} of the $3-$CQIC is based on our knowledge in regards to the $2-$user version, the $2-$CQIC. For any $2-$user interference channel (IC), classical or quantum, the best known coding strategy is based on Han and Kobayashi's technique \cite{198101TIT_HanKob} of message splitting via superposition coding, designed in the context of a $2-$user classical IC. Adopting their \cite{198101TIT_HanKob} strategy, Sen, in his earlier work \cite{201207ISIT_Sen}, proposed sequential decoding, designed the necessary decoding POVMs, proposed a novel non-commutative union bound \cite{202201SOSA_OdoVen, 201207ISIT_Sen} and proved achievability of the current known largest inner bound for the capacity of a $2-$CQIC in the asymptotic regime. More recently, inventing new techniques of \textit{tilting smoothing and augmentation} (TSA) \cite{201806arXiv_Sen} - a topic we shall dwell on later - Sen \cite{202103SAD_Sen} has designed new simultaneous decoding POVMs, analyzed its performance to derive an inner bound \cite{202102SAD_Sen} in the far more challenging and general one-shot regime. Sen's inner bounds in \cite{201207ISIT_Sen} and \cite{202102SAD_Sen} are indeed the classical-quantum (CQ) generalization of Han and Kobayashi's inner bound \cite{198101TIT_HanKob}. In summary, the best known coding strategy for a $2-$CQIC is message splitting via superposition coding \cite{198101TIT_HanKob} and the corresponding inner bounds are proven achievable by Sen's simultaneous \cite{202103SAD_Sen} and sequential decoding \cite{201207ISIT_Sen}. When we graduate to the $3-$CQIC, each Tx can also employ Marton's precoding \cite{197905TIT_Mar} to optimize multiplexing of different message parts to enable efficient interference decoding. Thus, one can combine, at least in principle, the techniques of \cite{198101TIT_HanKob}, \cite{197905TIT_Mar}, Sen's simultaneous decoding POVMs \cite{202103SAD_Sen}, his analysis steps \cite{202103SAD_Sen} with convex split \cite{201001TIT_HarJaiMcaRad,201709PRL_AnsDevVamJan,201909TIT_AnsJaiWar,201709QIP_Wil,202511TIT_CheGao} - the one-shot version of covering \cite{201512TIT_SavWil} to analyze precoding - and prove achievability of a natural generalization of Sen's inner bound to the $3-$CQIC both in the one-shot and asymptotic regime. Is this natural generalization of Sen's inner bound optimal for a general $3-$CQIC?
}

{One common thread connecting all the above strategies and, more broadly, information-theoretic studies is the all-pervasive use of (unstructured) \textit{independent and identically distributed} (IID) codes. Specifically, in designing coding strategies and proving inner bounds, the multiple codes employed in the strategy and each of their individual codewords are picked IID. Since the highly probable codes that dominate the average performance are \textit{bereft of any structural properties}, except for mere empirical ones, underlying this all-pervasive approach is an assumption that joint relationships or richer structural properties are redundant to achieve optimal throughput over multi-user q-channels. Another aspect is that studies in multi-user q-channels are focused on $2-$user channels. This restriction is possibly motivated by a belief that optimal coding strategies for the latter naturally generalize to yield optimal strategies for three or more user channels, the latter offering no new degrees of freedom.}

{In this work, we step beyond this conventional approach and demonstrate that multi-user q-channels possess new degrees of freedom that cannot be exploited via unstructured IID codes. In our first step (Sec.~\ref{Sec:3to1CQICAchivablerateregion}), we put forth a coding strategy based on jointly designed coset codes, wherein two constituent codes possess tightly coupled algebraic closure properties. See Fig.~\ref{Fig:FigCodeStructureCQIC(StepI)}. Crucially, we design a new decoding POVM that can decode bi-variate interference more efficiently than unstructured IID codes. To accomplish this, our new decoding POVM cannot decode into codes that are designed, but `functions of codebooks'. As we elaborate in due course, we face a specific challenge of simultaneous decoding not encountered and/or resolved by Sen's work. Resorting to conditional typical decoding POVMs, designing new tilting maps \cite{202103SAD_Sen}, we derive a new inner bound (Thm.~\ref{Thm:3CQICStageIRateRegion}). Identifying specific instances (Exs.~\ref{Ex:3CQICadditivenoncommutative}, \ref{Ex:nonadditiveinterference}), we analytically prove (Prop.~\ref{Prop:NoncomAddCstCds}, \ref{Prop:AchievabilityofcosetCdsnonadditiveexample}) that our derived inner bound is strictly larger than that achievable via any unstructured IID code based strategy.

Having thus established the need to go beyond unstructured IID codes and incorporate coset codes, we recognize (Rem.~\ref{Rem:nonadditiveexample}) that the latter does not substitute for the former. We therefore design a general coding strategy comprising of two layers (i) an unstructured IID code layer and (ii) a coset code layer with a specific inter-relationships (Fig.~\ref{Fig:3CQICStepIIICodeStructure}) between the codes. Crucially, we combine this with a decoding POVM that can \textit{simultaneously decode} into a \textit{combination of unstructured IID codes and `functions of codebooks'} as mentioned earlier. Characterizing its performance we derive a new inner bound (Thm.~\ref{Thm:3CQICStageIRateRegion}, \ref{Thm:3CQICStageIIRateRegion}) for the $3-$CQIC in the asymptotic regime in Sec.~\ref{Sec:SimultDecOfCosetcds3CQIC}. The latter inner bound subsumes the current known largest based on unstructured IID codes (Rem.~\ref{Rem:afterstep3}), and as demonstrated via Props.~\ref{Prop:NoncomAddCstCds}, \ref{Prop:AchievabilityofcosetCdsnonadditiveexample} strictly enlarges for specific $3-$CQICs. Thms.~\ref{Thm:3CQICStageIRateRegion}, \ref{Thm:3CQICStageIIRateRegion} can be naturally generalized to yield analogous inner bounds for bit carrying capacity of $3-$QICs {(Rem.~\ref{Rem:ExtInnerBnd})}. These are our \textbf{main contributions}.

{The core of our idea lies in the recognition that Rxs over a $3-$CQIC suffer from \textit{bi-variate} interferences owing to the presence of \underline{two} interfering transmissions that each Rx has to contend with. This motivates Rxs operating on a $3-$CQIC to decode \textit{bi-variate} interferences, or more generally bi-variate functions, \textit{efficiently}. As demonstrated in an earlier work \cite{202606ISIT_GouPad}, coding strategies based on unstructured IID codes are inherently incapable of efficiently decoding bi-variate functions. Jointly designed coset codes possessing \textit{inter-} and \textit{intra-}structural properties coupled with new decoding POVMs that exploit these algebraic structures outperform unstructured IID codes, as also evidenced in a series of classical works \cite{198101TIT_HanKob,202010TIT_SenLimKim,200710TIT_NazGas,201603TIT_PadSahPra,201804TIT_PadPra}.}

{It is evident from past research \cite{197201PPI_Hol,199801TIT_Hol,199707PhyRev_SchWes,200107TIT_Win,201206TIT_FawHaySavSenWil,201207ISIT_Sen,202103SAD_Sen,202102SAD_Sen} that lifting up coding techniques from a classical channel to a general q-channel, analyzing performance and characterizing corresponding inner bounds for the latter is strewn with several challenges. It took over two decades to elevate solution for the capacity of the point-to-point (PTP) pure state CQ channel \cite{197201PPI_Hol} to the general PTP CQ channel \cite{199801TIT_Hol,199707PhyRev_SchWes}. The problem of designing and analyzing simultaneous decoding is an important case in point. While this is an exercise in a classical information theory graduate course, lifting this up to design and analyze a quantum simultaneous decoder, even in the asymptotic regime, took over fifteen years \cite{202103SAD_Sen,202102SAD_Sen,201806arXiv_Sen} since Winter's recognition of the challenge \cite[Sec.~VIII]{200107TIT_Win}. Evidently, this pursuit has required, and resulted in new ideas of TSA \cite{202102SAD_Sen,201806arXiv_Sen,202103SAD_Sen}. Another example is that of binning \cite{197905TIT_Mar}. The challenge of analyzing binning has resulted in mis-steps \cite{201207ISIT_Sen} (see \cite[Sec.~I]{201512TIT_SavWil}) resulting in Savov and Wilde's technique of overcounting \cite{201512TIT_SavWil}. In fact, owing to the latter's limitations \cite{202408SalHayHay}, this challenge remains unresolved.

{Designing and analyzing an efficient coding strategy based on coset codes for communicating over a general $3-$CQIC entails addressing all of the above challenges in a more involved setting with jointly designed coset codes. Firstly, exploiting jointly designed codes requires decoding the bi-variate function efficiently. This entails decoding the function output, while being oblivious to its arguments. To elaborate, the decoding POVM must decode into a new collection of sequences (Rem.~\ref{Rem:CrucialChangeInTheDecoding}), obtained by applying the bi-variate function to all pairs of the original two coset codes, single-letter wise. The decoding POVM must therefore be designed to decode into `functions of codebooks' - an aspect unaddressed in previous works involving unstructured IID codes, including \cite{202103SAD_Sen,202102SAD_Sen}. The subspaces into which tilting must be performed, and the tilting maps themselves need to be respecified to enable bi-variate function decoding. Secondly, as mentioned earlier and elaborated in Rem.~\ref{Rem:nonadditiveexample}, we are further required to design and analyze a decoder POVM that simultaneously decodes into a combination of 'functions of coset codes' and unstructured IID codes. Thirdly, coset code strategies invariably entail binning. {While use of unstructured IID codes lets us choose codewords IID from any target distribution, imposing algebraic closure forces codewords to be only \textit{pairwise independent} and \textit{uniformly distributed}. Therefore, we have to employ binning \cite{197905TIT_Mar} to sieve out codewords whose empirical distribution matches the target distribution.}  As elaborated in \cite[Sec.~3 prior to Defn.~1]{202408SalHayHay}, the overcounting technique is not amenable in a three receiver scenario.} {Taking this opportunity, we adopt a likelihood encoder to analyze binning. This enables us to seamlessly separate the analysis of binning and TSA at the decoder (Rem.~\ref{Rem:BinningandChannelCodingAnalysis}). In conjunction with these, we identify the necessary analytical tools to combine these different elements in a manner that facilitates an information-theoretic study. Fourthly, characterizing inner bounds corresponding to arbitrary finite fields and generic single-letter distributions is a key aspect of our work. Our study of the non-commuting non-additive $3-$CQIC (Ex.~\ref{Ex:nonadditiveinterference}) culminating in Prop.~\ref{Prop:AchievabilityofcosetCdsnonadditiveexample} portrays the utility and need for a broader theory involving structured codes possessing algebraic closure properties. Indeed, while the additive structure of Ex.~\ref{Ex:3CQICadditivenoncommutative} might lead one to conclude rate gains are only for additive scenarios, Prop.~\ref{Prop:AchievabilityofcosetCdsnonadditiveexample} puts such thoughts to rest.}

Our work builds on considerable prior findings\cite{200107TIT_Win,201512TIT_SavWil,201206TIT_FawHaySavSenWil,201902TIT_AnsJaiWar,201901JMP_AnsJaiWar,202306TIT_LedLeuSidSmiSmo}. Holevo \cite{199801TIT_Hol}, Schumacher and Westmoreland \cite{199707PhyRev_SchWes} laid the first foundation stones by characterizing the capacity of a PTP CQ channel. Subsequent notions of typical subspaces and powerful operator inequalities \cite{200307TIT_HayNag} set the stage to lift up classical coding techniques to multi-terminal q-channels. Reporting challenges with a quantum simultaneous decoder, Winter \cite{200107TIT_Win} formulated notions of a \textit{gentle} (or \textit{tender} \textit{measurement}) to characterize the capacity of a CQ multiple access channel (CQMAC) \cite{200107TIT_Win}.  Centrality of a simultaneous decoder for network q-channels, particularly for one-shot studies, the reported challenges spurred much investigation. Fawzi et.~al.'s~\cite{201206TIT_FawHaySavSenWil} two Rx simultaneous decoder and Sen's \cite{201207ISIT_Sen} sequential decoder contributed alternate approaches. Solving a long-standing challenge by inventing techniques of TSA, Sen's findings \cite{202103SAD_Sen,201806arXiv_Sen} have enabled him to prove the achievability of natural CQ analogues of best-known inner bounds for several multi-terminal q-channels \cite{202102SAD_Sen} in the general one-shot regime. Note that our work focuses exclusively on an asymptotic study. This is owed to the fact that linear codes of block-length $1$ are trivial, and one-shot coset code studies entail a new approach.

We briefly state the significance of our findings. Unstructured IID code based strategies form the defacto coding strategies \cite{200107TIT_Win,201512TIT_SavWil,201206TIT_FawHaySavSenWil,202102SAD_Sen,201902TIT_AnsJaiWar,201604JPA_BocCaiNot} in network quantum information theory (QIT). In this work, we step beyond this conventional approach and design new strategies based on jointly designed coset codes, derive new inner bounds that subsume all current known unstructured IID code based inner bounds and strictly enlarge upon the same for identified examples. A key challenge is to design and analyze simultaneous decoding of coset and unstructured IID codes. While Sen \cite{202103SAD_Sen,202102SAD_Sen} has proposed a new solution to perform simultaneous decoding, his techniques remain to be new, have not yet been employed more broadly, and is restricted to simultaneous decoding of unstructured IID codes. Our contributions must be viewed within this context. We enhance Sen's technique to performs simultaneous decoding of coset and unstructured IID codes. These findings, in conjunction with our study of $3-$CQBCs \cite{202503arXivBC_GouPad}, maybe viewed as building a new framework of coset code based strategies for network QIT.

\med\textbf{Organization} : {Since we are combining techniques that are either relatively new \cite{202103SAD_Sen}, or outside mainstream quantum information theory - such as coset codes - we present our findings in three pedagogical steps to facilitate comprehension. Following preliminaries in Sec.~\ref{Sec:Preliminaries}, we elaborate on the main ideas underlying coset codes by designing our coset coding strategies in the context of three carefully chosen examples in Sec.~\ref{Sec:Roleofcosetcodesandstrictsuboptimalityforthe3CQIC}. Building on Rem.~\ref{Rem:nonadditiveexample}, we present our first step that combines all elements in Sec.~\ref{Sec:3to1CQICAchivablerateregion}. Owing to a fewer number of codes, we are able to detail all elements of our proof in this step. The characterized inner bound therein is also proven to be strictly larger for identified non-commutative additive and non-additive examples. Step II in Sec.~\ref{SubSec:StageICodeStrategy3CQIC} presents the full suite of coset codes used to tackle all bi-variate interferences and a corresponding inner bound (Thm.~\ref{Thm:3CQICStageIRateRegion}). Finally, as needed, we enlarge the inner bound presented in Step II by including IID unstructured codes in Step III to characterize the final desired inner bound in Thm.~\ref{Thm:3CQICStageIIRateRegion}. Throughout, we have stressed on detailed and clearly justified proofs.}

\section{Preliminaries and Problem Statement}
\label{Sec:Preliminaries}
We adopt standard notation, specifically as in \cite{BkWilde_2017}, supplemented with the following. For $K\in \mathbb{N}$, $[K] \define \left\{1,\ldots,K \right\}$. For prime $\Prime \in \integers$, $\CalF_{\Prime}$ will denote a finite field of size $\Prime$ with $\oplus$ denoting field addition in $\CalF_{\Prime}$ (i.e., mod$-\Prime$). For a Hilbert space $\CalH$, $\mathcal{L}(\CalH),\CalP(\CalH)$ and $\CalD(\CalH)$ denote the collection of linear, positive and density operators acting on $\CalH$ respectively.

We let an \underline{underline} denote an appropriate aggregation of objects. For example, $\ulineCalX \define \CalX_{1}\times \CalX_{2} \times \CalX_{3}$, $\ulinex \define (x_{1},x_{2},x_{3}) \in \ulineCalX$ and in regards to Hilbert spaces $\CalH_{Y_{i}}: i \in [3]$, we let $\CalH_{\ulineY} \define \otimes_{i=1}^{3}\CalH_{Y_{i}}$. Let $*$ denote the binary convolution: $p * q \coloneqq p(1-q) + (1-p)q$. For $p \in [0,1]$, $h_b(p) \define -p\log_2 p -(1-p)\log_2(1-p)$ denote the binary entropy function. For $A \in \mathcal{L}(\mathcal{H})$, $\|A\|_{1}$ denotes the trace norm and $\|A\|_{\infty}$ denotes the operator norm. A positive operator valued measurement (POVM) is a collection of positive operators that sum to the identity operator. When we say a POVM $\theta_{\mathcal{A}} \define \{\theta_a : a \in \mathcal{A}\}$, we assume that $\theta_a$ is positive for $a \in \mathcal{A}$ and $\sum_{a \in \mathcal{A}} \theta_a=I$ where $\mathcal{A}$ is a finite set. We abbreviate probability mass function, conditional typical projector, and unconditional typical projector as PMF, C-Typ-Proj and U-Typ-Proj respectively.

Consider a (generic) \textit{$3-$CQIC} $(\rho_{\ulinex} \in \CalD(\CalH_{\ulineY}): \ulinex \in \ulineCalX,\kappa_{j}:j \in [3])$ specified through (i) three finite sets $\CalX_{j}: j \in [3]$, (ii) three Hilbert spaces $\CalH_{Y_{j}}: j \in [3]$, (iii) a collection $( \rho_{\ulinex} \in \CalD(\CalH_{\ulineY} ): \ulinex \in \ulineCalX )$ of density operators and (iv) three non-negative valued cost functions $\kappa_{j}:\CalX_{j} \rightarrow [0,\infty) : j \in [3]$. The cost function is assumed to be additive, i.e., cost expended by Tx $j$ in preparing the state $\otimes_{t=1}^{n}\rho_{x_{1t}x_{2t}x_{3t}}$ is $\olinekappa_{j}^{n} \define \frac{1}{n}\sum_{t=1}^{n}\kappa_{j}(x_{jt})$. Reliable communication on a $3-$CQIC entails identifying a code.
\begin{definition}
A \textit{$3-$CQIC code} $c=(n,\ulineCalM,\ulinee,\uline{\mu})$ of block-length $n$ consists of three (i) message index sets $\CalM_{j}: j \in [3]$, (ii) encoder maps $e_{j}: \CalM_{j} \rightarrow \CalX_{j}^{n}: j \in [3]$ and (iii) POVMs $\mu_{j} \define \{ \mu_{j,m_j}: \CalH_{j}^{\otimes n} \rightarrow \CalH_{j}^{\otimes n} : m_j \in \CalM_{j} \} : j \in [3]$. The average error probability of the $3-$CQIC code $(n,\ulineCalM,\ulinee,\ulinemu)$ is
\begin{eqnarray}
 \label{Eqn3CQIC:AvgErrorProb}
 \mathbf{P}(\ulinee,\ulinemu) \define 1-\frac{1}{|\CalM_{1}||\CalM_{2}||\CalM_{3}|}\sum_{\ulinem \in \ulineCalM}\tr\left\{\left(\mu_{1,m_1} \otimes \mu_{2,m_2} \otimes \mu_{3,m_3} \right) \left( \otimes_{t=1}^n \rho_{x_{1t}(m_1)x_{1t}(m_2)x_{1t}(m_3)}\right) \right\}.
 \nonumber
\end{eqnarray}
where $\ulinemu \define \otimes_{j=1}^3 \mu_j$ and $(x_{jt}(m_j) : 1\leq t \leq n) = x_{j}^{n}(m_{j}) \define e_{j}(m_{j})$ for $j \in [3]$. The average cost per symbol of transmitting message $\ulinem \in \ulineCalM$ is
$\uline{\tau}(\uline{e}|\ulinem) \define \left( \olinekappa_{j}^{n}(e_{j}(m_{j})): j \in [3] \right)$, and the average cost per symbol of $3-$CQIC code is $\ulinetau(\ulinee) \define \frac{1}{|\ulineCalM|}\sum_{\ulinem \in \ulineCalM}\ulinetau(\ulinee|\ulinem)$.
\end{definition}
\begin{definition}
A rate-cost vector $(R_{1},R_{2},R_{3},\tau_{1},\tau_{2},\tau_{3}) \in [0,\infty)^{6}$ is \textit{achievable} if there exists a sequence of $3-$CQIC code $(n,\ulineCalM^{(n)},\ulinee^{(n)},\uline{\mu}^{(n)})$ for which $\displaystyle\lim_{n \rightarrow \infty}\mathbf{P}(\ulinee^{(n)},\uline{\mu}^{(n)}) = 0$,
\begin{eqnarray}
 \label{Eqn3CQIC:3CQICAchievability}
 \lim_{n \rightarrow \infty} \frac{1}{n}\log |\CalM_{j}^{(n)}| = R_{j}, \mbox{ and }\lim_{n \rightarrow \infty} \ulinetau(\ulinee)_{j} \leq \tau_{j} :j \in [3].
 \nonumber
\end{eqnarray}
The capacity-cost region $\mathscr{C}$ of the $3-$CQIC is the set of all achievable rate-cost vectors and $\mathscr{C}(\ulinetau) \define \{ \ulineR:(\ulineR,\ulinetau) \in \mathscr{C}\}$.
\end{definition}
The novelty of our proposed coding strategy is based on \textit{coset codes} - an ensemble of codes possessing algebraic closure property. Specifically, we employ \textit{nested coset codes} (NCC) which we define below.
\begin{definition}
\label{Def:CosetCds}
    An $(n,k,g,b^n)$ coset code built over a finite field $\CalF_{\Prime}$ is specified via (i) a generator matrix $g \in \CalF_{\Prime}^{k \times n }$ and (ii) a bias $b^n \in \CalF_{\Prime}^n$. We let $u^n(a) \define a g \oplus b^n$ for $a \in \CalF_{\Prime}^k$ denote the codewords of the coset code $(n,k,g,b^n)$, where as mentioned above, $\oplus$ denotes field addition in $\mathcal{F}_{\Prime}$. The rate of this code is $\frac{k}{n} \log(\Prime)$.
\end{definition}
\begin{definition}
\label{Defn:NestedCosetCode}
An $(n,k,l,g_{I},g_{O/I}, b^n)$ NCC over $\CalF_{\Prime}$  consists of (i) generator matrices $ g_{I} \in \CalF_{\Prime}^{k\times n}$ and $ g_{O/I} \in \CalF_{\Prime}^{l \times n}$, and
(ii) bias vector $ b^n \in \CalF_{\Prime}^n$. We let $u^n(a,m) \define a g_{I}\oplus m g_{O/I} \oplus b^n$, for $(a,m) \in \CalF_{\Prime}^k \times \CalF_{\Prime}^l$ denote codewords of the coset code. The collection $\left(u^{n}(a,m)\in \CalF_{\Prime}^{n} : a \in \CalF_{\Prime}^{k}\right)$ is referred to as the coset corresponding to message $m \in \CalF_{\Prime}^{l}$.
\end{definition}
\section{Strict suboptimality of unstructured IID codes for non-commutative $3-$CQICs}
\label{Sec:Roleofcosetcodesandstrictsuboptimalityforthe3CQIC}
Our goal here is to \textit{explain} through a set of examples \textit{how and why} coset codes can yield strictly higher rates over $3-$CQICs. 
\begin{comment}{
The examples are carefully chosen so as to illustrate the advantages 
Building from a commutative Ex.~\ref{Ex:3CQICRoleOfCosetCds} and an associated discussion, we put forth a non-commutative Ex.~\ref{Ex:3CQICadditivenoncommutative} for which we demonstrate that a coset code strategy involving a bi-variate interference decoding can yield strictly larger rate regions.
}\end{comment}
Our first non-commutative example - Ex.~\ref{Ex:3CQICadditivenoncommutative} - is an 'additive' one and illustrates how coset codes can enable optimal decoding of a bi-variate additive interference. Our second non-commutative Ex.~\ref{Ex:nonadditiveinterference} establishes that higher throughput can be obtained even for non-`additive' $3-$CQICs. In addition to portraying rate gains for `non-additive' $3-$CQICs,  Ex.~\ref{Ex:nonadditiveinterference} serves to guide (Rem.~\ref{Rem:nonadditiveexample}) the structure of a general inner bound using coset codes. For pedagogical reasons we begin with a simple commutative Ex.~\ref{Ex:3CQICRoleOfCosetCds}.
\begin{figure}
\begin{minipage}{0.45\textwidth}
\centering
\includegraphics[width=1.7in]{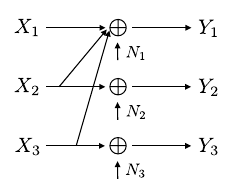}
\caption{Equivalent Classical 3-user IC of Ex. \ref{Ex:3CQICRoleOfCosetCds}}
    \label{Fig:3to1CQICadditive}
\end{minipage}
\hfill
\begin{minipage}{0.5\textwidth}
    \centering
    \vspace{0.3in}
    \resizebox{2.3in}{!}{
        \begin{tabular}{|c|c|c|c|}
            \hline
            $X_2$ & $X_3$ & $X_2 \oplus_3 X_3$ & $X_2 \lor X_3$  \\ \hline
            0 & 0 & 0 & 0 \\ \hline
            0 & 1 & 1 & 1 \\ \hline
            1 & 0 & 1 & 1 \\ \hline
            1 & 1 & 2 & 0 \\ \hline
        \end{tabular}
    }
    %\vspace{0.35in}
    \caption{Ternary addition of binary inputs embedded in the ternary field, with $p(X_2=2)=p(X_3=2)=0$.}
\label{Fig:nonadditiveexample}
\end{minipage}
\vspace{-0.15in}
\end{figure}
\begin{Notation}
\label{notation:examples}
For $x\in \{0,1\}$, $\delta \in (0,\frac{1}{2})$, define $\sigma_{\delta}(x)=(1-\delta)\ketbra{1-x}+\delta\ketbra{x}$. For $\varphi \in (0,\frac{\pi}{2})$, define $\gamma(x) = \ketbra{0} \mathds{1}{\{x=0\}}+\ketbra{v_{\varphi}}\mathds{1}{\{x=1\}}$ where $\ket{v_{\varphi}} = \left[\cos(\varphi)~\sin(\varphi)\right]^{t}$. Let $\mathcal{C}_j(\uline{\tau}) \define \max\{R_j : (R_1,R_2,R_3) \in \mathscr{C}(\uline{\tau})\}$ and $C_j \define \mathcal{C}_j(\infty)$ denote the interference-free capacity of user $j$ with and without input-cost constraints, respectively, for $j \in [3]$.
\end{Notation}

Through each of the Examples \ref{Ex:3CQICRoleOfCosetCds}, \ref{Ex:3CQICadditivenoncommutative} and \ref{Ex:nonadditiveinterference}, we investigate the achievability of the rate triple $(\CalC_1,\CalC_2,\CalC_3)$ under the condition
\begin{eqnarray}
    \CalC_1 + \CalC_2 + \CalC_3 > C_1. \label{Eqn:commonconditionfortheexamples}
\end{eqnarray}
%\vspace{-0.1in}
\begin{example}
    \label{Ex:3CQICRoleOfCosetCds}
    Consider a $3-$CQIC with binary inputs and qubit output spaces, i.e., $\CalX_j=\{0,1\}$ and $\CalH_{Y_j}=\mathbb{C}^2 $ for $ j \in [3]$. The channel is defined as $\rho_{\ulinex} \define  \sigma_{\delta_1}(x_1 \oplus x_2 \oplus x_3) \otimes \sigma_{\delta_2}(x_2) \otimes \sigma_{\delta_3}(x_3)$ for $\ulinex \in \uline{\CalX}$.
Inputs of users 2 and 3 are not constrained, i.e., $\kappa_{j}(0)= \kappa_{j}(1)=0$ for $j=2,3$. User 1's input is constrained with respect to a Hamming cost function, i.e., $\kappa_1(x)=x : x \in\{0,1\}$, to an average cost of $ \tau \in (0,\frac{1}{2})$ per symbol.
\end{example}

For simplicity, let $\delta_2=\delta_3=\delta$. Since $\rho_{\ulinex}$ for $\ulinex \in \mathcal{\ulineX}$ are commuting, Ex.~\ref{Ex:3CQICRoleOfCosetCds} can be identified via a 3-user classical binary IC (Fig.~\ref{Fig:3to1CQICadditive}) with $Y_{1}=X_{1}\oplus X_{2} \oplus X_{3} \oplus N_{1}, Y_{k} = X_{k} \oplus N_{k}$ where $N_{1}\sim$ Ber$(\delta_{1})$, $N_{k}\sim$ Ber$(\delta)$ for $k=2,3$. Input $X_{1}$ is constrained to a Hamming cost $\tau_{1}$, and $X_{2},X_{3}$ are unconstrained. We will argue, as in  that $(\mathcal{C}_1, \mathcal{C}_2, \mathcal{C}_3)$, under the condition \eqref{Eqn:commonconditionfortheexamples}, is not achievable using unstructured IID codes, but is achievable using coset codes as long as $C_1 > \mathcal{C}_1 + \max\{\mathcal{C}_2, \mathcal{C}_3\}$ (note that here $\mathcal{C}_2 = \mathcal{C}_3$). Despite being similar to that in \cite{201603TIT_PadSahPra}, we provide this argument here to facilitate pedagogy.

Let's fix user $k$'s rate $R_{k}=\CalC_{k}$ for $k=2,3$. This forces $X_{2},X_{3}$ to be independent Ber$(\frac{1}{2})$ RVs, implying $I(X_{1};Y_1)=0$. This in turn implies that Rx $1$ \textbf{has to} decode interference, fully or partially, to achieve \textit{any} positive rate. How do unstructured IID codes enable Rx $1$ to decode interference?

The unstructured IID coding strategy entails that Tx $k \in \{2,3\}$ splits its information into $U_{k}$ and $X_{k}$, Rx $1$ decodes $U_{2},U_{3},X_{1}$, while Rx $k \in \{2,3\}$ decodes $U_{k},X_{k}$.
Suppose $p_{X_1} p_{U_2X_2} p_{U_3X_3}$ is the corresponding single-letter test channel induced by such a coding strategy.

For such a test channel, so long as $H(X_{k}|U_{k}) > 0$ for either $k\in \{2,3\}$, it can be shown \cite{201603TIT_PadSahPra} that $H(X_{2} \oplus X_{3} | U_{2}, U_{3}) > 0$, implying that Rx~1 cannot achieve $\CalC_{1}$. Indeed, note that the latter is the interference-free, cost-constrained capacity of Rx~1, which cannot be achieved if Rx~1 ignores the full interference $X_2 \oplus X_3$. If \eqref{Eqn:commonconditionfortheexamples} holds, it can be shown that there exists \textbf{no} PMF $p_{X_{1}}p_{X_{2}U_{2}}p_{X_{3}U_{3}}$ for which $p_{X_{1} X_{2}}= \mbox{Ber}(\frac{1}{2})\otimes \mbox{Ber}(\frac{1}{2})$ \textbf{and} $H(X_{2}|U_{2}) =H(X_{3}|U_{3})= 0$. Thus, \eqref{Eqn:commonconditionfortheexamples} precludes unstructured IID codes from achieving $(\CalC_{1},\CalC_{2},\CalC_{3})$. How can linear codes help under condition \eqref{Eqn:commonconditionfortheexamples}?

Observe that users $2$ and $3$'s channels are simple binary symmetric channels with crossover probability $\delta$ (BSC($\delta$)). Let $\lambda$ be a linear code of rate $1-h_{b}(\delta)$ that achieves capacity of BSC$(\delta)$. By employing \textit{cosets $\lambda_{2},\lambda_{3}$ of the same linear code} $\lambda$ for both users $2$ and $3$, the collection of all interference patterns - all possible sums of user $2$ and $3$'s codewords - is contained in \textit{another coset $\lambda_{2}\oplus \lambda_{3}$ of the same linear code}. Crucially, $\lambda_{2}\oplus \lambda_{3}$ is of the same rate $1-h_{b}(\delta)$. Rx $1$ just decodes into $\lambda_{2}\oplus \lambda_{3}$ and hunts for the interference $X_{2}\oplus X_{3}$. It can recover the \textit{sum of the codewords} even while being oblivious to the \textit{pair}, so long as $\CalC_1 + \max\{\CalC_2, \CalC_3\} = \CalC_1 + 1-h_{b}(\delta) < C_1$ holds.

Now, suppose $\delta_{2} \neq \delta_{3}$, we would have chosen the smaller code as the sub-coset of the larger code, thereby containing all possible mod$-2$ sums of the codewords within a coset of the larger code of rate $\max\{\CalC_{2},\CalC_{3}\}$. Then, \eqref{Eqn:commonconditionfortheexamples} does \textbf{not preclude} linear codes from achieving $(\CalC_{1},\CalC_{2},\CalC_{3})$. In summary, we conclude the following, whose formal proof can be found in \cite{201603TIT_PadSahPra}.

\begin{proposition}
\label{Prop:RoleofCosetCodes}
    Consider Ex.~\ref{Ex:3CQICRoleOfCosetCds} with
    $\tau, \delta_1, \delta_2, \delta_3$ satisfying \eqref{Eqn:commonconditionfortheexamples}. $(\CalC_1,\CalC_2,\CalC_3)$ is not achievable using any known unstructured IID coding strategy but is achievable using a coset codes strategy, as long as $C_1 > \CalC_1 + \max\{\CalC_2,\CalC_3\}$.
    \end{proposition}

\begin{example}
 \label{Ex:3CQICadditivenoncommutative}
Consider a $3-$CQIC with binary input sets and qubit output spaces, that is $\CalX_j=\{0,1\}$ and $\CalH_{Y_j}=\mathbb{C}^2 $ for $ j \in [3]$. The channel is defined as $\rho_{\ulinex} \define  \gamma(x_1 \oplus x_2 \oplus x_3) \otimes \sigma_{\delta_2}(x_2) \otimes \sigma_{\delta_3}(x_3)$, for $\ulinex \in \uline{\CalX}$.
Inputs of users 2 and 3 are not constrained, i.e., $\kappa_{j}(0)= \kappa_{j}(1)=0$ for $j=2,3$. User 1's input is constrained with respect to a Hamming cost function, i.e., $\kappa_1(x)=x : x \in\{0,1\}$, to an average cost of $ \tau \in (0,\frac{1}{2})$ per symbol.
\end{example}
Ex.~\ref{Ex:3CQICadditivenoncommutative} is a non-commutative one. Indeed, the states $\gamma(0),\gamma(1)$ do not share a common diagonalization basis. Let us first bound the above capabilities of an unstructured IID code based strategy.
\begin{proposition}
\label{Prop:strictsuboptimalityiidcodesforadditiveinterference}
    Consider the $3-$CQIC in Ex.~\ref{Ex:3CQICadditivenoncommutative} under condition \eqref{Eqn:commonconditionfortheexamples}. $(\CalC_1,\CalC_2,\CalC_3)$ is not achievable using any known unstructured coding strategy.
\end{proposition}
\begin{proof}
A similar train of arguments as used for Ex.~\ref{Ex:3CQICRoleOfCosetCds} will enable us to prove this statement. See Appendix \ref{App:additiveinterference} for a proof.
\end{proof} 

Our prior discussion in regards to Ex.~\ref{Ex:3CQICRoleOfCosetCds} should indicate to the reader that we could obtain rate gains if Tx $2$ and $3$ employ coset codes corresponding to the uniform distribution and Rx $1$ employs a POVM that enables it to decode the sum of the codewords chosen by Tx $2$ and $3$. We refer the reader to Prop.~\ref{Prop:NoncomAddCstCds} where it is formally stated and proven that $(\CalC_1,\CalC_2,\CalC_3)$ is achievable using a coset code strategy. Prop.~\ref{Prop:NoncomAddCstCds} in conjunction with Prop.~\ref{Prop:strictsuboptimalityiidcodesforadditiveinterference} therefore establishes that coset code strategies can strictly outperform unstructured IID codes over non-commutative $3-$CQICs. We now go one more step.

\begin{comment}{
\begin{proposition}
\label{Prop:NoncomAddCstCds}
    Consider the $3-$CQIC in Ex.~\ref{Ex:3CQICadditivenoncommutative} under condition \eqref{Eqn:commonconditionfortheexamples}. The rate triple $(\CalC_1,\CalC_2,\CalC_3)$ is achievable using coset codes as long as $C_1> \mathcal{C}_1 + \max\{\mathcal{C}_2,\mathcal{C}_3\} $.
\end{proposition}
}\end{comment}

Observe that establishing the above rate gains required our coset code strategy to achieve rates corresponding to uniform distribution induced on finite fields. This requirement aligns well with conventional information-theoretic proof techniques since a randomly chosen coset code (or linear code) induces uniform distribution on the underlying finite field and hence is capable of achieving rates corresponding to the uniform distribution. A general coset code based strategy must, at the very least, enable a Rx to decode the sum of codewords from coset codes corresponding to the uniform distribution. Does it suffice to achieve rates corresponding to the uniform distribution on the finite fields? The following example, in conjunction with the ensuing discussion demonstrates achieving rates corresponding to non-uniform distributions induced over finite fields and designing POVMs to decode the sum of chosen codewords can enable us to achieve rate gains over a much larger class of $3-$CQICs specifically even over `non-additive' $3-$CQICs.

\begin{comment}{Having proven unstructured IID codes do not achieve $(\CalC_1,\CalC_2,\CalC_3)$ under condition \eqref{Eqn:commonconditionfortheexamples}, is this rate triple achievable? In Sec.~\ref{Sec:3to1CQICAchivablerateregion}, we design a coding strategy based on coset codes, characterize performance, and derive an inner bound. We then show in Prop.~\ref{Prop:NoncomAddCstCds} that the rate triple $(\CalC_1,\CalC_2,\CalC_3)$ is achievable using the designed coding strategy. Prop.~\ref{Prop:NoncomAddCstCds} therefore establishes that coset codes can yield strictly higher rates. We now go further and enquire whether coset codes can yield higher throughput even if the interference is `non-additive'? In the following example, we consider an `extreme non-additive'\footnote{Since there are only two non-trivial bi-variate functions with binary inputs - mod$-2$ addition and logical OR - with the first being additive, the second may be referred to as an `extreme' non-additive function. Since Ex.~\ref{Ex:nonadditiveinterference} establishes gains for the logical OR, we refer to it as an `extreme non-additive' scenario.} scenario and using Prop.~\ref{Prop:AchievabilityofcosetCdsnonadditiveexample} we prove that coset codes can yield higher throughput even if the interference is non-additive.}\end{comment}

\begin{example}
\label{Ex:nonadditiveinterference}
    Consider a $3-$CQIC with binary inputs and qubit output spaces, i.e., $\CalX_j=\{0,1\}$ and $\CalH_{Y_j}= \mathbb{C}^2$ for $j \in [3]$. The channel is defined as $\rho_{\ulinex} \define \gamma\left(x_1 \oplus (x_2 \lor x_3)\right) \otimes \sigma_{\delta_2}(x_2) \otimes \sigma_{\delta_3}(x_3)$ for $\ulinex \in \uline{\CalX}$, where $\lor$ denotes the logical OR. User j's input is constrained with respect to a Hamming cost function, i.e., $\kappa_j(x)=x : x\in \{0,1\}$, to an average cost of $\tau_j \in (0,\frac{1}{2})$ for $j \in [3]$ per symbol.
    \end{example}

In addition to being non-commutative, the above $3-$CQIC is also `non-additive'. Can coset code strategies yield higher rates? Let us once again begin by bounding on the above capabilities of any unstructured coding based strategy.
\begin{proposition}
\label{Prop:nonadditiveinterference}
    Consider the $3-$CQIC in Ex.~\ref{Ex:nonadditiveinterference}
    under condition \eqref{Eqn:commonconditionfortheexamples}. $(\CalC_1,\CalC_2,\CalC_3)$ is not achievable using any known unstructured coding strategy.
    \end{proposition}
\begin{proof}
    See Appendix \ref{App:Nonadditive3to1CQIC} for a proof.
\end{proof}
\vspace{-0.05in}
Can coset codes achieve this rate triple $(\mathcal{C}_1,\mathcal{C}_2,\mathcal{C}_3)$? Instead of viewing the input alphabets of users $2$ and $3$ to be binary, suppose we (i) consider $p_{X_{2}}(0)=p_{X_{3}}(0)=p_{X_{2}}(1)=p_{X_{3}}(1)=\frac{1}{2}$ and $p_{X_{2}}(2)=p_{X_{3}}(2)=0$ with input alphabets $\CalX_{2}=\CalX_{3} = \{0,1,2\}$ of users $2$ and $3$ `viewed' as the ternary field, (ii) build coset codes over the ternary field for users $2$ and $3$ corresponding to the above non-uniform distribution while ensuring no codewords possess the symbol $2$, and enable (iii) Rx $1$ to decode the ternary field sum of the chosen `ternary' codewords. Then note that, since $H(X_2 \lor X_3 | X_2 \oplus_3 X_3)=0$ (see Fig.~\ref{Fig:nonadditiveexample}), Rx 1 can recover the full interference. Indeed, the recovered ternary addition of the chosen codewords can be mapped to the interference pattern - the logical OR of the chosen codewords. Since $p_{X_{2}}(2)=p_{X_{3}}(2)=0$, if the codewords of the code induce this chosen non-uniform PMF they should be able to input the same on the $3-$CQICs. We are led to the following.

\begin{comment}{Observe that, if we view $X_2$, $X_3$ as living on the ternary fields and enable Rx 1 to decode $X_2 \oplus_3 X_3$ - the ternary field addition - then note that $H(X_2 \lor X_3 | X_2 \oplus_3 X_3)=0$ (see Fig.~\ref{Fig:nonadditiveexample}). Therefore, decoding the ternary addition can enable Rx 1 to recover the full interference. We are led to the following.}\end{comment}

\begin{remark}
\label{Rem:nonadditiveexample}
{As is evident from the above discussion, reinforced via a formal proof provided in Prop.~\ref{Prop:AchievabilityofcosetCdsnonadditiveexample}. As we will see in Sec.~\ref{Sec:3to1CQICAchivablerateregion}, one can achieve $(\mathcal{C}_1,\mathcal{C}_2,\mathcal{C}_3)$ for Ex.~\ref{Ex:nonadditiveinterference} if we can build coset codes over arbitrary finite fields and achieve rates corresponding to \underline{non-uniform distribution}. Next, we have to be able to decode bi-variate interference at all three Rxs, requiring each Tx to encode parts of their message using coset codes. Thm.~\ref{Thm:3CQICStageIRateRegion} presents such a coding strategy and a corresponding inner bound, wherein each Rx decodes bi-variate interference and the induced single-letter distribution is arbitrary. This does not suffice since each Rx over a $3-$CQIC encounters a combination of uni-variate and bi-variate interferences. The former is more efficiently decoded via unstructured IID codes and the latter via coset codes. We will therefore have to design a two-layer coding strategy that combines both unstructured IID codes and coset codes to manage uni-variate and bi-variate interference efficiently. The need for this two-layer coding strategy is also evident from an earlier related work of Ahlswede and Han \cite[Sec.~VI]{198305TIT_AhlHan} in the context of a source coding problem, which has been built upon in \cite[Thm.~4]{201603TIT_PadSahPra}.}
\end{remark}
Building on the above remark, our goal through the rest of the article is to design and analyze a two-layer coding strategy comprising both coset and unstructured IID codes. While the former enables efficient decoding of bi-variate interferences, the latter assists in decoding uni-variate interference components efficiently. This general two-layer coding strategy is developed in three pedagogical steps. Our first step, described in Sec.~\ref{Sec:3to1CQICAchivablerateregion}, will be to design and analyze a coding strategy that manages only bi-variate interference suffered by a single Rx. Specifically, only Rx 1 will decode bi-variate interference, which is facilitated by Tx 2 and 3 encoding their messages using coset codes (see Fig.~\ref{Fig:3to1CQICcodingstrategy}).

\section{A Coset Codes Strategy to Manage bi-variate Interference at a Single Receiver}
\label{Sec:3to1CQICAchivablerateregion}
\begin{comment}{As we discussed in Sec.~\ref{Sec:Introduction}, we present our new inner bounds in three steps, with each step aimed at describing a specific element. The first step is chosen so as to provide a a complete proof in a simplified setting that combines all the new elements - simultaneous decoding of unstructured and coset codes using TSA, and its combination with a likelihood encoder. The derived inner bound in Thm.~\ref{Thm:3to1CQICRateRegion} also enables us to prove that coset codes are strictly more efficient for additive and non-additive, non-commutative Ex.~\ref{Ex:3CQICadditivenoncommutative} and \ref{Ex:nonadditiveinterference} respectively. The inner bound presented in this first step also permits a compact description obtained via a Fourier-Motzkin elimination. Moreover, since the }\end{comment}

{The goal of the first step presented here is to provide a proof in a simplified scenario that \textit{combines} all the new elements - simultaneous decoding of unstructured and coset codes using TSA, and its combination with a likelihood encoder. The derived inner bound permits a compact description obtained via Fourier-Motzkin elimination as stated in Thm.~\ref{Thm:3to1CQICRateRegion}. Moreover, it enables us to prove that coset codes are strictly more efficient for additive and non-additive, non-commutative Ex.~\ref{Ex:3CQICadditivenoncommutative} and \ref{Ex:nonadditiveinterference} respectively.}

{The coding strategy we present in this step employs two codes - a nested coset code and an IID unstructured code - for users $2$ and $3$ respectively, and a single unstructured IID code for user $1$. See Fig.~\ref{Fig:3to1CQICcodingstrategy}, \ref{Fig:Codestructure3to1CQICThm1} wherein Rvs $U_{2}$ and $U_{3}$ are encoded using coset codes built on the same finite field and $X_{k}$ for $k \in [3]$ is encoded via unstructured IID codes. Rx $1$ has to simultaneously decode $U_{2}\oplus U_{3}$ and $X_{1}$, while Rxs $j$ simultaneously decode $U_{j},X_{j}$ for $j \in \{2,3\}$. An informed reader will recognize that the new nontrivial aspect of a proof revolves only around the design and analysis of Rx $1$'s simultaneous decoder. Indeed, (i) the latter involves decoding the sum of the codewords chosen by Rxs $2$ and $3$ - a new element, and (ii) simultaneous decoding of Rxs $2$ and $3$ can be analyzed in a standard fashion using the techniques of Fawzi et. al \cite[Sec.~V.B]{201206TIT_FawHaySavSenWil} which circumvents the need for Sen's TSA \cite{202103SAD_Sen}. To facilitate easy comprehension of the proof of Thm.~\ref{Thm:3to1CQICRateRegion} and, moreover, highlight combination of TSA with decoding the sum of codewords, we prove a simplified coding strategy in Lemma \ref{Lemma:3to1CQICRateRegion} wherein Rxs $2$ and $3$ do not split their message, i.e. $X_{j}=f_{j}(U_{j})$ for $j \in \{2,3\}$. Proof of Thm.~\ref{Thm:3to1CQICRateRegion} can be obtained by just combining the proof of Lemma \ref{Lemma:3to1CQICRateRegion} with the simultaneous decoder of Fawzi et. al \cite[Sec.~V.B]{201206TIT_FawHaySavSenWil} and its analysis. Proof of Lemma \ref{Lemma:3to1CQICRateRegion} involves less clutter owing to a reduced number of Rvs even while capturing the non-trivial Rx $1$'s simultaneous decoder and its analysis.} 
\begin{figure}
\centering
\includegraphics[width=3in]{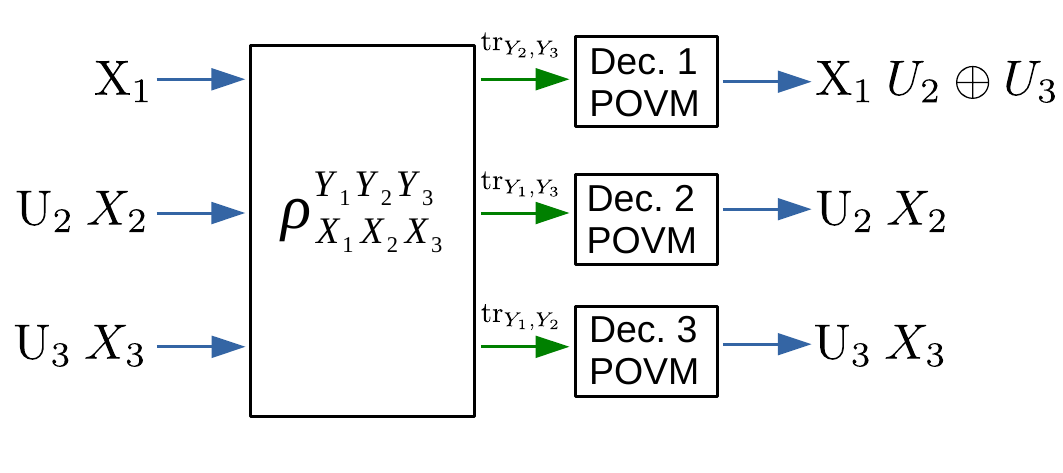}
    \vspace{-0.05in}
    \caption{Depiction of Rvs of the coding strategy.}
    \label{Fig:3to1CQICcodingstrategy}
    \vspace{-0.15in}
\end{figure}

\begin{comment}{The coding strategy that yields the inner bound in Thm.~\ref{Thm:3to1CQICRateRegion} employs two codes - a nested coset code and an IID unstructured code - each for users $2$ and $3$ respectively. An informed reader will recognize that the non-triviality aspect of our work is in analyzing Rx $1$'s error event which involves simultaneously decoding the sum of the chosen codewords with and its message embedded in an unstructured code, for  As depicted in Fig.~\ref{Fig:Codestructure3to1CQIC}, the coding strategy in this step two codebooks for users $2$ and $3$ To assist an interested reader in the proof, we prove the main elements of Thm.~\ref{Thm:3to1CQICRateRegion}.}
\end{comment}

\begin{lemma}
\label{Lemma:3to1CQICRateRegion}
A rate-cost triple $(R_1,R_2,R_3,\tau_1,\tau_2,\tau_3)$ is achievable if there exists (i) a finite field $\CalU_2 = \CalU_3 = \CalF_{\Prime}$, (ii) PMFs $p_{X_1}$, $p_{U_2}$, $p_{U_3}$ on $\mathcal{X}_1$, $\mathcal{U}_2$ and $\mathcal{U}_3$ respectively, and (iii) functions $f_j: \CalU_j \rightarrow \CalX_j$ for $j=2,3$, such that $\mathbb{E}[\kappa(X_1)] \leq \tau_1$, $\sum_{u_j} p_{U_j}(u_j) \kappa(f_j(u_j)) \leq \tau_j$,
\begin{eqnarray}
   R_1  &<& I(X_1;Y_1, U_{2} \oplus U_{3}), \nonumber \\
   R_j &<& I(U_j;Y_j) ,\nonumber \\
     R_j &<& I(U_{2} \oplus U_{3};Y_1, X_1) - H(U_2 \oplus U_3) + H(U_j),\mbox{ and }  \nonumber \\
     R_1 + R_j &<& I(U_{2} \oplus U_{3},X_1;Y_1) -H(U_2 \oplus U_3) + H(U_j), \mbox{ for } \: j=2,3 \:\: \mbox{holds},\nonumber 
\end{eqnarray}
 with all information quantities being computed with respect to the state
\begin{eqnarray}
\label{Eqn:Step1ProofSinlgeLetterState}
    \rho^{U_2 \oplus U_3 U_2 U_3 X_1 \ulineY} \define  \sum_{u,u_2,u_3,x_1} p_{X_1}(x_1) p_{U_2}(u_2) p_{U_3}(u_3) \mathds{1}{\{u=u_2 \oplus u_3\}} \ketbra{u~u_2~u_3~x_1} \otimes \rho^{Y_1Y_2Y_3}_{x_1,f_2(u_2),f_3(u_3)}. \nonumber
\end{eqnarray}
\end{lemma}

\noindent The above inner bound can be strictly larger for non-commutative $3-$CQICs. For a quick perusal of this fact the reader is referred to Sec.~\ref{Subsec:CosetCdsNonComEx}. We now proceed to a proof of the above lemma.
\begin{proof}
\label{Proof:Lemma}
As stated in the hypothesis, choose a generic finite field and let $\Prime$ denote its cardinality. Set $\CalU_{2}=\CalU_{3}=\mathcal{F}_{\Prime}$ be this finite field of cardinality $\upsilon$. Choose generic functions $f_j: \mathcal{U}_j \rightarrow \mathcal{X}_j$, for $j=2,3$ and generic PMFs $p_{X_{1}},p_{U_{2}}$ and $p_{U_{3}}$ on $\CalX_{1}, \CalU_{2}$ and $\CalU_{3}$ respectively, that satisfy the cost constraints $\mathbb{E}[\kappa(X_1)] \leq \tau_1$, $\sum_{u_j} p_{U_j}(u_j) \kappa(f_j(u_j)) \leq \tau_j$ and set PMF $p_{U_2U_3\ulineX}=p_{X_1}p_{U_2}p_{U_3} \mathds{1}{\{X_2 = f_2(U_2)\}} \mathds{1}{\{X_3 = f_3(U_3)\}} \in \CalD(\CalU_{2}\times \CalU_{3}\times \ulineCalX)$. Having chosen these objects, fix the same throughout the rest of the discussion and let the joint quantum state be defined in \eqref{Eqn:Step1ProofSinlgeLetterState} with respect to this PMF $p_{U_2U_3\ulineX}$. Let $(R_{1},R_{2},R_{3}) \in [0,\infty)^{3}$ be a rate triple satisfying all the bounds stated in the hypothesis of the lemma statement with respect to this joint single-letter quantum state in \eqref{Eqn:Step1ProofSinlgeLetterState}. For every rate triple $(R_{1},R_{2},R_{3})$ satisfying the above bounds, there exist auxiliary parameters $B_{2}>0, B_{3}>0$ such that $R_{1}>0, R_{j}>0$ for $j=2,3$ and
 \begin{eqnarray}
\label{Eqn:Thm1PreFMBounds1}
B_{j} &>& \log(\Prime) -H(U_j), \nonumber \\
B_{j} + R_{j}&<&I(U_j;Y_j) + \log(\Prime) - H(U_j) , \nonumber \\ 
R_1 + B_{j} + R_{j} &<& I(X_1,U_2 \oplus U_3;Y_1)  + \log(\Prime) - H(U_2 \oplus U_3), \nonumber \\
B_{j} + R_{j} &<& I(U_2 \oplus U_3;Y_1|X_1) + \log(\Prime)- H(U_2 \oplus U_3) ,  \mbox{ for }  j=2,3, \mbox{ and } \nonumber \\
R_{1} &<& I(X_{1};Y_{1}|U_2 \oplus U_3). \nonumber 
\end{eqnarray}
This existence can be verified by a standard Fourier Motzkin elimination (FME) \cite[Appendix D]{BkNITElGamalKim_2011} technique. Specifically, by eliminating $B_{2},B_{3}$ by the FME technique we obtain the bounds stated in the lemma statement. Having identified the parameters that define our coding strategy, we now describe the code structure.

\begin{figure}
\centering
\includegraphics[width=6in]{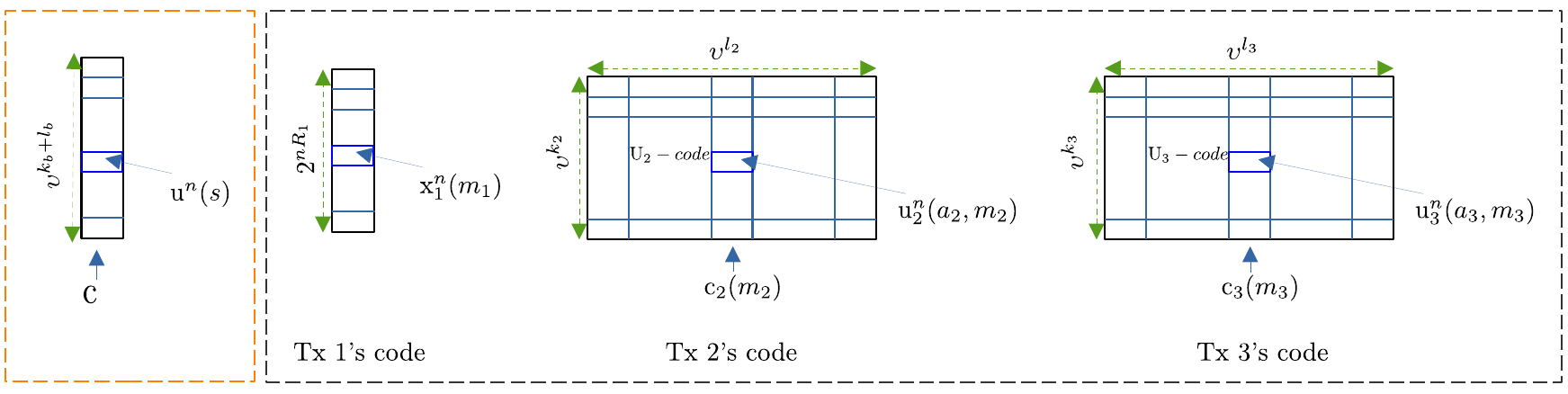}
    \caption{The coding strategy involves three codes, depicted in the grey box - two of them being NCCs built on the finite field $\mathcal{F}_{\Prime}$ and an unstructured IID code built on $\CalX_{1}$. Rx $1$, in addition to decoding into its code, decodes into the collection of `codewords' in the orange dotted box. Specifically, the collection of the codewords in the orange dotted box is obtained by adding all pairs of codewords in the $U_{2}$ and $U_{3}$ codes.
    }
    \label{Fig:Codestructure3to1CQIC}
    \vspace{-0.15in}
\end{figure}

\med\textbf{Code structure}: The coding strategy comprises three codes - one for each user. See Fig.~\ref{Fig:Codestructure3to1CQIC}. User $1$ employs an unstructured code $c_1 \define \{x_1^n(m_1): m_1 \in [2^{nR_1}]\}$. User $2$ and $3$ employ NCCs built over the same finite field $\mathcal{F}_{\Prime}$. More importantly, as we will elaborate below the two NCCs are chosen such that the smaller of the two is a subcoset of the larger. For $j=2,3$, set
\begin{eqnarray}
\label{Eqn:Step1ProofBiggerOfkPlusl}
\mbox{{ $k_j \define \lceil\frac{nB_j}{\log(\Prime)} \rceil$} and $l_j \define \lfloor\frac{nR_j}{\log(\Prime)}\rfloor$ and define } \sfb, \sfs \mbox{ such that }\{\sfb,\sfs\}=\{2,3\} \mbox{ and } k_{\sfb}+l_{\sfb} \geq k_{\sfs}+l_{\sfs}.
\end{eqnarray}
For $j=2,3$, user $j$'s code is specified through (i) a NCC $(n,k_j,l_j,g_I^j,g_{O|I}^j,b_j^n)$ built over the finite field $\CalF_{\Prime}$ with parameters $g_I^j \in \CalF_{\Prime}^{k_j \times n}$, $g_{O|I}^j \in \CalF_{\Prime}^{l_j \times n}$, bias $b_j^n\in \CalF_{\Prime}^n$, $k_j := \lceil\frac{nB_j}{\log(\Prime)} \rceil$ and $l_j := \lfloor\frac{nR_j}{\log(\Prime)}\rfloor$ and (ii) a binning map $a_j: \mathcal{F}_{\Prime}^{l_j} \rightarrow \mathcal{F}_{\Prime}^{k_j}$. Let $g_{j}\define \left[\!\! \begin{array}{c}g_{I}^{j}\\g_{O/I}^{j}\end{array}\!\!\right]$ for $j=2,3$. These generator matrices are identified such that $g_{\sfb}\define \left[\!\! \begin{array}{c}g_{\sfs}\\g_{\Delta}\end{array}\!\!\right]$. This choice of the generator matrices ensures that the smaller of {these} NCCs is a subcoset of the larger NCC. This will also be imposed by the distribution of the corresponding random codebooks with respect to which we analyze the average error probability. For $j=2,3$, let $u^n_j(a_j(m_j),m_j)$  denote the chosen codeword from the coset $c_j(m_j) \define \{u^n_j(a_j(m_j),m_j) : a_j \in \mathcal{F}_{\Prime}^{k_j}\}$ corresponding to message $m_j$.
\begin{comment}
    The coding strategy comprises of three codes. See Fig.~\ref{Fig:Codestructure3to1CQIC}, \colRed{Firstly, consider two NCCs $(n,k_j,l_j,g_I^j,g_{O|I}^j,b_j^n): j = 2,3$ built over the finite field $\CalF_{\Prime}$ with parameters $g_I^j \in \CalF_{\Prime}^{k_j \times n}$, $g_{O|I}^j \in \CalF_{\Prime}^{l_j \times n}$ and bias $b_j^n\in \CalF_{\Prime}^n$. Let $c_j(m_j)=\{u_j^n(a_j,m_j) : a_j \in \CalF_{\Prime}^{k_j}\}$ be the coset corresponding to message $m_j$ where $u_j^n(a_j,m_j)=a_j g_I^j \oplus m_j g_{O|I}^j \oplus b_j^n$ for $j = 2,3$.
Note that the smaller of these NCCs must be a subcoset of the larger code. This ensures closure under finite-field addition.} Let $R_j=\frac{l_j}{n} \log(\Prime)$ for $j=2,3$. For each $m_j \in \CalF_{\Prime}^{l_j}$, we choose $a_j \in \CalF_{\Prime}^{k_j}$ according to
    \[p(A_j(m_j)=a_j|c_j(m_j))=\frac{r_{U_j^n}(u_j^n(a_j,m_j))}{\sum_{\Tilde{a}_j} r_{U_j^n}(u_j^n(\Tilde{a}_j,m_j))},\]
where $r_{U_j^n}(u_j^n(a_j,m_j))=\frac{p_{U_j}^n(u^n_j(a_j,m_j))}{q_{U_j}^n(u^n_j(a_j,m_j))}$ and $q_{U_j}$ is the uniform distribution on $\CalF_{\Prime}$.
\end{comment}

\begin{comment}
    Secondly, consider an unstructured IID code $c_1=\{x_1^n(m_1): m_1 \in 2^{nR_1}\}$ generated randomly and independently where each codeword is distributed according to $p_{X_1}^n$.
\end{comment}
\begin{remark}
 \label{Rem:InterAndIntraStructure}
 We highlight the intra- and inter-structural properties of the codes. Firstly, $c_{2},c_{3}$, being coset code, possess certain algebraic closure properties bereft in unstructured code strategies. Secondly, the rows of $g_{\sfs}$ being contained in the rows of $g_{\sfb}$ implies that by adding all codewords of user $2$'s NCC to all codewords of user $3$'s NCC, we obtain a collection of vectors of cardinality at most $\Prime^{k_{\sfb}+l_{\sfb}}$ and not $\Prime^{k_{1}+l_{1}+k_{2}+l_{2}}$. We crucially exploit this sum containment in the decoding specified below.
\end{remark}

\med\textbf{Encoding}: Upon observing $m_1$ the Tx $1$ sends the codeword $x^n_1(m_1)$. For $j = 2,3$, upon observing $m_j$ the Tx $j$ sends $f_j^n(u_j^n(a_j(m_j),m_j))$. Here, note that $f_{j}(\cdot)$ is evaluated letter-by-letter.

\med \textbf{Decoding POVM}: We first describe the decoding POVM for Rx 1. Rx 1 decodes the message $m_1$ and the \textit{sum of the chosen $U_{2}-,U_{3}-$codewords} simultaneously.
Naturally, $m_{1}$ is deciphered by decoding into $c_{1}$. In regards to the sum of the chosen $U_{2}-,U_{3}-$codewords, the decoder identifies the `sum coset code'  $c$ with parameters $(n,k_{\sfb}+l_{\sfb},g_{\sfb},b^n)$, where $b^n=b_2^n \oplus b_3^n$.
For $s \in \mathcal{F}_{\Prime}^{k_{\sfb}+l_{\sfb}}$, we define $u^n(s) \define s g_{\sfb} \oplus b^n$ to represent a  generic codeword in this `sum coset code'.\footnote{Strictly speaking, one must refer to codewords of this sum coset code as vectors since none of the Txs choose any codeword in this sum coset code explicitly.} If one were to think classically for a moment, Rx $1$ having received $y_{1}^{n}$ looks for a pair $(x^{n}(m_{1}), u^{n}(s)) \in c_{1} \times c$ that is jointly typical with the received $y_{1}^{n}$ with respect to the joint PMF $ p_{X_{1},U_{2}\oplus U_{3},Y_{1}}$. In the following, we construct the simultaneous decoding POVM of Rx $1$ to accomplish this in our CQ setting.
\begin{comment}
%with $a \define a_2 \oplus a_3 \in \mathcal{F}_{\Prime}^k$, $m \define m_2 \oplus m_3 \in \mathcal{F}_{\Prime}^l$, $l \define \max\{l_2,l_3\}$ and $k \define \max\{k_2,k_3\}$.    
\end{comment}

\begin{remark}
 \label{Rem:CrucialChangeInTheDecoding}
 We direct a reader's attention to the key differences from conventional unstructured IID code strategies that are evident from the above description. Firstly, the joint algebraic closure relationship induced between $c_{2}$ and $c_{3}$ by choosing one of them to be a sub-coset of the other. Secondly, the Rx $1$ constructs a new collection $c$ of codewords into which it decodes. This collection is used by neither Txs. Furthermore, we have to combine these elements with the need to alter the POVMs via TSA to accomplish simultaneous decoding.
\end{remark}

If simultaneous decoding required no newer elements, we would just be constructing conditional typical projectors corresponding to state $\rho_{x_{1},u}^{Y_{1}}$ in \eqref{Eqn:3to1CQICDecoder1States1}. However, as Sen has discovered, simultaneous decoding will require identifying auxiliary spaces and tilting the projectors. We therefore now employ Sen’s TSA technique \cite{202103SAD_Sen} to appropriately tilt our identified decoding POVM as described below. We should note here that TSA is performed to handle unions and intersections of typical subspaces, which can overlap in complex ways. Tilting maps rotate such intersecting subspaces in orthogonal directions, thereby increasing their separation. Specifically, for three possible error events corresponding to distinct subspaces, we define two tilting maps for the first two subspaces, representing the case where a single message is decoded incorrectly, while the third subspace, corresponding to errors in both messages, remains untilted. {To enable this construction, we first enlarge the original Hilbert space by adding extra dimensions to enable the tilts. We then augment the space to ensure the smoothing property \cite{202103SAD_Sen}.}
\begin{comment}{—that is, the difference between the expectation of the original state and the expectation of the tilted state over the random codebook \footnote{When we say `average of $\cdots$' we mean the expectation of the tilted state over the random codebook.} is small.}\end{comment}

Consider two auxiliary finite sets $\CalD_1$ and $\CalD_2$ along with corresponding Hilbert space ${\CalD_{1}}$ and ${\CalD_{2}}$ of dimension $|\CalD_{1}|$ and $|\CalD_{2}|$ respectively. We define the extended space $\CalH_{Y_1}^e$ as follows
\[\left(\CalH_{Y_1}^e\right)^{\otimes n} \define \left(\CalH_{Y_1}^G\right)^{\otimes n} \oplus \left( \left(\CalH_{Y_1}^G\right)^{\otimes n} \otimes \CalD_1^{\otimes n} \right) \oplus \left(\left(\CalH_{Y_1}^G\right)^{\otimes n} \otimes \CalD_2^{\otimes n}\right)\]
where $\CalH_{Y_1}^G=\CalH_{Y_1} \otimes \mathbb{C}^2$.
In the sequel, we let $\boldsymbol{\CalH_{Y_1}}= \CalH_{Y_1}^{\otimes n}$, $\boldsymbol{\CalH_{Y_1}^G}=\left(\CalH_{Y_1}^G\right)^{\otimes n}$,
$\boldsymbol{\CalH_{Y_1}^{e}}=\left(\CalH_{Y_1}^{e}\right)^{\otimes n}$ and $\boldsymbol{\CalD_{j}}= \CalD_{j}^{\otimes n} : j \in [2]$. The $G$ in the superscript here and throughout in the following denotes the Gelfand-Naimark extension. For $j \in [2]$ and $d_{j}^{n} \in \CalD_{j}^{n}$ we define the tilting map $\CalT_{ d_{j}^{n}, \eta} : \boldsymbol{\CalH_{Y_1}^G}\rightarrow
\boldsymbol{\CalH_{Y_1}^G} \oplus \left(\boldsymbol{\CalH_{Y_1}^G}  \otimes   \boldsymbol{\CalD_{j}}\right)
$ as
\begin{eqnarray}
\CalT_{ d_{j}^{n}, \eta}(\ket{h}) = \frac{1}{\sqrt{1+\eta^2} } \left(\ket{h} +\eta \ket{h} \otimes  \ket{d_{j}^{n}}\right),
\nonumber
\end{eqnarray}
where $\eta$ will be chosen appropriately in the sequel. 
Let $\Pi_{x_1^n,u^n}^{Y_1}, \Pi_{x_1^n}^{Y_1}$ and $\Pi_{u^n}^{Y_1}$ be the C-Typ-Proj with respect to the states $ \rho^{Y_1}_{x_1^n,u^n} \define  \left(\rho^{Y_1}_{x_1,u}\right)^{\otimes n }$, $\rho^{Y_1}_{x_1^n} \define \left(\rho^{Y_1}_{x_1}\right)^{\otimes n}$ and $\rho^{Y_1}_{u^n} \define \left(\rho^{Y_1}_{u}\right)^{\otimes n}$  respectively, and let $\Pi^{Y_1}$ be the U-Typ-Proj with respect to the state $\left(\rho^{Y_1}\right)^{\otimes n}$, where 
\begin{comment}
Consider the classical-quantum state 
\begin{eqnarray}
\label{Eqn:3to1CQICCQstateDec1}
    \rho^{U_2 \oplus U_3U_2U_3X_1Y_1} \define \sum_{u,u_2,u_3, x_1} p_{X_1}(x_1) p_{U_2}(u_2) p_{U_3}(u_3) \mathds{1}\{u=u_2 \oplus u_3\} \ketbra{u,u_2,u_3, x_1} \otimes \rho_{x_1,f_2(u_2),f_3(u_3)}^{Y_1}. 
\end{eqnarray}
where $U \define U_2 \oplus U_3$.    
We define the following associated density operators, 
\end{comment}
\begin{eqnarray}
\rho^{Y_1}_{x_1,u} &\define& \sum_{u_2,u_3} p_{U_2|U_2 \oplus U_3}(u_2|u) p_{U_3|U_2 \oplus U_3 U_2}(u_3|u,u_2) \rho_{x_1,f_2(u_2),f_3(u_3)}^{Y_1}, \label{Eqn:3to1CQICDecoder1States1} \\
\rho^{Y_1}_{x_1} &\define& \sum_{u_2,u_3} p_{U_2}(u_2) p_{U_3}(u_3) \rho_{x_1,f_2(u_2),f_3(u_3)}^{Y_1}, \label{Eqn:3to1CQICDecoder1States2}\\
\rho^{Y_1}_{u} &\define& \sum_{x_1,u_2,u_3} p_{X_1}(x_1) p_{U_2|U_2 \oplus U_3}(u_2|u) p_{U_3|U_2 \oplus U_3 U_2}(u_3|u,u_2) \rho_{x_1,f_2(u_2),f_3(u_3)}^{Y_1}, \mbox{ and}\label{Eqn:3to1CQICDecoder1States3} \\
\rho^{Y_1} &\define& \sum_{x_1,u_2,u_3} p_{X_1}(x_1) p_{U_2}(u_2) p_{U_3}(u_3) \rho_{x_1,f_2(u_2),f_3(u_3)}^{Y_1}. \label{Eqn:3to1CQICDecoder1States4}
\end{eqnarray}    

Consider the following POVM elements 
\begin{eqnarray}
    G^1_{x_1^n,u^n} = \Pi_{u^n}^{Y_1} \: \Pi^{Y_1}_{x_1^n,u^n} \: \Pi_{u^n}^{Y_1}, \:\: G^2_{x_1^n,u^n} = \Pi_{x_1^n}^{Y_1} \: \Pi_{x_1^n,u^n}^{Y_1} \: \Pi_{x_1^n}^{Y_1}, \mbox{ and }  G^3_{x_1^n,u^n} = \Pi^{Y_1} \: \Pi_{x_1^n,u^n}^{Y_1} \: \Pi^{Y_1},\label{Eqn:3to1CQICtheoriginalpovmelements}
\end{eqnarray}
\noindent By Gelfand–Naimark’s Thm.~\cite[Thm.~3.7]{BkHolevo_2019}, there exists orthogonal projectors $\olineG^{\mathcal{J}}_{x_1^n,u^n} \in \CalP(\boldsymbol{\CalH_{Y_1}^G} )$ that yields identical measurement statistics on states in $\boldsymbol{\CalH_{Y_1}^G} $that $G^{\mathcal{J}}_{x_1^n,u^n}$ gives on states in $\boldsymbol{\CalH_{Y_1}}$ for $\mathcal{J} \in \{1,2,3\}$.
Let 
\begin{eqnarray}
\olineB_{x_1^n,u^n}^{\mathcal{J}}=I_{\boldsymbol{\CalH_{Y_1}^G}}-\olineG^{\mathcal{J}}_{x_1^n,u^n}, \mbox{ for }  {\mathcal{J} \in \{1,2,3\}} \label{Eqn:3to1CQICcomplementprojector}
\end{eqnarray}
be the complement projector. Now, consider
\begin{eqnarray}
\beta_{x_1^n,d_1^n,u^n,d_2^n}^{\mathcal{J}}=\CalT_{d^n_{\mathcal{J}^c},\eta}\left(\olineB^{\mathcal{J}}_{x_1^n,u^n}\right), \mbox{ for }  \mathcal{J} \in\{1,2\}, \mbox{ and }   
\beta^3_{x_1^n,d_1^n,u^n,d_2^n}=\olineB^3_{x_1^n,u^n} \label{Eqn:3to1CQICTiltingofprojectors}
\end{eqnarray}
where $\beta^{\mathcal{J}}_{x_1^n,d_1^n,u^n,d_2^n}$ represent the tilted projector along direction $d_{\mathcal{J}^c}^{n}$ for $\mathcal{J} \in \{1,2\}$, and $\beta^3_{x_1^n,d_1^n,u^n,d_2^n}$ represent the untilted projector.
Next, we define $\beta^{*}_{x_1^n,d_1^n,u^n,d_2^n}$ as the projector in  $\boldsymbol{\CalH_{Y_1}^{e}}$ whose support is the union of the supports of $\beta^{\mathcal{J}}_{x_1^n,d_1^n,u^n,d_2^n}$ for all $\mathcal{J} \in \{1,2,3\}$. Let $\Pi_{\boldsymbol{\CalH_{Y_1}^G}}$ be the orthogonal projector in $\boldsymbol{\CalH_{Y_1}^{e}}$ onto $\boldsymbol{\CalH_{Y_1}^G}$. We define the square root measurement \cite{BkWilde_2017,BkHolevo_2019} $\{\mu_{m_1,s}^{Y_1} : (m_1,s) \in [2^{nR_1}] \times \mathcal{F}_{\Prime}^{k_{\sfb}+ l_{\sfb}}\}$ as
\begin{eqnarray}
\mu_{m_1,s}^{Y_1} &\define& \left(\sum_{\widehat{m}_1,\widehat{s}} \gamma^{*}_{(x_1^n,d_1^n)(\widehat{m}_1),(u^n,d_2^{n})(\widehat{s})}\right)^{-\frac{1}{2}}
\gamma^{*}_{(x_1^n,d_1^n)(m_1),(u^n,d_2^{n})(s)}  \left(\sum_{\widehat{m}_1,\widehat{s}} \gamma^{*}_{(x_1^n,d_1^n)(\widehat{m}_1),(u^n,d_2^{n})(\widehat{s})}\right)^{-\frac{1}{2}},
\mbox{ and } \nonumber \\ \mu^{Y_1}_{-1} &\define& I - \sum_{m_1} \sum_{s} \mu^{Y_1}_{m_1}, \mbox{ where } \nonumber \\
\gamma^{*}_{(x_1^n,d_1^n)(m_1),(u^n,d_2^{n})(s)} &\define&\left(I_{\boldsymbol{\CalH_{Y_1}^{e}}}-  \beta^{*}_{(x_1^n,d_1^n)(m_1),(u^n,d_2^{n})(s)}\right) \Pi_{\boldsymbol{\CalH_{Y_1}^G} }\left(I_{\boldsymbol{\CalH_{Y_1}^{e}}}-  \beta^{*}_{(x_1^n,d_1^n)(m_1),(u^n,d_2^{n})(s)}\right). \label{Eqn:3to1CQICPovmelementfordecoder1}  
\end{eqnarray}

\med\textit{Decoding POVM for Rxs $2$ and $3$}: The decoding POVM for Rxs $2$ and $3$ are identical. Therefore, we only describe one of them. Rx 2 is only interested in message $m_2$. Consequently, to decode $m_2$, we use the NCC $(n,k_2,l_2,g_I^2,g_{O|I}^2,b_2^n)$ used by Tx 2. 
\begin{comment}
Consider the classical-quantum state  
\begin{eqnarray}
\label{Eqn:3to1CQICCQstateDec2}
&& \rho^{U_2U_3X_1Y_2} \define \sum_{u_2,u_3,x_1} p_{X_1}(x_1) p_{U_2}(u_2) p_{U_3}(u_3) \ketbra{u_2,u_3,x_1} \otimes \rho^{Y_2}_{x_1,f_2(u_2),f_3(u_3)}.
\end{eqnarray}
We define the following associated density operators 
\end{comment}
Let $\Pi_{u_2^n}^{Y_2}$ and $\Pi^{Y_2}$ be the typical projectors with respect to the states $\rho_{u_2^n}^{Y_2} = \left(\rho_{u_2}^{Y_2}\right)^{\otimes n }$ and $\left(\rho^{Y_2} \right)^{\otimes n }$ respectively, where 
\begin{eqnarray}
    \rho_{u_2}^{Y_2} &\define& \sum_{u_3,x_1} p_{U_3}(u_3) p_{X_1}(x_1) \rho_{x_1,f_2(u_2),f_3(u_3)}^{Y_2}, \mbox{ and}\label{Eqn:3to1CQICassociateddensityoperatorsforuser2.1} \\ 
     \rho^{Y_2} &\define& \sum_{u_2,u_3,x_1} p_{U_2}(u_2) p_{U_3}(u_3) p_{X_1}(x_1) \rho_{x_1,f_2(u_2),f_3(u_3)}^{Y_2} \label{Eqn:3to1CQICassociateddensityoperatorsforuser2}.
\end{eqnarray}
We define the square root measurement \cite{BkWilde_2017,BkHolevo_2019} $\{\mu_{m_2}^{Y_2} : m_2 \in [2^{nR_2}]\}$ as  
\begin{eqnarray}
    &&\hspace{-0.25in}\mu_{m_2}^{Y_2} \define \left( \sum_{\widehat{m}_2,\widehat{a}_2} \Pi^{Y_2} \Pi^{Y_2}_{u_2^n(\widehat{a}_2,\widehat{m}_2)} \Pi^{Y_2} \right)^{-\frac{1}{2}} \left(\sum_{a_2} \Pi^{Y_2} \Pi^{Y_2}_{u_2^n(a_2,m_2)} \Pi^{Y_2} \right)  \left( \sum_{\widehat{m}_2,\widehat{a}_2} \Pi^{Y_2} \Pi^{Y_2}_{u_2^n(\widehat{a}_2,\widehat{m}_2)} \Pi^{Y_2} \right)^{-\frac{1}{2}}\!\!\!, \mbox{ and }\mu^{Y_2}_{-1} \define I - \sum_{m_2} \mu_{m_2}^{Y_2}.\nonumber 
\end{eqnarray}

\med \textbf{Error analysis}:
As is standard, we derive an upper bound on the error probability of a good code by averaging the error probability over an ensemble of codes.  We begin by specifying the codebook distribution, i.e. the random code distribution, with respect to which this averaging is performed. Towards that end, recall that $g_{j} = \left[\!\! \begin{array}{c}g_{I}^{j}\\g_{O/I}^{j}\end{array}\!\!\right]$ for $j=2,3$ and these generator matrices satisfy $g_{\sfb}\define \left[\!\! \begin{array}{c}g_{\sfs}\\g_{\Delta}\end{array}\!\!\right]$, where $\sfb$ is defined as in \eqref{Eqn:Step1ProofBiggerOfkPlusl}.  From the code structure, encoding and decoding it is evident that our coding strategy is completely specified via the following objects : $c_{1}=(x_{1}^{n}(m_{1}): m_{1} \in [2^{nR_{1}}] ), g_{\sfb},b_{j}^{n}$ for $j \in \{2,3\}$, $(d_{1}^{n}(m_{1}): m_{1} \in [2^{nR_{1}}])), (d_{2}^{n}(s): s \in \CalF_{\Prime}^{k_{\sfb}+l_{\sfb}}))$ and finally maps $a_{j}:\CalF_{\Prime}^{l_{j}}\rightarrow \CalF_{\Prime}^{k_{j}}$ for $j \in \{2,3\}$. It suffices therefore to specify the joint distribution of these objects. For any choice of the arguments in their respective range spaces, let
\begin{eqnarray}
 &&P\left(
 \begin{array}{c}
\left( X_{1}^{n}(m_{1}) = x_{1}^{n}(m_{1}): m_{1} \in [2^{nR_{1}}] \right) ,G_{b}=g_{b}, \left( B_{j}^{n}=b_{j}^{n}:j=2,3 \right), \left(D_{1}^{n}(m_{1})=d_{1}^{n}(m_{1}) : m_{1} \in [2^{nR_1}]\right), \\\left(D_2^n(s)=d_2^n(s) : s \in \mathcal{F}_{\Prime}^{k_{\sfb}  + l_{\sfb}}\right),  \left(A_{j}(m_{j}) = a_{j}(m_{j}) :m_{j} \in \mathcal{F}_{\Prime}^{l_j}:j=2,3 \right)
 \end{array}
 \right)
 \nonumber\\
 &=& \left[\prod_{m_{1}=1}^{2^{nR_{1}}}p^n_{X_{1}}(x^n_{1}(m_{1}))\right]\frac{1}{\Prime^{k_{\sfb}l_{\sfb}}}\frac{1}{\Prime^{2n}}
 \left[\prod_{m_{1}=1}^{2^{nR_{1}}}\frac{1}{|\CalD_{1}|^{n}}\right] \left[\prod_{s=1}^{\Prime^{k_{\sfb}+l_{\sfb}}}\frac{1}{|\CalD_{2}|^{n}} \right]  \left[\prod_{j=2}^{3}\prod_{m_{j}=1}^{2^{nR_j}}\frac{p^n_{U_{j}}({a_{j}(m_{j})g_{I}^{j}\oplus m_{j}g_{O/I}^{j}\oplus b_{j}^{n} })}{\sum_{\alpha_{j} \in \Prime^{k_{j}}}p^n_{U_{j}}(\alpha_{j}g_{I}^{j}\oplus m_{j}g_{O/I}^{j}\oplus b_{j}^{n}) }\right] 
\label{Eqn:DistStepI}
\end{eqnarray}
specify our joint random codebook distribution. In addition, throughout the rest of the proof we define
\begin{eqnarray}
 \label{Eqn:Step1ProofLikelihoodEncRat}
 r_{U_{j}}^{n}(u_{j}^{n}) \define \frac{p_{U_{j}}^{n}(u_{j}^{n})}{q_{U_{j}}^{n}(u_{j}^{n})}, \mbox{ where }{q_{U_{j}}^{n}(u_{j}^{n})}\define \frac{1}{\Prime^{n}}\mbox{ is the uniform distribution, and hence} \nonumber \\
 P\left(A_{j}(m_{j})=a_{j}\Big|C_{j}(m_{j})=\left( u_j^n(a_j,m_j) : a_j \in \mathcal{F}_{\Prime}^{k_j}\right)\right) = \frac{r_{U_{j}}^{n}(u_{j}^{n}(a_{j},m_{j}))}{\sum_{\tildea_{j}} r_{U_{j}}^{n}(u_{j}^{n}(\tildea_{j},m_{j}))}.
\end{eqnarray}

We now derive an upper bound on the average error probability of a good code, by averaging the error probability of every code with respect to the above specified codebook distribution. Note that, upon receiving the quantum state $\rho^{Y_1}_{x_1^n(m_1),f_2^n(u_2^n(a_2(m_2),m_2)),f_3^n(u_3^n(a_3(m_3),m_3))}$, Rx $1$ prepares the auxiliary state $\ketbra{0}$, concatenates the same with the received state and measures this concatenated state via the square root measurement \cite{BkWilde_2017,BkHolevo_2019} $\{\mu^{Y_1}_{m_1,s}: (m_1,s) \in [2^{nR_1}] \times \mathcal{F}_{\Prime}^{k_{\sfb}+ l_{\sfb}}\}$ on the combined state $\left(\rho^{Y_1}_{x_1^n(m_1),f_2^n(u_2^n(a_2(m_2),m_2)),f_3^n(u_3^n(a_3(m_3),m_3))} \otimes \ketbra{0}\right)$. Hence, the average error probability of the code is given by
\begin{eqnarray}
    &&\hspace{-0.25in}\mathbf{P}(\ulinee,\uline{\mu})=\frac{1}{|\mathcal{\ulineM}|} \frac{1}{\Prime^{k_{\sfb} + l_{\sfb}}} \sum_{\ulinem} \sum_{s}\tr \left\{ \begin{aligned}
        &\left( I - \mu^{Y_1}_{m_1,s} \otimes \mu_{m_2}^{Y_2} \otimes \mu_{m_3}^{Y_3}\right)  \left( \rho_{x_1^n(m_1),f_2^n(u_2^n(a_2(m_2),m_2)),f_3^n(u_3^n(a_3(m_3),m_3))}^{Y_1Y_2Y_3} \otimes \ketbra{0}\right)  
    \end{aligned}\right\}. \nonumber 
    \end{eqnarray}
    Let $J$ and $T$ be operators on the quantum systems A and B, respectively, satisfying $0 \leq J \leq I^A$ and $0 \leq T \leq I ^B$. Then, observe that 
    \begin{eqnarray}
       I^A \otimes I^B - J \otimes T &=& (I^A-J) \otimes T + I^A \otimes (I^B -T) \nonumber \\
       &\leq& (I^A - J) \otimes I^B + I^A \otimes I^B - (I^A \otimes T). \nonumber 
    \end{eqnarray}
    Therefore, we have
    \begin{eqnarray}
        I^{Y_1Y_2Y_3} - \mu^{Y_1}_{m_1,s} \otimes \mu_{m_2}^{Y_2} \otimes \mu_{m_3}^{Y_3} &\leq&  \left(I^{Y_1Y_2Y_3} - \mu^{Y_1}_{m_1,s} \otimes I^{Y_2} \otimes I^{Y_3} \right) + \left(I^{Y_1Y_2Y_3} - I^{Y_1} \otimes  \mu_{m_2}^{Y_2} \otimes I^{Y_3}\right) \nonumber \\
        &&+ \left(I^{Y_1Y_2Y_3} -I^{Y_1} \otimes I^{Y_2} \otimes  \mu_{m_3}^{Y_3}\right).
        \nonumber 
        \end{eqnarray}
        Using the above inequality, we obtain
   \begin{eqnarray}
\mathbf{P}\left( \ulinee, \uline{\mu} \right) &\leq& \frac{1}{|\mathcal{\ulineM}|} \frac{1}{\Prime^{k_{\sfb} + l_{\sfb}}} \sum_{\ulinem} \sum_{s} \tr \left\{  \left( I - \mu^{Y_1}_{m_1,s} \right)    \left( \rho_{x_1^n(m_1),f_2^n(u_2^n(a_2(m_2),m_2)),f_3^n(u_3^n(a_3(m_3),m_3))}^{Y_1} \otimes \ketbra{0}\right)  \right\} \nonumber \\
&+& \frac{1}{|\mathcal{\ulineM}|} \sum_{\ulinem} \tr \left\{ \left( I - \mu_{m_2}^{Y_2} \right)  \rho_{x_1^n(m_1),f_2^n(u_2^n(a_2(m_2),m_2)),f_3^n(u_3^n(a_3(m_3),m_3))}^{Y_2}   \right\} \nonumber \\
&+&\frac{1}{|\mathcal{\ulineM}|} \sum_{\ulinem} \tr \left\{  \left( I -  \mu_{m_3}^{Y_3}\right)  \rho_{x_1^n(m_1),f_2^n(u_2^n(a_2(m_2),m_2)),f_3^n(u_3^n(a_3(m_3),m_3))}^{Y_3} \right\} \label{Eq:3to1CQICErrorprobabilty}
\end{eqnarray}
\med\textit{\underline{Rx $1$'s Error Analysis}}: We begin by analyzing the first term in \eqref{Eq:3to1CQICErrorprobabilty}. We have  
\begin{comment}
    %Observe that $\mu_{m_1}^{Y_1} \geq \Gamma^{Y_1}_{m_1,S} $, where
%\begin{eqnarray} 
%&& \Gamma^{Y_1}_{m_1,S} \define \left(\sum_{\widehat{m}_1,\widehat{s}} \gamma^{*}_{(x_1^n,d_1^n)(\widehat{m}_1),(u^n,d_2^{n})(\widehat{s})}\right)^{-\frac{1}{2}}
%\gamma^{*}_{(x_1^n,d_1^n)(m_1),(u^n,d_2^{n})(S)} \left(\sum_{\widehat{m}_1,\widehat{s}} \gamma^{*}_{(x_1^n,d_1^n)(\widehat{m}_1),(u^n,d_2^{n})(\widehat{s})}\right)^{-\frac{1}{2}} , \nonumber 
%&&A(m)=A_2(m_2) \oplus A_3(m_3), \quad \mbox{and} \quad m=m_2 \oplus m_3 \nonumber.
%\end{eqnarray}
%Therefore, we have 
\end{comment}
\begin{eqnarray}
&&\frac{1}{|\mathcal{\ulineM}|} \frac{1}{\Prime^{k_{\sfb} + l_{\sfb}}} \sum_{\ulinem} \sum_{s} \tr \left\{  \left( I - \mu^{Y_1}_{m_1,s} \right)    \left( \rho_{x_1^n(m_1),f_2^n(u_2^n(a_2(m_2),m_2)),f_3^n(u_3^n(a_3(m_3),m_3))}^{Y_1} \otimes \ketbra{0}\right)  \right\} \nonumber \\
&=& \frac{1}{|\mathcal{\ulineM}|} \frac{1}{\Prime^{k_{\sfb} + l_{\sfb}}} \sum_{\ulinem} \sum_{s} \sum_{a_2, a_3}  \mathds{1}{\left\{a_2(m_2)=a_2\right\}} \mathds{1}{\left\{a_3(m_3)=a_3\right\}} \nonumber \\
&&\tr \left\{  \left( I - \mu^{Y_1}_{m_1,s} \right)  \left( \rho_{x_1^n(m_1),f_2^n(u_2^n(a_2,m_2)),f_3^n(u_3^n(a_3,m_3))}^{Y_1} \otimes \ketbra{0}\right)  \right\} \nonumber \\ 
&=& t_1 + t_2, \nonumber \mbox{ where}\\
t_1&\define& \frac{1}{|\mathcal{\ulineM}|} \frac{1}{\Prime^{k_{\sfb} + l_{\sfb}}}\sum_{\ulinem} \sum_{s}\sum_{a_2, a_3}  \left[\mathds{1}{\left\{a_2(m_2)=a_2\right\}} \mathds{1}{\left\{a_3(m_3)=a_3\right\}}-\frac{r_{U_2}^n(u_2^n(a_2,m_2))}{\Prime^{k_2}}\frac{r_{U_3}^n(u_3^n(a_3,m_3))}{\Prime^{k_3}}\right] \nonumber \\ &&\tr \left\{  \left( I - \mu^{Y_1}_{m_1,s} \right)  \left( \rho_{x_1^n(m_1),f_2^n(u_2^n(a_2,m_2)),f_3^n(u_3^n(a_3,m_3))}^{Y_1} \otimes \ketbra{0}\right)  \right\} \label{Eqn:3to1CQICSourceCodingBnd},  \mbox{ and}\\ 
t_2&\define& \frac{1}{|\mathcal{\ulineM}|} \frac{1}{\Prime^{k_{\sfb} + l_{\sfb}}} \frac{1}{\Prime^{k_2 +k_3}} \sum_{\ulinem} \sum_{s} \sum_{a_2, a_3}   r_{U_2}^n(u_2^n(a_2,m_2)) r_{U_3}^n(u_3^n(a_3,m_3)) \nonumber \\ &&\tr \left\{  \left( I - \mu^{Y_1}_{m_1,s} \right)  \left( \rho_{x_1^n(m_1),f_2^n(u_2^n(a_2,m_2)),f_3^n(u_3^n(a_3,m_3))}^{Y_1} \otimes \ketbra{0}\right)  \right\} \nonumber.
\end{eqnarray}

\noindent The last equality is obtained by adding and subtracting the product 
\begin{eqnarray}
\frac{r_{U_2}^n(u_2^n(a_2,m_2))}{\Prime^{k_2}}\frac{r_{U_3}^n(u_3^n(a_3,m_3))}{\Prime^{k_3}} \label{Eqn:Ratio}    
\end{eqnarray}
which decomposes the expression into two terms. 
\begin{remark}
\label{Rem:BinningandChannelCodingAnalysis}
We highlight the above technique of adding and subtracting the term in \eqref{Eqn:Ratio}, which enables us to decompose the error into two terms, one which is analyzed through a covering \cite{CuffPhDThesis} and the second through a conventional channel coding analysis. This enables us to break the independence and channel code analysis yields the right bound via the divergence. This approach essentially decouples the binning analysis and the channel coding analysis, thus leaving the channel coding analysis oblivious to the presence of binning.    
\end{remark}

\noindent Bounding $t_1$ on the above in Prop.~\ref{Prop:Dec1Cuffterm} below is performed by a covering argument \cite{CuffPhDThesis} referred to therein as `cloud mixing'. 

\begin{proposition}
\label{Prop:Dec1Cuffterm}
For any $\epsilon \in (0,1)$, and for all sufficiently small $\delta$ and sufficiently large $n$, we have $\mathbb{E}[t_1] \leq \epsilon$ if 
\begin{eqnarray}
    \frac{k_j}{n}\log(\Prime) > \log(\Prime) - H(U_j), \mbox{ for } j=2,3. \nonumber 
\end{eqnarray}
\end{proposition}
\begin{proof}
    See Appendix \ref{App:Dec1Cuffterm} for a proof.
\end{proof}

Our next task is to bound $t_{2}$ on the above. To do this, we employ an alternate `proxy' state. Specifically, for each codeword triplet, we substitute the original received state ($\rho_{x_1^n,u^n}^{Y_1} \otimes \ketbra{0}$) by a specific `tilted state'. Since the chosen tilted state is close in $\mathbb{L}_{1}-$norm to the original received state, the effect of this substitution on the error probability can be suppressed. Towards identifying this tilted state, we define a new tilting map
$\CalT_{d_{1}^{n},d_{2}^{n}, \eta} : \boldsymbol{\CalH_{Y_1}^G}\rightarrow
\boldsymbol{\CalH_{Y_1}^e}$ as
\begin{eqnarray}
\CalT_{d_{1}^{n},d_{2}^{n}, \eta}(\ket{h}) = \frac{1}{\sqrt{1+2\eta^2} } \left(\ket{h} +\eta \ket{h} \otimes  \ket{d_{1}^{n}}+\eta \ket{h} \otimes  \ket{d_{2}^{n}}\right),
\nonumber
\end{eqnarray} 
to tilt the state in all directions. As discussed earlier, this tilting is chosen carefully so as to ensure that the tilted state remains close to the original state in $\mathbb{L}_{1}$ norm. Prop.~\ref{Prop:3to1CQICclosnessofstates} establishes this thereby guaranteeing that the two states induce approximately the same measurement outcome statistics. Let
\begin{eqnarray}
    \theta_{x_1^n,d_1^n,u^n,d_2^n} \define \CalT_{d_1^n,d_2^n,\eta}\left( \rho_{x_1^n,u^n}^{Y_1} \otimes \ketbra{0}\right), \label{Eqn:3to1CQICtiltedstate}
\end{eqnarray}
be the tilted state along all directions, where  $\CalT_{d_1^n,d_2^n,\eta}$ acts on each pure state in the mixture individually.
\begin{proposition}
\label{Prop:3to1CQICclosnessofstates}
    For n sufficiently large, we have 
    \begin{eqnarray}
        \norm{\theta_{x_1^n,d_1^n,u^n,d_2^n} - \left( \rho_{x_1^n,u^n}^{Y_1} \otimes \ketbra{0}\right) }_1 \leq 4 \eta. \nonumber 
    \end{eqnarray}
\end{proposition}
\begin{proof}
    The proof is provided in Appendix \ref{App:3to1CQICClosenessOfStates}.
\end{proof}

\noindent To bound the term $t_2$, we evaluate conditional expectation over $U_2^n(a_2,m_2)$ and $U_3^n(a_3,m_3)$. This yields
\begin{eqnarray}
 \mathbb{E}\left[t_2\right]&=& \frac{1}{|\mathcal{\ulineM}|} \frac{1}{\Prime^{k_2 + k_3}} \frac{1}{\Prime^{k_{\sfb} + l_{\sfb}}} \sum_{\ulinem} \sum_{a_2, a_3} \sum_{s}  \sum_{u_2^n,u_3^n} \sum_{u^n} q_{U_2}^n(u_2^n) q_{U_3}^n(u_3^n)   r_{U_2}^n(u_2^n(a_2,m_2)) r_{U_3}^n(u_3^n(a_3,m_3))  \nonumber \\ 
 &&\mathds{1}{\{u^n=u_2^n \oplus u_3^n\}}\tr \left\{  \left( I - \mu^{Y_1}_{m_1,s} \right)  \left( \rho_{X_1^n(m_1),f_2^n(u_2^n),f_3^n(u^n_3)}^{Y_1} \otimes \ketbra{0}\right)  \right\} \nonumber \\
   &\overset{(a)}{=}& \frac{1}{|\mathcal{\ulineM}|} \frac{1}{\Prime^{k_2 + k_3}} \frac{1}{\Prime^{k_{\sfb} + l_{\sfb}}} \sum_{\ulinem} \sum_{a_2, a_3}  \sum_{s} \sum_{u_2^n,u_3^n} \sum_{u^n}  p_{U_2}^n(u_2^n) p_{U_3}^n(u_3^n)  \mathds{1}{\{u^n=u_2^n \oplus u_3^n\}} \nonumber \\
   &&\tr \left\{  \left( I - \mu^{Y_1}_{m_1,s} \right)  \left( \rho_{X_1^n(m_1),f_2^n(u_2^n),f_3^n(u^n_3)}^{Y_1} \otimes \ketbra{0}\right)  \right\} \nonumber \\
   &\overset{(b)}{=}& \frac{1}{|\mathcal{\ulineM}|} \frac{1}{\Prime^{k_2 + k_3}} \frac{1}{\Prime^{k_{\sfb} + l_{\sfb}}} \sum_{\ulinem} \sum_{a_2, a_3} \sum_{s} \sum_{u_2^n,u_3^n} \sum_{u^n} p_{U_2 \oplus U_3}^n(u^n) p_{U_2|U_2 \oplus U_3}^n(u_2^n|u^n) p_{U_3|U_2 \oplus U_3 U_2}^n(u_3^n|u^n,u_2^n)   \nonumber\\
   &&\tr \left\{  \left( I - \mu^{Y_1}_{m_1,s} \right)  \left( \rho_{X_1^n(m_1),f_2^n(u_2^n),f_3^n(u^n_3)}^{Y_1} \otimes \ketbra{0}\right)  \right\} \nonumber \\
   &\overset{(c)}{=}& \frac{1}{|\mathcal{M}_1|}  \frac{1}{\Prime^{k_{\sfb} + l_{\sfb}}} \sum_{m_1} \sum_{s}  \sum_{u^n} p_{U_2 \oplus U_3}^n(u^n) \tr \left\{  \left( I - \mu^{Y_1}_{m_1,s} \right)  \left( \rho_{X_1^n(m_1),u^n}^{Y_1} \otimes \ketbra{0}\right)  \right\} \nonumber \\
   &\overset{(d)}{\leq}& t_{2.1} + t_{2.2}, \mbox{ where }  \nonumber \\ 
   t_{2.1}&\define& \frac{1}{|\mathcal{M}_1|}  \frac{1}{\Prime^{k_{\sfb} + l_{\sfb}}} \sum_{m_1} \sum_{s}  \sum_{u^n} p_{U_2 \oplus U_3}^n(u^n) \tr \left\{  \left( I - \mu^{Y_1}_{m_1,s} \right)   \theta_{(X_1^n,D^n_1)(m_1),u^n,D^n_2(s)} \right\} \nonumber,   \mbox{ and } \\
    t_{2.2}&\define& \frac{1}{|\mathcal{M}_1|}  \frac{1}{\Prime^{k_{\sfb} + l_{\sfb}}} \sum_{m_1} \sum_{s}  \sum_{u^n} p_{U_2 \oplus U_3}^n(u^n) \norm{\theta_{(X_1^n,D^n_1)(m_1),u^n,D^n_2(s)} - \left(\rho_{X_1^n(m_1),u^n}^{Y_1} \otimes \ketbra{0}\right) }_1 \nonumber.
\end{eqnarray}
In the above chain of equalities/inequalities, (a) is obtained by using $q_{U_2}^n(u^n_2)r_{U_2}^n(u_2^n)=p_{U_2}^n(u^n_2)$ and $q_{U_3}^n(u^n_3)r_{U_3}^n(u_3^n)=p_{U_3}^n(u^n_3)$, (b) follows from 
\begin{eqnarray}
p_{U_2}^n(u_2^n) p_{U_3}^n(u_3^n)  \mathds{1}\{u^n=u_2^n \oplus u_3^n\}=p_{U_2 \oplus U_3}^n(u^n) p_{U_2|U_2 \oplus U_3}^n(u_2^n|u^n) p_{U_3|U_2 \oplus U_3 U_2}^n(u_3^n|u^n,u_2^n), \nonumber    
\end{eqnarray}
(c) is obtained from the definition of $\rho^{Y_1}_{X_1^n(m_1), u^n}$ in \eqref{Eqn:3to1CQICDecoder1States1} and (d) follows from
substituting the state $\left(\rho_{X_1^n(m_1),u^n}^{Y_1} \otimes \ketbra{0}\right)$ with the tilted state $\theta_{(X_1^n,D^n_1)(m_1),u^n,D^n_2(s)}$ defined in \eqref{Eqn:3to1CQICtiltedstate}. We simply bound the term $t_{2.2}$ by $4 \eta$ using Prop.~\ref{Prop:3to1CQICclosnessofstates}. To upper-bound the term $t_{2.1}$, we employ Hayashi-Nagaoka inequality \cite{200307TIT_HayNag}. We obtain, 
\begin{eqnarray}
 && \quad \quad \quad \quad \quad  t_{2.1} \leq 2 \: t_{2.1.1} + 4 \left( t_{2.1.2} + t_{2.1.3} + t_{2.1.4}\right), \mbox{ where}\nonumber\\ 
t_{2.1.1}&\define& \frac{1}{|\mathcal{M}_1|} \frac{1}{\Prime^{k_{\sfb} + l_{\sfb}}} \sum_{m_1} \sum_{s}  \sum_{u^n} p_{U_2 \oplus U_3}^n(u^n) \tr \left\{  \left( I - \gamma^{*}_{(X_1^n,D^n_1)(m_1),u^n,D_2^n(s)} \right)   \theta_{(X_1^n,D^n_1)(m_1),u^n,D^n_2(s)} \right\}, \nonumber  \\
  t_{2.1.2} &\define& \frac{1}{|\mathcal{M}_1|} \frac{1}{\Prime^{k_{\sfb} + l_{\sfb}}} \sum_{m_1} \sum_{\Tilde{m}_1 \neq m_1}\sum_{s}  \sum_{u^n} p_{U_2 \oplus U_3}^n(u^n) \tr \left\{  \gamma^{*}_{(X_1^n,D^n_1)(\Tilde{m}_1),u^n,D_2^n(s)}   \theta_{(X_1^n,D^n_1)(m_1),u^n,D^n_2(s)} \right\},\nonumber \\
  t_{2.1.3} &\define& \frac{1}{|\mathcal{M}_1|} \frac{1}{\Prime^{k_{\sfb} + l_{\sfb}}} \sum_{m_1} \sum_{s} \sum_{\Tilde{s} \neq s}  \sum_{u^n} p_{U_2 \oplus U_3}^n(u^n) \tr \left\{  \gamma^{*}_{(X_1^n,D^n_1)(m_1),(U^n,D_2^n)(\Tilde{s})}   \theta_{(X_1^n,D^n_1)(m_1),u^n,D^n_2(s)} \right\},\nonumber \\
  t_{2.1.4} &\define& \frac{1}{|\mathcal{M}_1|} \frac{1}{\Prime^{k_{\sfb} + l_{\sfb}}} \sum_{m_1} \sum_{\Tilde{m}_1 \neq m_1} \sum_{s} \sum_{\Tilde{s} \neq s} \sum_{u^n}  p_{U_2 \oplus U_3}^n(u^n) \tr \left\{ \gamma^{*}_{(X_1^n,D_1^n)(\Tilde{m}_1),(U^n,D_2^n)(\Tilde{s})} \theta_{(X_1^n,D_1^n)(m_1),u^n,D_2^n(s)}\right\}. \nonumber \\
    \nonumber 
\end{eqnarray}
In the following propositions, we provide the rate constraints required to bound these error terms.
\begin{proposition}
\label{Prop:Dec1firsttermHay}
    For any $\epsilon \in (0,1)$, and for all sufficiently small $\delta, \eta >0$ and sufficiently large $n$, we have $\mathbb{E}[t_{2.1.1}] \leq \epsilon$.
\end{proposition}
\begin{proof}
    The proof is provided in Appendix \ref{App:Dec1FirsttermHay}.
\end{proof}
\begin{proposition}
    \label{Prop:Dec1SecondtermHay}
    For any $\epsilon \in (0,1)$, and for all sufficiently small $\delta,\eta >0$ and sufficiently large $n$, we have $\sum_{i=2}^{4}\mathbb{E}[t_{2.1.i}] \leq \epsilon$ if the following inequalities holds.
\begin{eqnarray}
R_1  &<&  I(X_1;Y_1|U_2 \oplus U_3),\nonumber \\
\frac{(k_{\sfb}+l_{\sfb})}{n} \log(\Prime)  &<&  I(U_2 \oplus U_3;Y_1|X_1)  + \log(\Prime) - H(U_2 \oplus U_3),  \mbox{ and}\nonumber \\ 
        R_1 + \frac{(k_{\sfb}+l_{\sfb})}{n} \log(\Prime)  &<& I(X_1,U_2 \oplus U_3;Y_1)  +\log(\Prime) - H(U_2 \oplus U_3), \nonumber 
    \end{eqnarray}
    where all the mutual information quantities are computed with respect to the state 
    \begin{eqnarray}
    \rho^{U_2 \oplus U_3U_2U_3X_1Y_1} \define \sum_{u,u_2,u_3, x_1} p_{X_1}(x_1) p_{U_2}(u_2) p_{U_3}(u_3) \mathds{1}\{u=u_2 \oplus u_3\}  \ketbra{u,u_2,u_3, x_1} 
    \otimes \rho_{x_1,f_2(u_2),f_3(u_3)}^{Y_1}. \nonumber
\end{eqnarray}
\end{proposition}
\begin{proof}
    The proof is provided in Appendix \ref{App:Dec1SecondtermHay}.
\end{proof}
This completes analysis of error at Rx $1$. We now proceed to analyzing the second and third terms in \eqref{Eq:3to1CQICErrorprobabilty}.
\med\textit{\underline{Rx $2$ and $3$'s Error Analysis}}: Since Rx $2$ and $3$'s decoding POVMs and error probability terms are identical, we provide the analysis for Rx $2$ $-$ the second term in \eqref{Eq:3to1CQICErrorprobabilty}. Observe that 
\begin{eqnarray}
&&\frac{1}{|\mathcal{\ulineM}|} \sum_{\ulinem} \tr \left\{ \left( I - \mu_{m_2}^{Y_2} \right)  \rho_{x_1^n(m_1),f_2^n(u_2^n(a_2(m_2),m_2)),f_3^n(u_3^n(a_3(m_3),m_3))}^{Y_2}   \right\} \nonumber \\ 
&\leq& \frac{1}{|\mathcal{\ulineM}|} \sum_{\ulinem} \tr \left\{ \left( I - \Gamma_{m_2}^{Y_2} \right)  \rho_{x_1^n(m_1),f_2^n(u_2^n(a_2(m_2),m_2)),f_3^n(u_3^n(a_3(m_3),m_3))}^{Y_2}   \right\} \nonumber\\
&=& \frac{1}{|\mathcal{\ulineM}|} \sum_{\ulinem} \sum_{a_2, a_3}   \mathds{1}{\left\{a_2(m_2)=a_2\right\}} \mathds{1}{\left\{a_3(m_3)=a_3\right\}} \tr \left\{ \left( I - \Gamma_{m_2}^{Y_2} \right)  \rho_{x_1^n(m_1),f_2^n(u_2^n(a_2,m_2)),f_3^n(u_3^n(a_3,m_3))}^{Y_2}   \right\} \nonumber\\
&=& t_1 + t_2, \mbox{ where }  \nonumber\\
t_1&\define& \frac{1}{|\mathcal{\ulineM}|} \sum_{\ulinem} \sum_{a_2, a_3}  \left[\mathds{1}{\left\{a_2(m_2)=a_2\right\}} \mathds{1}{\left\{a_3(m_3)=a_3\right\}}-\frac{r_{U_2}^n(u_2^n(a_2,m_2))}{\Prime^{k_2}}\frac{r_{U_3}^n(u_3^n(a_3,m_3))}{\Prime^{k_3}}\right] \nonumber \\ &&\tr \left\{ \left( I - \Gamma_{m_2}^{Y_2} \right)  \rho_{x_1^n(m_1),f_2^n(u_2^n(a_2,m_2)),f_3^n(u_3^n(a_3,m_3))}^{Y_2}   \right\},\mbox{ and} \nonumber \\
t_2&\define&\frac{1}{|\mathcal{\ulineM}|} \frac{1}{\Prime^{k_2 + k_3}} \sum_{\ulinem} \sum_{a_2, a_3}  r_{U_2}^n(u_2^n(a_2,m_2)) r_{U_3}^n(u_3^n(a_3,m_3))  \tr \left\{ \left( I - \Gamma_{m_2}^{Y_2} \right)  \rho_{x_1^n(m_1),f_2^n(u_2^n(a_2,m_2)),f_3^n(u_3^n(a_3,m_3))}^{Y_2}   \right\}. \nonumber 
\end{eqnarray}
The first inequality follows from  $\mu_{m_2}^{Y_2} \geq\Gamma_{m_2}^{Y_2}$, where 
\begin{eqnarray}
  &&\hspace{-0.25in}\Gamma_{m_2}^{Y_2} \define \left( \sum_{\widehat{m}_2,\widehat{a}_2} \Pi^{Y_2} \Pi^{Y_2}_{u_2^n(\widehat{a}_2,\widehat{m}_2)} \Pi^{Y_2} \right)^{-\frac{1}{2}} \left( \Pi^{Y_2} \Pi^{Y_2}_{u_2^n(a_2(m_2),m_2)} \Pi^{Y_2} \right)  \left( \sum_{\widehat{m}_2,\widehat{a}_2} \Pi^{Y_2} \Pi^{Y_2}_{u_2^n(\widehat{a}_2,\widehat{m}_2)} \Pi^{Y_2} \right)^{-\frac{1}{2}}, \nonumber 
    \end{eqnarray}
and the last equality follows from adding and subtracting the product 
\begin{eqnarray}
    \frac{r_{U_2}^n(u_2^n(a_2,m_2))}{\Prime^{k_2}} \frac{r_{U_3}^n(u_3^n(a_3,m_3))}{\Prime^{k_3}}. \nonumber 
\end{eqnarray}    
We bound the term $t_1$ using the same arguments as we used to bound the term in \eqref{Eqn:3to1CQICSourceCodingBnd}.
\begin{proposition}
    For any $\epsilon \in (0,1)$, and for all sufficiently small $\delta$ and sufficiently large $n$, we have $\mathbb{E}[t_1] \leq \epsilon$ if
    \begin{eqnarray}
        \frac{k_j}{n} \log(\Prime) > \log(\Prime) - H(U_j), \mbox{ for } j=2,3. \nonumber
    \end{eqnarray}
\end{proposition}
\begin{proof}
    The proof follows the same arguments as in Appendix \ref{App:Dec1Cuffterm}.
\end{proof}
To bound the term $t_2$, we employ Hayashi-Nagaoka inequality \cite{200307TIT_HayNag}, we obtain 
\begin{eqnarray}
   && \quad \quad \quad \quad \quad  t_2 \leq 2 \: t_{2.1} + 4 \left( t_{2.2} +t_{2.3} \right),  \mbox{ where} \nonumber\\ 
t_{2.1}&\define&\frac{1}{|\mathcal{\ulineM}|} \frac{1}{\Prime^{k_2 + k_3}} \sum_{\ulinem} \sum_{a_2, a_3}  r_{U_2}^n(u_2^n(a_2,m_2)) r_{U_3}^n(u_3^n(a_3,m_3)) \nonumber \\ && \tr \left\{ \left( I - \Pi^{Y_2} \Pi^{Y_2}_{u_2^n(a_2,m_2)} \Pi^{Y_2} \right)  \rho_{x_1^n(m_1),f_2^n(u_2^n(a_2,m_2)),f_3^n(u_3^n(a_3,m_3))}^{Y_2}   \right\}, \nonumber \\
  t_{2.2}&\define&\frac{1}{|\mathcal{\ulineM}|} \frac{1}{\Prime^{k_2 + k_3}} \sum_{\ulinem} \sum_{a_2, a_3}  \sum_{\Tilde{a}_2 \neq a_2}r_{U_2}^n(u_2^n(a_2,m_2)) r_{U_3}^n(u_3^n(a_3,m_3)) \nonumber \\ &&\tr \left\{ \Pi^{Y_2} \Pi^{Y_2}_{u_2^n(\Tilde{a}_2,m_2)} \Pi^{Y_2}  \rho_{x_1^n(m_1),f_2^n(u_2^n(a_2,m_2)),f_3^n(u_3^n(a_3,m_3))}^{Y_2}   \right\}, \nonumber \\
t_{2.3}&\define&\frac{1}{|\mathcal{\ulineM}|} \frac{1}{\Prime^{k_2 + k_3}} \sum_{\ulinem} \sum_{a_2, a_3}  \sum_{\Tilde{m}_2 \neq m_2} \sum_{\Tilde{a}_2}r_{U_2}^n(u_2^n(a_2,m_2)) r_{U_3}^n(u_3^n(a_3,m_3)) \nonumber \\
&& \tr \left\{ \Pi^{Y_2} \Pi^{Y_2}_{u_2^n(\Tilde{a}_2,\Tilde{m}_2)} \Pi^{Y_2}  \rho_{x_1^n(m_1),f_2^n(u_2^n(a_2,m_2)),f_3^n(u_3^n(a_3,m_3))}^{Y_2} \right\}. \nonumber 
\end{eqnarray}
As is evident to an informed reader, the first term above $t_{2.1}$ falls exponentially with the typicality parameter, placing no constraints on the rate. Each of the terms $t_{2.2},t_{2.3}$ falls exponentially if the rate constraint stated in Prop.~\ref{Prop:Dec2SecondtermHay} is satisfied. We refer the reader to the corresponding appendices for a proof.
\begin{proposition}
    \label{Prop:Dec2FirsttermHay}
    For any $\epsilon \in (0,1)$, and for all sufficiently small $\delta$ and sufficiently large $n$, we have $\mathbb{E}[t_{2.1}] \leq \epsilon$.
\end{proposition}
\begin{proof}
    The proof is provided in Appendix \ref{App:Dec2FirsttermHay}.
\end{proof}
\begin{proposition}
\label{Prop:Dec2SecondtermHay}
    For any $\epsilon \in (0,1)$, and for all sufficiently small $\delta$ and sufficiently large $n$, we have $\mathbb{E}[t_{2.2}] + \mathbb{E}[t_{2.3}] \leq \epsilon$, if the following bounds hold
    \begin{eqnarray} 
    \frac{(k_2+l_2)}{n}\log(\Prime)  <  I(U_2;Y_2) +\log(\Prime) - H(U_2) , \nonumber
    \end{eqnarray}
    where all the mutual information quantities are computed with respect to the state
\begin{eqnarray}
\label{Eqn:3to1CQICCQstateDec2}
&&\hspace{-0.25in}\rho^{U_2U_3X_1Y_2} \define \sum_{u_2,u_3,x_1} p_{X_1}(x_1) p_{U_2}(u_2) p_{U_3}(u_3) \ketbra{u_2,u_3,x_1} \otimes \rho^{Y_2}_{x_1,f_2(u_2),f_3(u_3)}.
\end{eqnarray}
\end{proposition}
\begin{proof}
    The proof is provided in Appendix \ref{App:Dec2SecondtermHay}.
\end{proof}
{We have thus bounded the error probability at Rx $2$ - the second term in \eqref{Eq:3to1CQICErrorprobabilty}. As stated earlier, error analysis of Rx $3$ - the third term in \eqref{Eq:3to1CQICErrorprobabilty} is identical to that of the second term and is therefore omitted.} This completes the proof.
\end{proof}

\begin{theorem}
\label{Thm:3to1CQICRateRegion}
A rate-cost triple $(R_1,R_2,R_3,\tau_1,\tau_2,\tau_3)$ is achievable if there exists (i) a finite field $\CalU_2 = \CalU_3 = \CalF_{\Prime}$ (ii) PMFs $p_{X_1}$, $p_{U_2X_2}$, $p_{U_3X_3}$ such that $\mathbb{E}[\kappa(X_j)] \leq \tau_j : j \in [3]$,
\begin{eqnarray}
  R_1  &<& \min\{ 0,H(U_{j})-H(U_{2}\oplus U_{3}|Y_{1}):j=2,3\}+I(X_1;Y_1,U_{2} \oplus U_{3}), \nonumber \\
  R_j &<& I(X_j, U_j;Y_j),\nonumber \\
    R_j &<& I(U_{2} \oplus U_{3};Y_1|X_1) - H(U_2 \oplus U_3) + \min\{I(X_2, Y_2|U_2) +H(U_2), I(X_3,Y_3|U_3) + H(U_3)\}, \mbox{ and} \nonumber \\
    R_1 + R_j &<& I(X_{j};Y_{j}|U_{j})+I(X_{1};U_{2}\oplus U_{3},Y_{1})+H(U_{j})-H(U_{2}\oplus U_{3}|Y_{1}): j=2,3 \mbox{ holds},\nonumber
\end{eqnarray}
 with all information quantities being computed with respect to the state
\begin{eqnarray}
    \rho^{U_2 \oplus U_3 U_2 U_3 \ulineX \ulineY} \define \sum_{u,u_2,u_3,\ulinex} p_{X_1}(x_1) p_{U_2X_2}(u_2,x_2) p_{U_3X_3}(u_3,x_3) \mathds{1}{\{u=u_2 \oplus u_3\}} \ketbra{u,u_2,u_3,\ulinex} \otimes \rho^{Y_1Y_2Y_3}_{x_1x_2x_3}. \nonumber 
\end{eqnarray}
\end{theorem}

\begin{figure}
\centering
\includegraphics[width=4.8in]{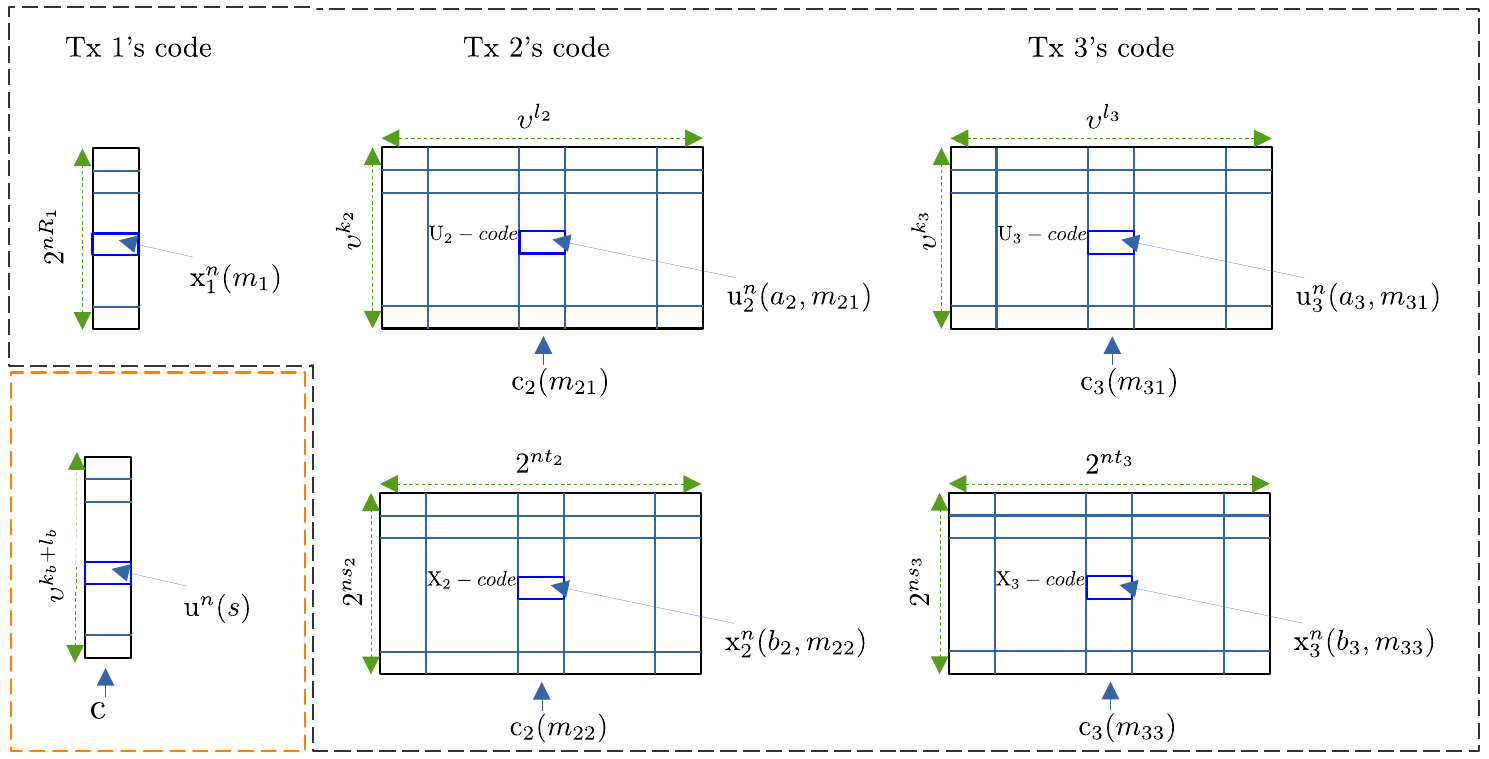}
    \caption{The coding strategy involves five codes depicted in the grey box - two of them being NCCs built on the finite field $\mathcal{F}_{\Prime}$ and three unstructured IID code built on $\CalX_{j}$ for $j \in [3]$. 
    Rx $1$, in addition to decoding into its code, decodes into the collection of `codewords' in the orange dotted box. Specifically, the collection of the codewords in the orange dotted box is obtained by adding all pairs of codewords in the $U_{2}$ and $U_{3}$ codes.}
    \label{Fig:Codestructure3to1CQICThm1}
    \vspace{-0.15in}
\end{figure}

\begin{proof}
In comparison to the proof of Lemma \ref{Lemma:3to1CQICRateRegion}, the only differences (new elements) in this proof relate to users $2$ and $3$. Specifically, instead of using a single coset code for encoding their message, Txs $2,3$ split their message into two parts, encode one of the parts via a coset code, as in the proof of Lemma \ref{Lemma:3to1CQICRateRegion}, and the other via an unstructured IID code built over $\CalX_{j}:j=2,3$. See Fig.~\ref{Fig:Codestructure3to1CQICThm1}. Rxs $2,3$ employ a simultaneous decoder to decode into their codebooks. For this, we employ the technique of Fawzi et.~al.~\cite{201206TIT_FawHaySavSenWil} to perform simultaneous decoding as we are required to only decode into two codebooks. In the interest of brevity, we do not provide these analysis steps here. We refer the reader to \cite[Proof of Thm.~1]{202503arXivBC_GouPad} provided in the context of a $3-$user quantum broadcast channel. The analysis steps therein can be straightforwardly adopted here.
\end{proof}

\subsection{Coset codes are strictly more efficient for Ex.~\ref{Ex:3CQICadditivenoncommutative} and Ex.~\ref{Ex:nonadditiveinterference}}
\label{Subsec:CosetCdsNonComEx}

{We demonstrate that the rate triple $(\mathcal{C}_1,\mathcal{C}_2,\mathcal{C}_3)$ is achievable for Ex.~\ref{Ex:3CQICadditivenoncommutative} and \ref{Ex:nonadditiveinterference} using coset codes. Thus proving that the latter is strictly more efficient for the additive, non-additive, non-commutative instances. We begin by stating a fact that will be used in the proofs of the following two propositions. Recall the notation defined in \eqref{notation:examples}.}
\begin{fact}
\label{fact:examples}
    Suppose we have a state $\sigma^{AB}= \alpha \ketbra{0}^A \otimes \ketbra{0}^B + (1- \alpha) \ketbra{1}^A \otimes \ketbra{v_{\varphi}}^B$, where $\alpha \in [0,1]$. It is easy to verify that  
    \begin{eqnarray}
        I(A;B) = h_b(f(1- \alpha)),  \mbox{ where }  f(t)=\frac{1 + \sqrt{1-4t(1-t)\sin(\varphi)^2}}{2},  \mbox{ for }   t \in [0,1].
        \nonumber
    \end{eqnarray}
\end{fact}

\begin{proposition}
\label{Prop:NoncomAddCstCds}
    Consider the $3-$CQIC in Ex.~\ref{Ex:3CQICadditivenoncommutative} under condition \eqref{Eqn:commonconditionfortheexamples}. The rate triple $(\CalC_1,\CalC_2,\CalC_3)$ is achievable using coset codes as long as $C_1> \mathcal{C}_1 + \max\{\mathcal{C}_2,\mathcal{C}_3\} $.
\end{proposition}
\begin{proof}
    It can be verified that $\CalC_j =1 - h_b(\delta_j)$ for $j=2,3$. From Fact \ref{fact:examples}, we have $C_1=h_b\left(\frac{1 + \cos(\varphi)}{2}\right)$ and $\CalC_1 = h_b\left(\frac{1 + \sqrt{1-4\tau(1-\tau)\sin(\varphi)^2}}{2}\right)$.
We only need to identify an appropriate PMF $p_{X_1}p_{U_2X_2}p_{U_3X_3}$ such that $(\CalC_1,\CalC_2,\CalC_3)$ is achievable using coset codes. Let $\CalU_2=\CalU_3=\CalF_2$. Define the PMF $p_{X_1}p_{U_2X_2}p_{U_3X_3}$ on $\CalU_2 \times \CalU_3 \times \mathcal{\ulineX}$ such that (i) $p_{X_1}(1)=\tau$, and (ii) $p_{U_jX_j}(1,1)= 1 - p_{U_jX_j}(0,0)=\frac{1}{2}$ for $j=2,3$. 
It is straightforward to verify that the PMF $p_{X_1}p_{U_2X_2}p_{U_3X_3}$ satisfies the input cost constraint and that the rate triple $(\CalC_1,\CalC_2,\CalC_3)$ satisfies the bounds in Thm.~\ref{Thm:3to1CQICRateRegion}.
\end{proof}
\begin{proposition}
\label{Prop:AchievabilityofcosetCdsnonadditiveexample}
    Consider the $3-$CQIC in Ex.~\ref{Ex:nonadditiveinterference} under condition \eqref{Eqn:commonconditionfortheexamples}. The rate triple $(\CalC_1,\CalC_2,\CalC_3)$ is achievable using coset codes, as long as 
    \begin{eqnarray}
        \mathcal{C}_1 + \mathcal{C}_2 +\mathcal{C}_3 < \vartheta, \nonumber  
    \end{eqnarray}
    where $\vartheta \define \min\{h_b(\tau_j) + (1-\tau_1)(1-\tau_2) \log \left( (1-\tau_1)(1-\tau_2) \right) + (\tau_1 * \tau_2) \log(\tau_1*\tau_2) + \tau_1 \tau_2 \log(\tau_1 \tau_2) -h_b(f(\tau_1)) + h_b(f(\tau_1 * \beta)) : j=2,3\} + h_b(f(\tau_1))$ and $f(t)= \frac{1+\sqrt{1-4t(1-t)\sin(\varphi)^2}}{2}$, for $t \in [0,\frac{1}{2}]$.
\end{proposition}
\begin{proof}
    It can be verified that $\CalC_j =h_b(\tau_j * \delta_j) -h_b(\delta_j)$ for $j=2,3$. From Fact \ref{fact:examples}, we have $\CalC_1 = h_b(f(\tau_1))$ and $C_1=h_b(f(\tau_1 * \beta))$, where $\beta= \tau_2 + \tau_3 - \tau_2 \tau_3$ and $f(t)=\frac{1+\sqrt{1-4t(1-t)\sin(\varphi)^2}}{2}$ for $t \in [0,\frac{1}{2}]$.
We only need to identify an appropriate PMF $p_{X_1}p_{U_2X_2}p_{U_3X_3}$ such that $(\CalC_1,\CalC_2,\CalC_3)$ is achievable using coset codes. Let $\CalU_2=\CalU_3=\CalF_3$, and let $p_{X_1}p_{U_2X_2}p_{U_3X_3}$ be a PMF defined on $\CalU_2 \times \CalU_3 \times \mathcal{\ulineX}$, where (i) $p_{X_1}(1)=\tau_1$, and (ii) $p_{U_jX_j}(1,1)= 1 - p_{U_jX_j}(0,0)=\tau_j,p_{U_{j}}(2)=0$ for $j=2,3$. 
It is straightforward to verify that the PMF $p_{X_1}p_{U_2X_2}p_{U_3X_3}$ satisfies the input cost constraint and that the rate triple $(\CalC_1,\CalC_2,\CalC_3)$ satisfies the bounds in Thm.~\ref{Thm:3to1CQICRateRegion}, as long as $ \mathcal{C}_1 + \mathcal{C}_2 +\mathcal{C}_3 < \vartheta$.
\end{proof}

\section{Simultaneous Decoding of Coset Codes Over The $3-$CQIC}
\label{Sec:SimultDecOfCosetcds3CQIC}
In Sec.~\ref{Sec:3to1CQICAchivablerateregion}, we provided a proof that includes all the new elements in a simplified setting, which was the focus of our first step. We now proceed to obtain a generalized inner bound. As discussed in Remark \ref{Rem:nonadditiveexample}, we now describe a general coding strategy that employs IID codes to decode uni-variate components and coset codes to decode bi-variate components of the interference at \textit{each} Rx. We begin by introducing some notation.

\begin{comment}{As discussed in Sec.~\ref{Sec:Introduction}, we present this inner bound in two steps. Step II presented in Thm.~\ref{Thm:3CQICStageIRateRegion} enables Rxs decode bi-variate interference via simultaneous decoding based on TSA \cite{202103SAD_Sen}. We then characterize an inner bound in Thm.~\ref{Thm:3CQICStageIIRateRegion} obtained when we append the unstructured IID code as another layer.}\end{comment}

\begin{Notation}
    \label{Not:3CQICStep2}
    In any context, when used in conjunction, $i,j,k$ will denote distinct indices in $[3]$, hence $\{i,j,k\}=[3]$ and we let $\dbrackthree \define \{(1,2),(1,3),(2,1),(2,3),(3,1),(3,2)\}$.
\end{Notation}

\begin{figure}
\centering
\includegraphics[width=5in]{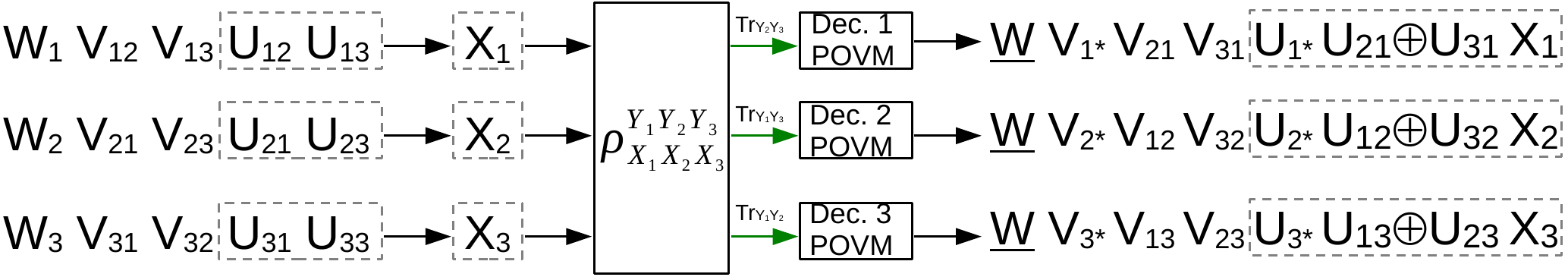}
    \caption{Depiction of all Rvs in the full blown coding strategy. In Sec.~\ref{SubSec:StageICodeStrategy3CQIC} (Step II) only Rvs in the gray dashed box are non-trivial, with the rest trivial.}
    \label{Fig:3CQICRVsInFullFinalStrategy}
    \vspace{-0.15in}
\end{figure}
Tx $j$ splits its message into $6$ parts $-$ public $W_{j}$, semi-private $V_{j*}\define (V_{ji},V_{jk})$, `bi-variate' $U_{j*} \define( U_{ji},U_{jk})$ and private $X_{j}$. See Fig.~\ref{Fig:3CQICRVsInFullFinalStrategy}. In this coding strategy, the support of $U_{ji}$ and $U_{jk}$ are the finite fields $\CalF_{\Prime_{i}}$ and $\CalF_{\Prime_{k}}$, respectively, while the supports of the remaining variables are arbitrary finite sets. Rx $k$ decodes all public parts $\ulineW \define (W_{1},W_{2},W_{3})$, semi-private parts $V_{k*}$, `bi-variate' $U_{k}^{\oplus}\define U_{ik}\oplus U_{jk}, U_{k*}$, and finally its private part $X_{k}$. We present this coding strategy and the corresponding inner bound in two steps - Step II in Sec.~\ref{SubSec:StageICodeStrategy3CQIC} and Step III in Sec.~\ref{SubSec:Step2CodingStrategy3CQIC}. \begin{comment}{Building on Step~1 in Sec.~\ref{Sec:3to1CQICAchivablerateregion}, Step~2 in Sec.~\ref{SubSec:StageICodeStrategy3CQIC} will manage bi-variate interference at all Rx, and Step~3 in Sec.~\ref{SubSec:Step2CodingStrategy3CQIC} will provide the full coding strategy and the corresponding inner bound.}\end{comment}
\begin{figure}
\centering
\includegraphics[width=4in]{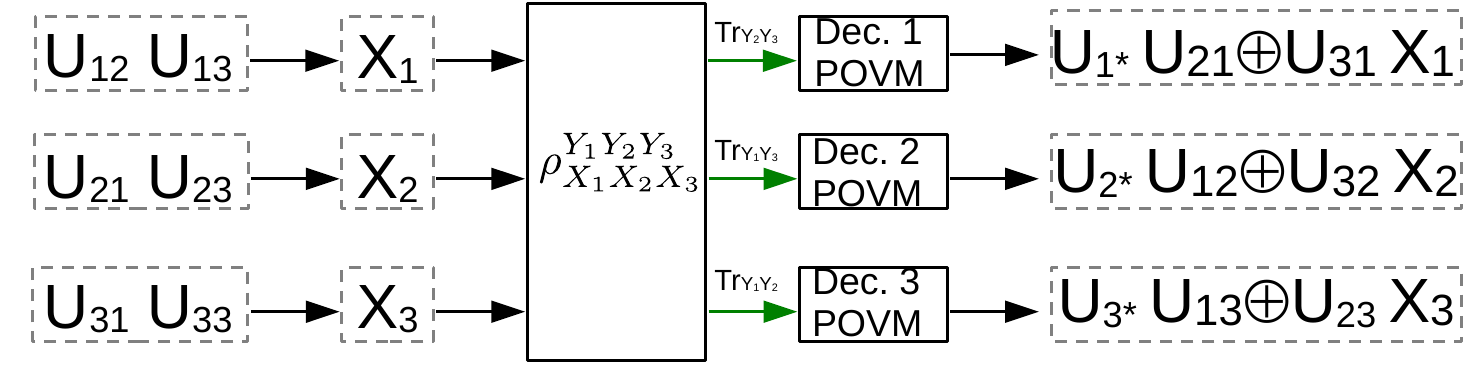}
    \caption{Depiction of the non-trivial Rvs active in Step II (Sec.~\ref{SubSec:StageICodeStrategy3CQIC}) of the coding strategy.}
    \label{Fig:3CQICRVs(step1)}
    \vspace{-0.15in}
\end{figure}

\subsection{Step II : Simultaneous Decoding of bi-variate Components}
\label{SubSec:StageICodeStrategy3CQIC}

In this second step, we activate only the `bi-variate' $\ulineU \define (U_{1*},U_{2*},U_{3*})$ and private parts $\ulineX \define (X_{1},X_{2},X_{3})$, choosing them to be non-trivial, with the rest $\ulineW=\phi,\ulineV \define (V_{1*} ,V_{2*}, V_{3*}) = \phi$ trivial (Fig.~\ref{Fig:3CQICRVs(step1)}). $\ulineX$ is coded using unstructured IID codes and $\ulineU$ is coded via coset codes.
\begin{theorem}
 \label{Thm:3CQICStageIRateRegion}
 Let $\hat{\alpha}_{S} \in [0,\infty)^{6}$ be the set of all rate-cost vectors $(\ulineR,\ulinetau)$ for which there exists for all $j \in [3]$, (i) finite fields $\SemiPrivateRVSet_{ji}=\CalF_{\Prime_{i}}, \SemiPrivateRVSet_{jk}=\CalF_{\Prime_{k}}$, (ii) PMF $p_{\ulineU\,\ulineX} =  p_{U_{1*}U_{2*}U_{3*}X_{1}X_{2}X_{3}}=\displaystyle\prod_{j=1}^{3}p_{U_{ji}U_{jk}X_{j}}$ on $\ulineCalU\times\ulineCalX$, (iii) non-negative numbers $S_{ji},S_{jk},T_{ji},T_{jk},K_{j},L_{j}$ such that $\Expectation\{\kappa_{j}(X_{j})\}\leq \tau_{j}$, $R_{j}=T_{ji}  +T_{jk}  +L_{j}$ and
\begin{eqnarray}
S_{A_{j}}-T_{A_{j}}+K_{j} &>& \sum_{a_{j} \in A_{j}} \log |\CalU_{a_{j}}|
\label{Eqn3CQIC:3CQICStep1SrcBnd1}
+ H(X_{j})- H(U_{A_{j}},X_{j}),\\
\label{Eqn3CQIC:3CQICStep1SrcBnd2}
S_{A_{j}}-T_{A_{j}} &>& \sum_{a \in A_{j}}\log |\CalU_{a}| - H(U_{A_{j}}),\\
\label{Eqn3CQIC:3CQICStep1ChnlBnd1}
S_{A_{j}} &<& \sum_{a \in A_{j}} \log |\CalU_{a}| - H(U_{A_{j}}|U_{A_{j}^{c}},U_{j}^{\oplus},X_{j},Y_{j}),
\\
S_{A_{j}}+S_{ij} &<& \sum_{a \in A_{j}} \log |\CalU_{a}| + \log \Prime_{j}
\label{Eqn3CQIC:3CQICStep1ChnlBnd2}
- H(U_{A_{j}},U_{j}^{\oplus}|U_{A_{j}^{c}},X_{j},Y_{j}),  \\
\label{Eqn3CQIC:3CQICStep1ChnlBnd3}
S_{A_{j}}+S_{kj} &<& \sum_{a \in A_{j}}\log |\CalU_{a}| + \log \Prime_{j}
- H(U_{A_{j}},U_{j}^{\oplus}|U_{A_{j}^{c}},X_{j},Y_{j}),\\
S_{A_{j}} +K_{j}+L_{j} &<&  \sum_{a \in A_{j}} \log |\CalU_{a}|+H(X_{j})
\label{Eqn3CQIC:3CQICStep1ChnlBnd4}
- H(U_{A_{j}},X_{j}|U_{A_{j}^{c}},U_{j}^{\oplus},Y_{j}), \\
S_{A_{j}}+K_{j}+L_{j}+S_{ij} &<& \sum_{a \in A_{j}} \log |\CalU_{a}| + \log \Prime_{j}+H(X_{j})
\label{Eqn3CQIC:3CQICStep1ChnlBnd5}
- H(U_{A_{j}},X_{j},U_{j}^{\oplus} |U_{A_{j}^{c}},Y_{j}), \\
S_{A_{j}}+K_{j}+L_{j}+S_{kj} &<& \sum_{a \in A_{j}}\log |\CalU_{a}| +\log \Prime_{j}+H(X_{j})
\label{Eqn3CQIC:3CQICStep1ChnlBnd6}
- H(U_{A_{j}},X_{j},U_{j}^{\oplus}|U_{A_{j}^{c}},Y_{j}),
\end{eqnarray}
holds for every $j \in [3]$, $A_{j} \subseteq \{ji,jk\}$, where $U_{j}^{\oplus}= U_{ij}\oplus U_{kj}$, $S_{A_{j}}
\define \sum_{\nu \in A_{j}}S_{\nu}, U_{A_{j}} = (U_{a_{j}}:a_{j}
\in A_{j})$, where all information quantities are computed with respect to the state 
\begin{eqnarray}
 \label{Eqn3CQIC:StageITestChnl}
\hspace{-0.02in}\Gamma^{\ulineU\!\!~\ulineU^{\oplus}\!\!~\ulineX\!\!~\ulineY}\define \!\!\sum_{\substack{u_{12},u_{13}\\u_{21},u_{23}\\u_{31},u_{32} }}\sum_{\substack{x_{1},x_{2},x_{3}\\u_{1}^{\oplus},u_{2}^{\oplus},u_{3}^{\oplus}}}\!\prod_{j=1}^{3}p_{U_{ji}U_{jk}X_{j}}(u_{ji},u_{jk},x_{j})\mathds{1}{\{u_{j}^{\oplus}=u_{ij}\oplus u_{kj}\}} \ketbra{u_{1*} u_{2*} u_{3*} u_{1}^{\oplus} u_{2}^{\oplus} u_{3}^{\oplus} \ulinex} \!\otimes\! \rho_{\ulinex}.\nonumber
\end{eqnarray}
Let $\alpha_{S}$ denote the convex closure of $\hat{\alpha}_{S}$. Then $\alpha_{S} \subseteq \ScrC$.
\end{theorem}
\begin{figure}
\centering
\includegraphics[width=4in]{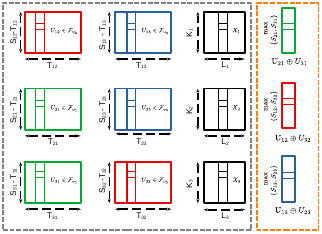}
    \caption{$9$ codes employed in the coding strategy are depicted in the grey box above. The $U_{ji} : ji \in \dbrackthree$ are coset codes built over finite fields. Codes with the same color are built over the same finite field, and the smaller of the two is a sub-coset of the larger. The black codes are built over the finite sets $\CalX_{j}$ using the conventional IID random code structure. Row $j$ depicts the codes of Tx, Rx $j$. Rx $j$, in addition to decoding into codes depicted in the row $j$, also decodes $U_{ij}\oplus U_{kj}$. The three collections of codewords in the orange dotted box on the right are employed only by the corresponding Rx. Specifically, the first collection of the codewords at the top of the orange dotted box is obtained by adding all pairs of codewords in the $U_{21}$ and $U_{31}$ codes.}
    \vspace{-0.15in}
    \label{Fig:FigCodeStructureCQIC(StepI)}
\end{figure}

\begin{remark}
 \label{Rem:Step1CodingThmRemarks}
 Inner bound $\alpha_{S}$ (i) subsumes \cite[Thm.~1]{202103arXiv_AnwPadPra3CQIC, 202107ISIT_AnwPadPra3CQIC}, (ii) is the CQ analogue of \cite[Thm.~3]{201603TIT_PadSahPra}, (iii) does \textit{not} include a time-sharing Rv and \textit{includes} the `don't care' inequalities \cite{200702ITA_KobHan,200807TIT_ChoMotGarElg} - \eqref{Eqn3CQIC:3CQICStep1ChnlBnd2}, \eqref{Eqn3CQIC:3CQICStep1ChnlBnd3} with $A_{j}=\phi$. Thus, $\alpha_{S}$ can be enlarged.
 \end{remark}
 \med\textit{\textbf{Discussion of the bounds}}: We \textit{explain} how (i) each bound arises and (ii) their role in driving error probability down. Given a message $m_{j} = (m_{ji},m_{jk},m_{jj})$ with three parts, Tx $j$ employs a likelihood encoder. On analyzing the error probability, the term corresponding to \eqref{Eqn:3to1CQICSourceCodingBnd} will yield 15 source coding bounds \cite{201604ITIT_SonCufPoo} that are identical to those one would obtain if Tx $j$ were attempting to find a triple of codewords jointly typical with respect to PMF $p_{U_{ji}U_{jk}X_{j}}$. \eqref{Eqn3CQIC:3CQICStep1SrcBnd1}, \eqref{Eqn3CQIC:3CQICStep1SrcBnd2} are indeed these source coding bounds. To recognize this, throughout this discussion, we rename the Rvs $Z_{1}=U_{ji}, Z_{2}=U_{jk}, Z_{3}=X_{j}$, PMFs $r_{\ulineZ}=r_{Z_{1}Z_{2}Z_{3}}=p_{U_{ji}U_{jk}X_{j}}$, $q_{\ulineZ}=q_{Z_{1}Z_{2}Z_{3}}= \frac{p_{X_{k}}}{|\CalU_{ji}|\cdot|\CalU_{jk}|}$ and bin sizes $B_{1}=S_{ji}-T_{ji},B_{2}=S_{jk}-T_{jk}, B_{3}=K_{j}$. Since $c_{ji} \define \{u_{ji}^n(a_{ji}, m_{ji}) : a_{ji} \in \mathcal{F}_{\Prime_i}^{s_{ji}-t_{ji}}, m_{ji} \in \mathcal{F}_{\Prime_i}^{t_{ji}}\}, c_{jk} \define \{u_{jk}^n(a_{jk}, m_{jk}) : a_{jk} \in \mathcal{F}_{\Prime_k}^{s_{jk}-t_{jk}}, m_{jk} \in \mathcal{F}_{\Prime_k}^{t_{jk}}\}$ are random coset codes \cite{202103arXiv_AnwPadPra3CQIC} with uniformly distributed codewords, and codewords of $X_{j}$ picked IID with respect to $\prod p_{X_{j}}$, any triple of codewords is distributed with PMF $q_{\ulineZ}^{n}$ (Fig.~\ref{Fig:FigCodeStructureCQIC(StepI)}). Standard source coding bounds to find one triple to be jointly typical with respect to $r_{\ulineZ}$ among $2^{n(B_{1}+B_{2}+B_{3})}$ triples are $B_{A} \stackrel{(A)}{>} D(r_{Z_{A}|Z_{A^{C}}}||q_{Z_{A}|Z_{A^{C}}}|r_{Z_{A^{C}}})$, for every $A \subseteq \{1,2,3\}$, where $Z_{A} = (Z_{\iota}:\iota \in A)$ and averaged divergence is as defined in \cite[Eqn.~2.4]{BKCsiszarKornerInfoTheory_2011}. The 7 bounds in \eqref{Eqn3CQIC:3CQICStep1SrcBnd1}, \eqref{Eqn3CQIC:3CQICStep1SrcBnd2} are just the 7 bounds $(A)$ for different choices of $A\subseteq \{1,2,3\}$. This is verified by direct substitution of $q_{\ulineZ},r_{\ulineZ}$. Imposing \eqref{Eqn3CQIC:3CQICStep1SrcBnd1}, \eqref{Eqn3CQIC:3CQICStep1SrcBnd2} guarantees Txs find a triple jointly typical with respect to $p_{U_{ji}U_{jk}X_{j}}$.
\noindent Now to channel coding bounds \eqref{Eqn3CQIC:3CQICStep1ChnlBnd1}-\eqref{Eqn3CQIC:3CQICStep1ChnlBnd6}. Our decoding analysis is a simple CQMAC decoding at each Rx. We shall therefore relate \eqref{Eqn3CQIC:3CQICStep1ChnlBnd1}-\eqref{Eqn3CQIC:3CQICStep1ChnlBnd6} to the bounds one obtains on a CQMAC with $4$ Txs. To the earlier renaming, we add $Z_{4}=U_{j}^{\oplus}\!=U_{ij}\oplus U_{kj}$, $Z_{5}=Y_{j}$, $\olineZ=Z_{1}Z_{2}Z_{3}Z_{4}Z_{5}$,
\begin{eqnarray}
  \label{Eqn3CQIC:Step1ExplanationChnlBnds}
  \ScrR^{\olineZ}\define \Gamma^{U_{ji}U_{jk}X_{j}U_{j}^{\oplus}Y_{j}} = \tr_{U_{i*}U_{k*}X_{i}X_{k}U_{i}^{\oplus}U_{k}^{\oplus}Y_{i}Y_{k}}\{\Gamma^{\ulineU\!\!~\ulineU^{\oplus}\!\!~\ulineX\!\!~\ulineY}\},
  \nonumber\\
    \ScrQ^{Z_{1}Z_{2}Z_{3}Z_{4}}\define \!\!\!\!\!\!\!\!\!\sum_{\substack{(z_{1},z_{2},z_{3},z_{4})\\\in \CalF_{\Prime_{i}}\times \CalF_{\Prime_{k}}\times \CalX_{j}\times \CalF_{\Prime_{j}}}}\!\!\!\!\!\!\!\!\!\frac{p_{X_{j}}(z_{3})\ketbra{z_{1}~z_{2}~z_{3}~z_{4}}}{|\CalU_{ji}|\cdot |\CalU_{jk}| \cdot| \CalU_{ij}|}
    \nonumber
 \end{eqnarray}
$C_{1}=S_{ji}, C_{2}=S_{jk},C_{3}=K_{j}+L_{j}, C_{4}=\max\{S_{kj},S_{ij}\}$. The $7$ bounds in \eqref{Eqn3CQIC:3CQICStep1ChnlBnd1}, \eqref{Eqn3CQIC:3CQICStep1ChnlBnd4} are in fact the $7$ bounds $C_{A} < D(\ScrR^{\olineZ}||\ScrR^{\olineZ_{A}} \otimes \ScrR^{\olineZ_{A^{C}}} )+D(\ScrR^{\olineZ_{A}}||\ScrQ^{\olineZ_{A}})$ obtained by different choices of $A \subseteq\{1,2,3\}$, and the 16 bounds in \eqref{Eqn3CQIC:3CQICStep1ChnlBnd2}, \eqref{Eqn3CQIC:3CQICStep1ChnlBnd3}, \eqref{Eqn3CQIC:3CQICStep1ChnlBnd5}, \eqref{Eqn3CQIC:3CQICStep1ChnlBnd6} are in fact the $8$ bounds $C_{A\cup\{4\}} < D(\ScrR^{\olineZ}||\ScrR^{\olineZ_{A\cup \{4\}}} \otimes \ScrR^{\olineZ_{A^{C}\setminus\{4\}}} )+D(\ScrR^{\olineZ_{A\cup\{4\}}}||\ScrQ^{\olineZ_{A\cup \{4\}}})$ obtained by different choices of $A \subseteq\{1,2,3\}$ but now replicated twice owing to the $\max$ in the definition of $C_{4}$. Having identified the bounds, let us interpret the upper-bound $D(\ScrR^{\olineZ}||\ScrR^{\olineZ_{A}} \otimes \ScrR^{\olineZ_{A^{C}}} )+D(\ScrR^{\olineZ_{A}}||\ScrQ^{\olineZ_{A}})$. Observe that, if $\ScrQ^{Z_{1}Z_{2}Z_{3}Z_{4}}=\ScrR^{Z_{1}Z_{2}Z_{3}Z_{4}}$, then the upper-bound will just have one term which is essentially the bound obtained on a CQMAC. The addition of the $D(\ScrR^{\olineZ_{A}}||\ScrQ^{\olineZ_{A}})$ is due to the fact that codewords of random code are \textit{not} distributed with PMF $p_{U_{ji}U_{jk}X_{j}U_{j}^{\oplus}}$, i.e., $\Gamma^{U_{ji}U_{jk}X_{j}}$, but with respect to $\ScrD^{Z_{1}Z_{2}Z_{3}Z_{4}}$. This explains \eqref{Eqn3CQIC:3CQICStep1ChnlBnd1}-\eqref{Eqn3CQIC:3CQICStep1ChnlBnd6}. Imposing the same guarantees correct decoding at each Rx.

\begin{table}
\renewcommand{\arraystretch}{1}
\setlength{\extrarowheight}{1pt}
\centering
\resizebox{7in}{!}{%
\begin{tabular}{|m{6cm}|m{8cm}|}
\hline
\textbf{Notation} & \textbf{Description} \\
\hline
$W_j$ & Public part of Tx $j$. \\
\hline
$V_{j*} = (V_{ji},V_{jk})$ &Semi-private part of Tx $j$. \\
\hline
$U_{j*} = (U_{ji},U_{jk})$ & bi-variate part of Tx $j$. \\
\hline
$X_j$ & Private part of Tx $j$. \\
\hline
$U_j^{\oplus} = U_{ij} \oplus U_{kj}$ & Sum (over the finite field) of bi-variate parts $U_{ij}$ and $U_{kj}$. \\
\hline
$\ulineW = (W_1,W_2,W_3)$ & Collection of public parts of all Txs. \\
\hline
$\ulineV =(V_{1*},V_{2*},V_{3*})$ & Collection of semi-private parts of all Txs. \\
\hline
$\ulineU =(U_{1*},U_{2*},U_{3*})$ & Collection of bi-variate parts of all Txs. \\
\hline
$\ulineX = (X_1,X_2,X_3)$ & Collection of private parts of all Txs. \\
\hline
$\ulineU^{\oplus} = (U_1^{\oplus}, U_2^{\oplus}, U_3^{\oplus})$ & Collection of summed bi-variate parts over all Txs. \\
\hline
$p_{\ulineU \ulineX} = \prod_{j=1}^3 p_{U_{j*}X_j}$ & Chosen test channel.\\
\hline
$m_j =(m_{ji},m_{jk}, m_{jj})$ & Message of Tx $j$. \\
\hline
$R_j = T_{ji}  + T_{jk}  + L_j$ & Total rate of Tx $j$. \\
\hline
$u_{ji}^n(a_{ji},m_{ji}) = a_{ji} g_{I}^{ji} \oplus m_{ji} g_{O|I}^{ji} \oplus b_{ji}^n$ & A generic codeword in coset indexed by message $m_{ji}$. \\
\hline
$c(m_{ji}) = \{u_{ji}^n(a_{ji},m_{ji}) : a_{ji} \in \mathcal{F}_{\Prime_i}^{s_{ji}-t_{ji}}\}$ & Coset code corresponding to message $m_{ji}$ over the finite field $\mathcal{F}_{\Prime_i}$. \\
\hline
$c(m_{jj}) = \{x_j^n(b_j,m_{jj}) : b_j \in [2^{nK_j}]\}$ & The bin corresponding to private message $m_{jj}$. \\
\hline
 $q_{U_{ji}}$ & Uniform distribution over the finite field $\mathcal{U}_{ji} = \mathcal{F}_{\Prime_i}$. \\
\hline
 $r^n_{U_{ji}U_{jk}X_j} = \frac{p_{U_{ji}U_{jk}X_j}^n}{q_{U_{ji}}^n q_{U_{jk}^n}^n p_{X_j}^n}$ &  Ratio to define the likelihood encoder.\\
\hline
\end{tabular}
}
\caption{Notation used in Step II: Simultaneous Decoding of Bi-variate Components}.
\label{Tab:3CQICNotationStep2}
\vspace{-0.2in}
\end{table}

\begin{proof}
Consider (i) $\SemiPrivateRVSet_{ji} = \CalF_{\Prime_{i}}$ for $ji \in \llbracket 3 \rrbracket$, and (ii) PMF $p_{\ulineU ~\!\!\ulineX} =  p_{U_{1*}U_{2*}U_{3*}X_{1}X_{2}X_{3}}=\displaystyle\prod_{j=1}^{3}p_{U_{ji}U_{jk}X_{j}}$ on $\ulineCalU\times\ulineCalX$ as described in the hypothesis. Let $(R_1,R_2,R_3)$ be a rate triple for which there exists non negative numbers $S_{ji},T_{ji}$, for $ji \in \llbracket 3 \rrbracket$, and $K_{j},L_{j}$, for $j \in [3]$ such that $\Expectation\{\kappa_{j}(X_{j})\}\leq \tau_{j}$, $R_{j}=T_{ji} +T_{jk} +L_{j}$, and the bounds in (\ref{Eqn3CQIC:3CQICStep1SrcBnd1})-(\ref{Eqn3CQIC:3CQICStep1ChnlBnd6}) hold. Throughout this proof, we only state the proof steps - code structure, encoding, decoding and error analysis - in the context of Tx, Rx $1$. The complete proof is obtained by just replicating these steps in the context of Tx, Rx $2$ and Tx, Rx $3$ with appropriate index changes. Furthermore, since the error analysis at each Rx is essentially the error analysis of an effective $4-$user CQMAC, we present the error analysis only for this $4-$user CQMAC.
Before presenting the detailed code structure, encoding, decoding, and error analysis, we summarize the key notation used throughout Step II. Table \ref{Tab:3CQICNotationStep2} lists all Rvs, PMF, messages, and rates introduced in this section. This table can be used as a reference while reading this section.

\noindent \textbf{Code structure:} Tx 1 splits its message $m_1$ into three parts, $m_1=(m_{12},m_{13},m_{11})$. See Fig.~\ref{Fig:FigCodeStructureCQIC(StepI)}. The semi-private parts $m_{12},m_{13}$ are encoded using coset codes built over $\CalU_{12},\CalU_{13}$ respectively, and $m_{11}$ is encoded using a conventional IID random code $\CalC_1$ built over $\CalX_{1}$. Rvs $U_{1*}\define (U_{12},U_{13})$ represent information of Tx 1 encoded in the semi-private parts $(m_{12},m_{13})$ and Rv $X_{1}$ represents the information of Tx 1 encoded in the private part $m_{11}$. Note $U_{12},U_{13},X_{1} \in \CalF_{\Prime_{2}} \times \CalF_{\Prime_{3}} \times \CalX_1$.
Specifically, the message $m_{11}$ is encoded using (i) $c_{11} \define \{x_{1}^{n}(b_1,m_{11}) : b_1 \in [2^{n K_{1}}],  m_{11} \in [2^{nL_{1}}] \}$ constructed over $\CalX_{1}$  and (ii) a binning map $b_j : [2^{nR_1}] \rightarrow [2^{nK_1}]$. Let $x_1^n(b_1(m_1), m_{11})$ be the chosen codeword from the bin $c_{11}(m_{11}) \define \{x_{1}^{n}(b_1,m_{11}) : b_1 \in [2^{n K_{1}}] \}$ corresponding to message $m_{11}$. For $i \in \{2,3\}$, the message $m_{1i}$ is encoded using (i) a NCC over $\CalF_{\Prime_i}$, denoted by $(n,s_{1i}-t_{1i},t_{1i},g^{1i}_{I}, g^{1i}_{O/I}, b_{1i}^n)$, where $g^{1i}_{I} \in \mathcal{F}_{\Prime_i}^{(s_{1i}-t_{1i})\times n}$, $g^{1i}_{O/I} \in \mathcal{F}_{\Prime_i} ^{t_{1i} \times n}$, $b_{1i}^n \in \mathcal{F}_{\Prime_i}^n$, $s_{1i} \define \left \lceil \frac{n(S_{1i} + T_{1i})}{\log(\Prime_i)} \right \rceil$, and $t_{1i} \define \left \lfloor \frac{nT_{1i}}{\log(\Prime_i)} \right \rfloor$, and (ii) a binning map $a_{1i} : [2^{nR_1}] \rightarrow [2^{n(S_{1i} + T_{1i})}]$. We emphasize that the coset codes built over $\CalU_{1i}$, $\CalU_{ki}$ intersect for $(i,k) \in \{(2,3),(3,2)\}$ (see Fig.~\ref{Fig:FigCodeStructureCQIC(StepI)}). In other words, the smaller of these two codes is a sub-coset of the larger. This is accomplished by ensuring that the rows of the generator matrix of the larger code contain all the rows of the generator matrix of the smaller code i.e., for $i \in \{2,3\}$ the generator matrices $g_{i1} = \left[\!\! \begin{array}{c}g_{I}^{i1}\\g_{O/I}^{i1}\end{array}\!\!\right]$ satisfy $g_{\sfb}\define \left[\!\! \begin{array}{c}g_{\sfs}\\g_{\Delta}\end{array}\!\!\right]$ where $\sfb, \sfs$ are defined such that 
$\{\sfb, \sfs\}=\{21, 31\} \mbox{ and } s_{\sfb} + t_{\sfb} \geq s_{\sfs} + t_{\sfs}$.
For $i \in \{2,3\}$, let $u_{1i}^n(a_{1i}(m_1),m_{1i})$ denote the chosen codeword from the coset $c_{1i}(m_{1i}) \define \{u_{1i}^n(a_{1i}, m_{1i}) : a_{1i} \in \mathcal{F}_{\Prime_i}^{(s_{1i} - t_{1i})}\}$
corresponding to message $m_{1i}$.
We remind the reader that we have only described the codes built at Tx $1$, with respective indices. The described codes in the context of Tx $1$ is replicated for the other two Txs $2$ and $3$ as depicted in Fig.~\ref{Fig:FigCodeStructureCQIC(StepI)}.

\noindent \textbf{Encoding Rule:} 
Upon observing $m_1$, Tx $1$ identifies the encoded codewords 
\begin{eqnarray}
\left(u_{12}^n(a_{12}(m_1),m_{12}), u_{13}^n(a_{13}(m_1),m_{13}), x_1^n(b_1(m_1),m_{11})\right), \nonumber 
\end{eqnarray}
and inputs $x_1^n(b_1(m_1),m_{11})$ on the channel.

\begin{comment}
For each message $m_1=(m_{12},m_{13},m_{11})$, the Tx~1 chooses randomly a sequence of codewords from $c(m_{12}) \times c(m_{13}) \times c(m_{11})$ according to the PMF $P\left(A_{12}(m_{12})=a_{12},A_{13}(m_{13})=a_{13}, B(m_{11})=b_1 \vert c(m_{12}), c(m_{13}), c(m_{11})\right)=$
\begin{eqnarray}
\frac{r_{U_{1*}X_1}^n(u_{12}^n(a_{12},m_{12}), u_{13}^n(a_{13},m_{13}),x_1^n(b_1,m_{11}))}{\sum_{\tilde{a}_{12}, \tilde{a}_{13}, \tilde{b}_1} r_{U_{1*}X_1}^n(u_{12}^n(\tilde{a}_{12},m_{12}), u_{13}^n(\tilde{a}_{13},m_{13}),x_1^n(\tilde{b}_1,m_{11}))},
\quad \mbox{where }\nonumber     
\end{eqnarray}
$r_{U_{1*}X_1}^n \define \frac{p_{U_{12}U_{13}X_1}^n}{q_{U_{12}}^n q_{U_{13}}^n p_{X_1}^n}$ and $q_{U_{1i}}$ denotes the uniform distribution over $\CalF_{\Prime_i}$ for $i \in \{2,3\}$.
Let 
\begin{eqnarray}
    \left(U_{12}^n(A_{12}(m_{12}), m_{12}), U_{13}^n(A_{13}(m_{13}), m_{13}), X_1^n(B(m_{11}),m_{11}) \right) \nonumber 
\end{eqnarray} 
be the chosen sequence of codewords. We now proceed to the decoding POVM.    
\end{comment}
\noindent \textbf{Decoding POVM:} Rx 1 decodes a bi-variate component $u^n_{21}(a_{21}(m_{2}),m_{21}) \oplus u^n_{31}(a_{31}(m_{3}),m_{31})$ in addition to the codewords $ u^n_{12}(a_{12}(m_{1}),m_{12}), u^n_{13}(a_{13}(m_{1}),m_{13}), x^n_1(b_{1}(m_1),m_{11})$ corresponding to its message. Specifically, Rx 1 decodes the quartet
\begin{eqnarray}
    \left( u^n_{12}(a_{12}(m_{1}),m_{12}), u^n_{13}(a_{13}(m_{1}),m_{13}), x^n_1(b_{1}(m_1),m_{11}), u^n_{21}(a_{21}(m_{2}),m_{21}) \oplus u^n_{31}(a_{31}(m_{3}),m_{31}) \right) \nonumber 
\end{eqnarray}
using simultaneous decoding. In particular, the construction of our simultaneous decoding POVM adopts the \textit{tilting, smoothing, and augmentation} technique of Sen \cite{202103SAD_Sen}. Towards describing this, we begin by characterizing the effective CQMAC state
\begin{eqnarray}
\label{Eqn3CQIC:EffectCQMACStateWithOrigNotation}
\xi^{U_{1*}X_{1}U_{1}^{\oplus}Y_{1}} &=& \sum_{u_{1*},x_{1},u_{1}^{\oplus}} p_{U_{1*}X_{1}}(u_{1*},x_{1}) p_{U_{1}^{\oplus}}(u_{1}^{\oplus}) \ketbra{u_{1*}~x_{1}~u_{1}^{\oplus}} \otimes \xi_{u_{1*},x_{1},u_{1}^{\oplus}},\\
\mbox{ where  } 
\xi_{u_{1*}x_{1}u_{1}^{\oplus}}&=&\sum_{u_{2*}} \sum_{u_{3*}}  \sum_{x_2} \sum_{x_3} 
p_{U_{2*}U_{3*}X_2X_3|U_{1*}X_1 U_1^{\oplus}} (u_{2*}, u_{3*}, x_2,x_3|u_{1*},x_1,u_1^{\oplus}) \rho^{Y_1}_{x_{1} x_2 x_3}.
\nonumber 
\end{eqnarray}

\begin{figure}
\begin{minipage}{0.45\textwidth}
    \centering
    \includegraphics[width=\textwidth]{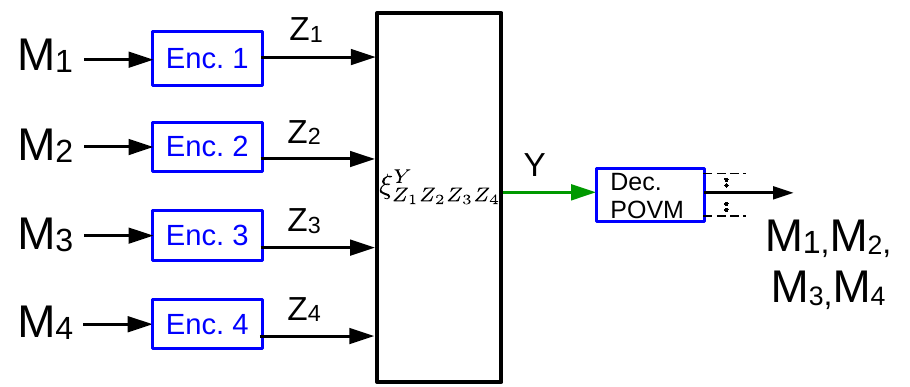}
    \caption{Communication over the effective $4-$user CQMAC}
    \label{Fig4CQMAC}
\end{minipage}
\hfill
\begin{minipage}{0.5\textwidth}
    \centering
    \resizebox{\textwidth}{!}{
        \begin{tabular}{|c|c||c|c|}
            \hline
            4-CQMAC & Original Notation & 4-CQMAC & Original Notation  \\
            \hline
            $m_{1}$ & $m_{12}$ & $\tilde{R}_1$ & $S_{12}$ \\
            \hline
            $m_{2}$ & $m_{13}$ & $\tilde{R}_2$ & $S_{13}$ \\
            \hline
            $m_3$ & $m_{11}$ & $\tilde{R}_3$ & $K_1+L_1$ \\
            \hline
            $m_4$ & $m_{21} \oplus m_{31}$ & $\tilde{R}_4$ & $\max\{S_{21},S_{31}\}$ \\
            \hline
            $Z_1$ & $U_{12}$ & $p_{Z_1Z_2Z_3Z_4}$ & $p_{U_{12}U_{13}X_1} p_{U_1^{\oplus}}$  \\
            \hline
            $Z_{2}$ & $U_{13}$ &$Y$ & $Y_1$ \\
            \hline
            $Z_3$ & $X_1$ & $\xi_{\ulinez}$ & $\xi_{u_{1*},x_1,u_1^{\oplus}}$   \\
            \hline 
            $Z_4$ & $U_1^{\oplus} = U_{21} \oplus U_{31}$ & $\ulineZ=(Z_1,Z_2,Z_3,Z_4)$ & $U_{12},U_{13},X_1,U_1^{\oplus}$  \\
            \hline
        \end{tabular}
    }
    \caption{Mapping between the 4-CQMAC and the original notation}
    \label{TabNotataion4CQMAC}
\end{minipage}
\vspace{-0.15in}
\end{figure}

Rx 1's error analysis is essentially the error analysis of the above CQMAC, where Rx attempts to decode \( U_{12}, U_{13}, X_{1} \), and \( U^{\oplus}_1 = U_{21} \oplus U_{31} \). We may, therefore, restrict our attention to the effective $4-$user MAC (see Fig.~\ref{Fig4CQMAC}) specified via \eqref{Eqn3CQIC:EffectCQMACStateWithOrigNotation}. To reduce clutter and simplify notation, we henceforth use underlined symbols to denote quadruples instead of triples. In particular, for any four objects $a_1,a_2,a_3,a_4$, we write $\ulinea \define (a_1, a_2, a_3, a_4)$. Accordingly, we rename $Z_{1}\define U_{12}$, $Z_{2} \define U_{13}$, $Z_{3} \define X_{1}$, $Z_{4} \define U_{21}\oplus U_{31}$, $\ulineZ \define (Z_{1}, Z_{2}, Z_{3} , Z_{4})$, a PMF
 $p_{\ulineZ}(\ulinez)= \sum_{u_{*1}}  p_{U_{1*} X_1 U_{*1} } (z_{1}, z_{2},z_{3}, u_{21},u_{31}) \mathds{1}\{z_{4}= u_{21} \oplus u_{31}\}$, $\tilde{R}_1 \define S_{12}$, $\tilde{R}_2 \define S_{13}$, $\tilde{R}_3 \define K_1 + L_1$, $\tilde{R}_4 \define \max\{S_{21}, S_{31}\}$ and $Y \define Y_1$ (see Fig.~\ref{TabNotataion4CQMAC}). Throughout the rest of the proof, we restrict our attention to the analysis of the error probability of the above $4-$user MAC on which the Rx attempts to decode $z_1, z_2, z_3$ and $z_4$.
Towards that end, we relabel the state in (\ref{Eqn3CQIC:EffectCQMACStateWithOrigNotation}) as
\begin{eqnarray}
 \xi^{\ulineZ Y} \define \sum_{z_{1},z_{2},z_{3},z_{4}} p_{\ulineZ}(\ulinez)\ketbra{z_{1}~z_{2}~z_{3}~z_{4}}\otimes \xi_{z_{1}z_{2}z_{3}z_{4}}. \nonumber
\end{eqnarray}
%\vspace{-0.1in}

\noindent To analyze the $4-$user CQMAC, we adopt Sen's \cite{202103SAD_Sen} technique of \textit{tilting, smoothing, and augmentation}. In this setting, there are $15=2^4-1$ possible error events, each corresponding to a different subspace, and these subspaces can overlap in a nontrivial way. Sen's \cite{202103SAD_Sen} idea is to enlarge the space so that we can tilt each error subspace in an orthogonal direction and increase their separation. We also augment the space to ensure the smoothing property of the states obtained by averaging. 
Following this approach, we define $14=(2^4-1)-1$ tilting maps, one for each error subspace except the last, where all four messages $(m_1,m_2,m_3,m_4)$ are wrong and remain untilted. In what follows, we present this approach in detail. First, we define the new enlarged and augmented space. Next, we define the tilting maps. 
\begin{comment}
%Then, we define the tilted state and we state how close it remains to the original state. Finally, we describe the tilted subspace and construct the final POVM, which can simultaneously decode all messages.
\end{comment}

\noindent For convenience, Table \ref{Tab:3CQICNotation} summarizes the notation for the auxiliary spaces, tilting maps, and projectors used in the construction of the simultaneous decoding POVM.
\begin{Notation}
\label{Not:proof3CQIC}
In the rest of this proof, $S$ \footnote{Throughout the remaining part of the proof, $S \subseteq [4]$ will denote a generic non-empty subset of $\{1,2,3,4\}$. We alert the reader to distinguish $S$ from the $S_{ij}$ that appears in the rates (see Fig.~\ref{Fig:FigCodeStructureCQIC(StepI)}). The latter quantity always appears with a subscript, while the former never appears with a subscript.} will always refer to a \textbf{non-empty} subset of $[4]$. The following objects are defined for non-empty subset of $S$.
\begin{comment}
Let $S$ denote\footnote{Throughout the remaining part of the proof, $S \subseteq [4]$ will denote a generic non-empty subset of $\{1,2,3,4\}$. We alert the reader to distinguish $S$ from the $S_{ij}$ that appears in the rates (see Fig.~\ref{Fig:FigCodeStructureCQIC(StepI)}). The latter quantity always appears with a subscript, while the former never appears with a subscript.} a \textbf{non-empty} subset of $[K]$. The complement of $S$ is denoted by $S^c = [K] \setminus S$. Unless stated otherwise, all subsets $S \subseteq [K]$ are assumed to be \textbf{non-empty}.        
\end{comment}
\end{Notation}
Consider four auxiliary finite sets  $\CalD_{i}$ for $i \in [4]$ along with corresponding four auxiliary Hilbert spaces ${\CalD_{i}}$ of dimension $|\CalD_{i}|$ for each $i \in [4]$.\footnote{$\CalD_{i}$ denotes both the finite set and the corresponding auxiliary Hilbert space. The specific reference will be clear from context.} Define the extended space $\CalH_{Y}^{e}$ as follows \footnote{Recall that $S \subseteq [4]$ denotes a generic non-empty subset of $\{1,2,3,4\}$. We once again alert the reader to distinguish $S$ from the $S_{ij}$ that appears in the rates (see Fig.~\ref{Fig:FigCodeStructureCQIC(StepI)}). The latter quantity always appears with a subscript, while the former never appears with a subscript.}
\begin{eqnarray}
    \nonumber %\label{Eqn:3CQICExtendedHilbertspace}
 \left(\CalH_{Y}^{e}\right)^{\otimes n} \define \CalH_{Y_{G}}^{\otimes n}
\oplus \bigoplus_{ S \subsetneq [4]} \left(\CalH_{Y_{G}}^{\otimes n}
 \otimes \CalD_{S}^{\otimes n} \right)\mbox{ where }\CalH_{Y_{G}}^{\otimes n} =\CalH_{Y}^{\otimes n} \otimes \complex^{2}, \CalD_{S}^{\otimes n} = \bigotimes_{s \in S} \CalD_{s}^{\otimes n}.
\end{eqnarray}

\begin{table}
\renewcommand{\arraystretch}{1}
\setlength{\extrarowheight}{1pt}
\centering
\resizebox{7in}{!}{%
\begin{tabular}{|l|l|l|}
\hline
\textbf{Notation} & \textbf{Description} & \textbf{Comment}\\
\hline
%$\CalD_i$ & Auxiliary finite set for index $i \in [4]$ & \\ \hline
$\CalD_i$ &  Auxiliary finite set and its corresponding Hilbert space, for $i \in [4]$& Used to ensure the smoothing property \\ \hline
$\CalH_Y$ & Channel output Hilbert space & The original output Hilbert space \\ \hline
$\CalH_{Y_G}$ & $\CalH_Y \otimes \mathbb{C}^2$ & Extended Hilbert space via Gelfand–Naimark theorem. \\ \hline
%$\CalH_Y^e$ & Extended output Hilbert space\\ \hline
$\boldsymbol{\CalH_Y}$ & $\CalH_Y^{\otimes n}$ & Tensor product of the channel output space.\\ \hline
$\boldsymbol{\CalH_{Y_G}}$ & $\CalH_{Y_G}^{\otimes n}$ & Tensor product of the Gelfand–Naimark extended space.\\ \hline
$\boldsymbol{\CalH_Y^e}$ & $\left(\CalH_Y^e\right)^{\otimes n} = \CalH_{Y_{G}}^{\otimes n} 
\oplus \bigoplus_{S \subsetneq [4]} \left(\CalH_{Y_{G}}^{\otimes n}
 \otimes \CalD_{S}^{\otimes n} \right)$  & Extended output Hilbert space. \\ \hline
$\boldsymbol{\CalD_S}$ & $\CalD_S^{\otimes n}$ & Tensor product of auxiliary spaces for subset $S$.\\ \hline
$\eta$ & Tilting parameter &  \\ \hline
$\CalT^S_{d_S^n,\eta}$ & $\frac{1}{\sqrt{\Omega(S,\eta)} } \left(\ket{h} +\eta^{ |S| } \ket{h} \otimes  \ket{d_{S}^{n}}\right)$ & Tilting isometry along direction $d_S^n$.\\ \hline
$G^S_{\ulinez^n}$ & $\Pi_{z_{S^{c}}^{n}} \Pi_{\ulinez^{n}}\Pi_{z_{S^{c}}^{n}}$ & \\ \hline
$\olineG^S_{\ulinez^n}$ & Orthogonal projector in $\mathcal{P}\left(\boldsymbol{\CalH_{Y_G}}\right)$ & \\ \hline
$\olineB^S_{\ulinez^n}$ & $I-\olineG^S_{\ulinez^n}$ & Complement projector of $\olineG^S_{\ulinez^n}$.  \\ \hline
$\beta^S_{\ulinez^n,\ulined^n}$ & $\CalT_{d^{n}_{S^c},\eta}^{S^c} ( \olineB_{\ulinez^{n}}^{S})$ & Tilted projector along auxiliary direction $d_{S^c}^n$. \\ \hline
$\beta^{[4]}_{\ulinez^n,\ulined^n}$& $\olineB_{\ulinez^{n}}^{[4]}$ & Untilted projector corresponding to the full set $[4]$. \\ \hline
$\beta^{*}_{\ulinez^n,\ulined^n}$ & Union-support projector over all $S\subseteq[4]$ & Combines supports of all $\beta^S_{\ulinez^n,\ulined^n}$. \\ \hline
$\Pi_{\boldsymbol{\CalH_{Y_G}}}$ & Projector onto $\boldsymbol{\CalH_{Y_G}}$& \\ \hline
$\gamma^{*}_{(\ulinez^n,\ulined^n)(\ulinem)}$ &  $\left(I_{\boldsymbol{\CalH_{Y}^{e}}}-  \beta^{*}_{(\ulinez^{n},\ulined^{n})(\ulinem)}\right) \Pi_{\boldsymbol{\CalH_{Y_G}} }\left(I_{\boldsymbol{\CalH_{Y}^{e}}}-  \beta^{*}_{(\ulinez^{n},\ulined^{n})(\ulinem)}\right)$ &\\ \hline
$\mu_{\ulinem}$ & $\left(\sum_{\uline{\widehat{m}}} \gamma^{*}_{(\ulinez^{n},\ulined^{n})(\widehat{\ulinem})}\right)^{-\frac{1}{2}}
\gamma^{*}_{(\ulinez^{n},\ulined^{n})(\ulinem)} \left(\sum_{\uline{\widehat{m}}} \gamma^{*}_{(\ulinez^{n},\ulined^{n})(\widehat{\ulinem})}\right)^{-\frac{1}{2}}$ & Decoding POVM element for message $\ulinem$. \\ \hline
\end{tabular}
}
\caption{Description of the elements used to construct a simultaneous decoding POVM.}
\label{Tab:3CQICNotation}
\vspace{-0.2in}
\end{table}

Note that $\dim(\CalD_{S}) = \prod_{s \in S}\dim(\CalD_{s})$. For $ S \subseteq [4]$, let  $\ket{d_{S}}=\otimes_{s \in S} \ket{d_s}$ be a computational basis vector of  ${\CalD_{S}}$, $Z_{S}:=\left(Z_{s}\right)_{s \in S}$, and let $\eta>0$ be chosen appropriately in what follows. In the sequel, we let $\boldsymbol{\CalH_{Y}}= \CalH_{Y}^{\otimes n}$,         $\boldsymbol{\CalH_{Y_G}}=\CalH_{Y_G}^{\otimes n}$,
$\boldsymbol{\CalH_{Y}^{e}}=(\CalH_{Y}^{e})^{\otimes n}$ and $\boldsymbol{\CalD_{S}}= \CalD_{S}^{\otimes n}$. For $S \subsetneq [4]$ we define the tilting map $\CalT^{S}_{ d_{S}^{n}, \eta} : \boldsymbol{\CalH_{Y_G}}\rightarrow
\boldsymbol{\CalH_{Y_G}} \oplus \left(\boldsymbol{\CalH_{Y_G}}  \otimes   \boldsymbol{\CalD_{S}}\right)
$ as
\begin{eqnarray}
\CalT^{S}_{ d_{S}^{n}, \eta}(\ket{h}) \define \frac{1}{\sqrt{\Omega(S,\eta)} } \left(\ket{h} +\eta^{ |S| } \ket{h} \otimes  \ket{d_{S}^{n}}\right), \mbox{ where }  \Omega(S,\eta) \define 1+ \eta^{2 |S|}.
\nonumber
\end{eqnarray}
 Next, we construct the decoding POVM. For $S \subseteq [4]$, we consider 
\begin{eqnarray}
\label{Eqn:3CQICOrigPOVM}
   G^{S}_{\ulinez^{n}}\define \Pi_{z_{S^{c}}^{n}} \Pi_{\ulinez^{n}}\Pi_{z_{S^{c}}^{n}}, \nonumber
\end{eqnarray}
where $\Pi_{\ulinez^{n}}$ is the C-Typ-Proj with respect to the state $\otimes_{t=1}^{n}\xi_{z_{1t}z_{2t}z_{3t}z_{4t}}$
and $\Pi_{z_{S^c}^{n} }$ is C-Typ-Proj with respect to the state $\otimes_{t=1}^{n}\left(\xi_{z_{st}:s \in S^c}= \otimes_{t=1}^{n}\sum_{z^n_S} p_{Z_S | Z_{S^c}}^n(z^n_{S}|z^n_{S^c}) \xi_{z_{1t}z_{2t}z_{3t}z_{4t}}\right)$.
%\begin{proposition}
%\label{Prop:3CQICProppropertiesoftheoriginalspovmelements}
 %   For all $\epsilon>0, \delta \in (0,1)$, $n$ sufficiently large and for $S \subseteq [4]$ we have 
  %  \begin{eqnarray}
   %     &&\sum_{\ulinez^n} p_{\ulineZ}^n(\ulinez^n) \tr\left( G^S_{\ulinez^n} \xi_{\ulinez^n} \right) \geq 1- \epsilon - 2 \sqrt{\epsilon} \nonumber \\
     %   && \sum_{\tilde{z}_S^n} \sum_{z^n_{S^c}} p_{\ulineZ}^n(\tilde{z}_S^n,z^n_{S^c}) \tr\left( G^{[S]}_{\tilde{z}^n_{S},z^n_{S^c}} \xi^{\otimes n}_{z^n_{S^c}}\right) \leq 2 ^{-n \left( I(Y; Z_S|Z_{S^c}) - 2 \delta \right)}\nonumber 
    %\end{eqnarray}
%\end{proposition}
%\begin{proof}
 %   The proof is analogous to the proof in Appendix \ref{App:3to1CQICProppropertiesoftheoriginalspovmelements}.
%\end{proof}
By Gelfand–Naimark’s Thm.~\cite[Thm.~3.7]{BkHolevo_2019}, there exists orthogonal projector $\olineG^{S}_{\ulinez^{n}}\in \CalP(\boldsymbol{\CalH_{Y_G}})$ that yields identical measurement statistic on states in $\boldsymbol{\CalH_{Y_G}}$ that $G^{S}_{\ulinez^{n}}$ gives on states in $\boldsymbol{\CalH_{Y}}$, for $S \subseteq [4]$. Let 
\begin{eqnarray}
\olineB_{\ulinez^{n}}^{S} \define I_{\boldsymbol{\CalH_{Y_G}}}-\olineG^{S}_{\ulinez^{n}},\mbox{ for }  S \subseteq [4] \label{Eqn:3CQICComplementprojector} 
\end{eqnarray}
be the complement projector. Now, consider
\begin{eqnarray}
 \beta^{S}_{\ulinez^{n},\ulined^{n}} \define \CalT_{d^{n}_{S^c},\eta}^{S^c} ( \olineB_{\ulinez^{n}}^{S}), \mbox{ for }  S \subsetneq [4] \mbox{ and } \beta^{[4]}_{\ulinez^{n},\ulined^{n}} \define  \olineB_{\ulinez^{n}}^{[4]}, \label{Eqn3CQIC:Tiltingofprojectors}
\end{eqnarray}
where $\beta^{S}_{\ulinez^{n},\ulined^{n}}$ denotes the tilted projector along direction $d_{S^c}^{n}$ for $S \subsetneq [4]$ and $\beta^{[4]}_{\ulinez^{n},\ulined^{n}}$ denotes the untilted projector.
Next, we define $\beta^{*}_{\ulinez^{n},\ulined^{n}}$ as the projector in $\boldsymbol{\CalH_{Y}^{e}}$ whose support is the union of the supports of$\beta^{S}_{\ulinez^{n},d_{S}^{n}}$ for all $S \subseteq [4]$.
Let $\Pi_{\boldsymbol{\CalH_{Y_G}}}$ be the orthogonal projector in $\boldsymbol{\CalH_{Y}^{e}}$ onto $\boldsymbol{\CalH_{Y_G}}$. 
We define the square root measurement \cite{BkWilde_2017,BkHolevo_2019}, to decode $\ulinem=(m_1,m_2,m_3,m_4)$, as 
\begin{eqnarray}
\mu_{\ulinem} &\define& \left(\sum_{\uline{\widehat{m}}} \gamma^{*}_{(\ulinez^{n},\ulined^{n})(\widehat{\ulinem})}\right)^{-\frac{1}{2}}
\gamma^{*}_{(\ulinez^{n},\ulined^{n})(\ulinem)} \left(\sum_{\uline{\widehat{m}}} \gamma^{*}_{(\ulinez^{n},\ulined^{n})(\widehat{\ulinem})}\right)^{-\frac{1}{2}}, \mbox{ and } \mu_{-1} \define I - \sum_{\ulinem} \mu_{\ulinem}, \nonumber  \\ \mbox{where }  
\gamma^{*}_{(\ulinez^{n},\ulined^{n})(\ulinem)} &\define& \left(I_{\boldsymbol{\CalH_{Y}^{e}}}-  \beta^{*}_{(\ulinez^{n},\ulined^{n})(\ulinem)}\right) \Pi_{\boldsymbol{\CalH_{Y_G}} }\left(I_{\boldsymbol{\CalH_{Y}^{e}}}-  \beta^{*}_{(\ulinez^{n},\ulined^{n})(\ulinem)}\right). \label{Eqn:3CQICgammadef}
\end{eqnarray}

\noindent \textbf{Distribution of the random code:} As is standard in information theory, to derive an upper bound on the average error probability, we average over the ensemble of all codes.
The distribution of the random code is analogous to \eqref{Eqn:DistStepI}, that is employed in step I analysis involving $3$ codes. This is naturally extended to the scenarios involving $9$ codes, so in essence the unstructured codebooks are picked independently, and the coset codes are picked such that the larger of the codes in the same field contains the smaller one. For more elaboration, please look at the distribution \eqref{Eqn:DistStepI} in step I.

\noindent \textbf{Error Analysis:} Based on the exposition provided in Remark \ref{Rem:BinningandChannelCodingAnalysis} ,we recognize that our approach enables us to separate the binning analysis and the channel coding analysis, thereby breaking the independence. We therefore assume no binning in the following analysis. 
As is standard in information theory, we derive an upper-bound on the probability of error. {We decompose the average error probability of the $4-$user CQMAC (Fig.~\ref{Fig4CQMAC}) into two parts. This is done by employing, as suggested in \cite{202103SAD_Sen}, an alternate `proxy' state. Specifically, for each codeword quadruple, we substitute the original received state ($\xi_{\ulineZ^n} \otimes \ketbra{0}$) by a specific `tilted state'. Since the tilted state is chosen such that is close in $\mathbb{L}_{1}-$norm to the original received state (Prop.~\ref{Prop:3CQICClosnessOFstates}), the effect of this substitution on the error probability can be suppressed. Towards identifying this tilted state, we define a new tilting map
\begin{eqnarray}
&&\hspace{-0.1in}\CalT_{ \ulined^{n}, \eta} :
\boldsymbol{\CalH_{Y_G}}  \rightarrow \boldsymbol{\CalH_{Y}^{e}}
\mbox{ as } \CalT_{ \ulined^{n}, \eta}(\ket{h}) \define \frac{1}{\sqrt{\Omega(\eta)} } \left(\ket{h} + \sum_{S \subsetneq [4]} \eta^{ |S| } \ket{h} \otimes \ket{d_{S}^{n}}\right), \mbox{ where } \Omega(\eta) \define 1+16 \eta^{2}+ 36 \eta^{4}+ 16 \eta^{6}. \nonumber
\end{eqnarray}
Let  
 \begin{eqnarray}
    \theta^{ \otimes n}_{\ulinez^n\ulined^n} \define \CalT_{ \ulined^{n}, \eta} \left\{\xi^{\otimes n}_{\ulinez^n}\otimes \ketbra{0} \right\}  \nonumber \label{Eqn:Tiltedstate}
 \end{eqnarray} 
be the tilted state with components in all appended auxiliary spaces. Here, the map $\CalT_{ \ulined^{n}, \eta}$ is linear and acts on each pure state in a mixture individually. 

\begin{proposition}
\label{Prop:3CQICClosnessOFstates}
    For $n$ sufficiently large, we have 
\begin{eqnarray}
     \norm{\theta_{\ulinez^{n}, \ulined^{n}}^{\otimes n} - \left(\xi_{\ulinez^{n}}^{\otimes n} \otimes \ket{0}\bra{0}\right) }_{1} \leq 12 \eta. 
\end{eqnarray}
\end{proposition}
\begin{proof}
    The proof is analogous to the proof in Appendix \ref{App:3to1CQICClosenessOfStates}.
\end{proof}
Having defined the alternate `proxy' tilted state, we have
\begin{eqnarray}
    &&\hspace{-0.25in}\mathbf{P}(\ulinee,\mu)= \frac{1}{|\mathcal{\ulineM}|} \sum_{\ulinem}\tr \left\{ \left(I-\mu_{\ulinem}\right)  \left(\xi_{\ulineZ}^n(\ulinem) \otimes \ketbra{0}\right)  \right\} \leq t_1 + t_2, \mbox{ where} \nonumber\\  
     &&\hspace{-0.25in}t_1 \define \frac{1}{|\mathcal{\ulineM}|} \sum_{\ulinem}\tr \left\{ \left(I-\mu_{\ulinem}\right)  \theta_{\ulineZ^n(\ulinem), \ulineD^n(\ulinem)}  \right\}, \mbox{ and } t_2\define \frac{1}{|\mathcal{\ulineM}|} \sum_{\ulinem} \norm{\theta_{\ulineZ^n(\ulinem), \ulineD^n(\ulinem)}-\left(\xi_{\ulineZ^n(\ulinem)} \otimes \ketbra{0}\right)}_1. \nonumber 
\end{eqnarray}}
The inequality follows from using the inequality $\tr\left(\Delta \rho \right) \leq \tr\left(\Delta \sigma\right) +  \norm{\rho-\sigma}_1$, with $0 \leq \Lambda, \rho, \sigma \leq I$.

We analyze the first term $t_1$, using the Hayashi–Nagaoka inequality \cite{200307TIT_HayNag}. This allows us to decompose $t_1$ as
\begin{eqnarray}
    && \hspace{3cm} t_1 = 2 \: t_{1.1} + 4 \sum_{S \subseteq [4]} t_{1.S}, \mbox{ where} \nonumber \\
    &&t_{1.1} \define \frac{1}{|\mathcal{\ulineM}|} \sum_{\ulinem}\tr \left\{  (I - \gamma^{*}_{\ulineZ^n(\ulinem),\ulineD^n(\ulinem)}) \theta_{\ulineZ^n(\ulinem), \ulineD^n(\ulinem)} \right\}, \mbox{ and} \nonumber \\
    && t_{1.S} \define \frac{1}{|\mathcal{\ulineM}|} \sum_{\ulinem} \sum_{\tilde{m}_S \neq m_S}\tr \left\{ \gamma^{*}_{Z_S^n(\tilde{m}_S), D^n_S(\tilde{m}_S), Z^n_{S^c}(m_{S^{c}}),D_{S^{c}}^n(m_{S^c})}\theta_{\ulineZ^n(\ulinem), \ulineD^n(\ulinem)} \right\}.\nonumber
 \end{eqnarray}
The following propositions summarize all the rate constraints that result from bounding these error terms.
\begin{proposition}
\label{Prop:4CQMACFirsttermHay}
For any $\epsilon \in (0,1)$, and for all sufficiently small $\delta, \eta > 0$, and sufficiently large $n$, we have $\mathbb{E}[t_{1.1}] \leq \epsilon$.
\end{proposition}
\begin{proof}
    The proof is provided in Appendix \ref{App:4CQMACFirsttermHay}.
\end{proof}

\begin{proposition}
\label{Prop:4CQMACSectermHay}
For any $\epsilon \in (0,1)$, and for all sufficiently small $\delta, \eta > 0$, and sufficiently large $n$, we have $    \mathbb{E}[t_{1.S}] \leq \epsilon,$ if
\begin{eqnarray}
\sum_{s \in S} \tilde{R}_s < I(Y;Z_S|Z_{S^c}), \mbox{ for }  S \subseteq [4]. \nonumber 
\end{eqnarray}
\end{proposition}
\begin{proof}
    The proof is provided in Appendix \ref{App:4CQMACSectermHay}.
\end{proof}

Next, from Prop.~\ref{Prop:3CQICClosnessOFstates}, we have for all $\epsilon \in (0,1)$ and for $\eta = \epsilon^{\frac{1}{5}}$, $\mathbb{E}[t_2] \leq 12 \: \epsilon^{\frac{1}{5}}$. The choice of $\eta$ comes from ensuring $\mathbb{E}[t_{1.1}] \leq \epsilon$, for $n$ sufficiently large, as shown in the proof of Prop.~\ref{Prop:4CQMACFirsttermHay}. 
Based on Remark \ref{Rem:BinningandChannelCodingAnalysis} and our exposition in the proof in Sec.~\ref{Sec:3to1CQICAchivablerateregion}, specifically the separation of binning and channel coding, this proof is complete.
\end{proof}

\subsection{Step III: Enlarging $\alpha_{S}$ via Unstructured IID Codes to $\alpha_{US}$}
\label{SubSec:Step2CodingStrategy3CQIC}
Incorporating the Han-Kobayashi technique \cite{198101TIT_HanKob} of decoding `uni-variate parts' coded via unstructured IID codes, we characterize an inner bound $\alpha_{US}$ that subsumes all currently known inner bounds. We provide a characterization of the same below. A proof of this inner bound is involved, but essentially it is a generalization of the proof provided for Lemma \ref{Lemma:3to1CQICRateRegion} and Thms.~\ref{Thm:3to1CQICRateRegion}, \ref{Thm:3CQICStageIRateRegion}. The code structure to prove achievability of the following inner bound is depicted in Fig.~\ref{Fig:3CQICStepIIICodeStructure}. The reader is also referred to Fig.~\ref{Fig:3CQICRVsInFullFinalStrategy} and the associated discussion found at the beginning of Sec.~\ref{Sec:SimultDecOfCosetcds3CQIC}.

\begin{figure}
    \centering
    \includegraphics[width=4.5in]{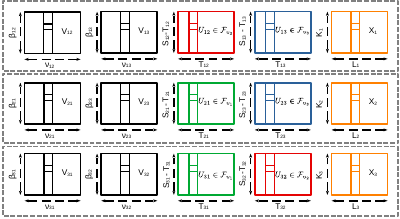}
    \caption{$15$ codes employed in the proof of Thm.~\ref{Thm:3CQICStageIIRateRegion}. Row $j$ depicts the codes of Tx, Rx $j$. The $U_{ji} : ji \in \dbrackthree$ are coset codes built over finite fields. Codes with the same color are built over the same finite field, and the smaller of the two is a sub-coset of the larger. The black codes are built over auxiliary finite sets $\CalV_{ji} : ji \in \dbrackthree$ using the conventional IID random code structure. The orange codes are conventional IID random codes built over input sets $\CalX_{1},\CalX_{2}$ and $\CalX_{3}$.}
    \label{Fig:3CQICStepIIICodeStructure}
    \vspace{-0.15in}
\end{figure}

\begin{comment}
{\begin{Notation}
   We let $\dbrackthree \define \{12,13,21,23,31,32\}$, $V_{1*}= (V_{12},V_{13})$ and similarly $V_{j*}= (V_{ji},V_{jk})$, $\CalV_{j*}=\CalV_{ji}\times \CalV_{jk}$. Similarly $U_{j*}= (U_{ji},U_{jk})$, $\CalU_{j*}=\CalU_{ji}\times \CalU_{jk}$. We let $\ulineU \define (U_{1*},U_{2*},U_{3*})=(U_{12},U_{13},U_{21},U_{23},U_{31},U_{32})$ and similarly for $V$. For any set $A$, $A^{c}$ denotes complement of $A$ with the underlying universe set being clear from context.
\end{Notation}}
\end{comment}
\begin{Notation}
  Let $\CalV_{j*}=\CalV_{ji}\times \CalV_{jk}$ and $\CalU_{j*}=\CalU_{ji}\times \CalU_{jk}$. For any set $A$, $A^{c}$ denotes the complement of $A$ with the underlying universe set being clear from context.   
\end{Notation}
\begin{theorem}
\label{Thm:3CQICStageIIRateRegion}
Let ${\alpha}_{US} \in [0,\infty)^{6}$ be the set of all rate-cost vectors $(\ulineR,\ulinetau)$ for which there exists for all $j \in [3]$, $ji \in \dbrackthree$ (i) finite sets $\CalV_{ji}$, (ii) finite fields $\SemiPrivateRVSet_{ij}=\CalF_{\upsilon_{j}}$, (iii) a PMF $p_{ \ulineV~\!\!\ulineU~\!\!\ulineX}=p_{V_{1*}V_{2*}V_{3*}U_{1*}U_{2*}U_{3*}X_{1}X_{2}X_{3}}$ on $\ulineCalV \times \ulineCalU \times\ulineCalX$ and (iv) nonnegative numbers $\beta_{ji},\nu_{ji},S_{ji},T_{ji}:ji \in \dbrackthree, K_{j},L_{j}:j\in [3]$, such that 
$R_{1}=\nu_{12}+\nu_{13}+T_{12}+T_{13}+L_{1},
R_{2}=\nu_{21}+\nu_{23}+T_{21} + T_{23}+L_{2},
R_{3}=\nu_{31}+\nu_{32}+T_{31}+T_{32}+L_{3}$, $\Expectation\left\{ \kappa(X_{j}) \right\} \leq \tau_{j}$,
\begin{eqnarray}
\label{Eqn3CQIC:SrcCodingBnd1}
\beta_{C_{j}}+S_{A_{j}} &>& \sum_{a \in A_{j}}\log|\CalU_{a}|+\sum_{c \in C_{j}}H(V_{c})-H(U_{A_{j}},V_{C_{j}})\\
\label{Eqn3CQIC:SrcCodingBnd2}
\beta_{C_{j}}+S_{A_{j}}+K_{j} &>& \sum_{a \in A_{j}}\log|\CalU_{a}|+\sum_{c \in C_{j}}H(V_{c})+H(X_{j})-H(U_{A_{j}},V_{C_{j}},X_{j})\\
\label{Eqn3CQIC:Chnlbnd1}
S_{A_{j}}+T_{A_{j}}+\beta_{C_{j}}+\nu_{C_{j}} &<& \sum_{a \in A_{j}} \! \log |\CalU_{a}| + \sum_{c \in C_{j}} H(V_{c}) \nonumber \\
&&- H(U_{A_{j}},V_{C_{j}}|U_{A_{j}^{c}},V_{C_{j}^{c}},U_{ij}\oplus U_{kj},X_{j},Y_{j})
\\
\label{Eqn3CQIC:Chnlbnd2}
S_{A_{j}} +T_{A_{j}} +\beta_{C_{j}} +\nu_{C_{j}}+S_{ij}+T_{ij} &<& \sum_{a \in A_{j}} \log |\CalU_{a}| + \log \upsilon_{j}+ \sum_{c \in C_{j}} H(V_{c}) \nonumber \\
&&- H(U_{A_{j}},V_{C_{j}},U_{ij}\oplus
U_{kj}|U_{A_{j}^{c}},V_{C_{j}^{c}},X_{j},Y_{j})
\\
\label{Eqn3CQIC:Chnlbnd4}
S_{A_{j}}+T_{A_{j}}+\beta_{C_{j}}+\nu_{C_{j}} +S_{kj}+T_{kj} &<& \sum_{a \in A_{j}}\log |\CalU_{a}| + \log \upsilon_{j}
+ \sum_{c \in C_{j}} H(V_{c}) \nonumber \\
&&- H(U_{A_{j}},V_{C_{j}},U_{ij}\oplus
U_{kj}|U_{A_{j}^{c}},V_{C_{j}^{c}},X_{j},Y_{j})
\\
\label{Eqn3CQIC:Chnlbnd6}
S_{A_{j}}+T_{A_{j}}+\beta_{C_{j}}+\nu_{C_{j}}+K_{j}+L_{j} &<& \sum_{a \in A_{j}}\log |\CalU_{a}|+H(X_{j})+ \sum_{c \in C_{j}} H(V_{c})
\nonumber \\
&&-H(U_{A_{j}},V_{C_{j}},X_{j}|U_{A_{j}^{c}},V_{C_{j}^{c}},U_{ij}\oplus
U_{kj},Y_{j})\\
\label{Eqn3CQIC:Chnlbnd8}
S_{A_{j}}+T_{A_{j}}+\beta_{C_{j}}+\nu_{C_{j}}+K_{j}+L_{j}+S_{ij}+T_{ij} &<& \sum_{a \in A_{j}}\log |\CalU_{a}| + \log \upsilon_{j}+ \sum_{c \in C_{j}} H(V_{c})
\label{Eqn3CQIC:Chnlbnd9}
+H(X_{j}) \nonumber \\ 
&&-H(U_{A_{j}},V_{C_{j}},X_{j},U_{ij}\oplus U_{kj}|U_{A_{j}^{c}},V_{C_{j}^{c}}Y_{j})
\\
\label{Eqn3CQIC:Chnlbnd10}
S_{A_{j}}+T_{A_{j}}+\beta_{C_{j}}+\nu_{C_{j}}+K_{j}+L_{j}+S_{kj}+T_{kj} &<& \sum_{a \in A_{j}}\log |\CalU_{a}| + \log \upsilon_{j}+H(X_{j})+ \sum_{c \in C_{j}} H(V_{c}) \nonumber\\
\label{Eqn3CQIC:Chnlbnd11}
&&- H(U_{A_{j}},V_{C_{j}},X_{j},U_{ij}\oplus
U_{kj}|U_{A_{j}^{c}},V_{C_{j}^{c}},Y_{j}),
\end{eqnarray}
hold for every ${A}_{j} \subseteq \left\{ ji,jk\right\},C_{j} \subseteq \left\{ ji,jk\right\}$ with
distinct indices $i,j,k$ in $\left\{ 1,2,3 \right\}$, where $S_{A_{j}} = \sum _{a \in A_{j}}S_{a}$,
$T_{{A}_{j}} \define \sum_{a \in {A}_{j}}T_{a}$, $\beta_{C_{j}} = \sum _{c \in C_{j}}\beta_{c}$, $\nu_{C_{j}} = \sum _{jk \in C_{j}}\nu_{jk}$, $U_{{A}_{j}} = (U_{a}:a \in {A}_{j})$, $V_{{C}_{j}} = (V_{c}:c \in {C}_{j})$, where all the information quantities are evaluated with respect to state
\begin{eqnarray}
\label{Eqn3CQIC:StageIITestChnl} \Phi^{\ulineV\!\!~\ulineU\!\!~\ulineU^{\oplus}\!\!~\ulineX\!\!~\ulineY}\define \begin{comment}\Gamma^{\ulineU_{1*}\ulineU_{2*}\ulineU_{3*}U_{1}^{\oplus}U_{2}^{\oplus}U_{3}^{\oplus}X_{1}X_{2}X_{3}\ulineY} \define {\Gamma^{U_{12}U_{13}U_{21}U_{23}U_{31}U_{32}X_{1}X_{2}X_{3}Y_{1}Y_{2}Y_{3}} \define \sum_{\ulineu,\ulinex}p_{\ulineU\ulineX}\ketbra{\ulineu~\ulinex}\otimes \rho_{\ulinex}}\end{comment}
\sum_{\substack{v_{1*},v_{2*},v_{3*}\\u_{1*},u_{2*},u_{3*} }}\sum_{\substack{x_{1},x_{2},x_{3}\\u_{1}^{\oplus},u_{2}^{\oplus},u_{3}^{\oplus} }}p_{\ulineV~\ulineU~\ulineX}(\ulinev,\ulineu,\ulinex)\mathds{1}{\left\{\substack{u_{ji}\oplus u_{jk}=u_{j}^{\oplus}\\u_{ki}\oplus u_{kj}=u_{k}^{\oplus}\\u_{ij}\oplus u_{ik}=u_{i}^{\oplus}}\right\}}\ketbra{\ulinev~\ulineu~  u_{1}^{\oplus}~ u_{2}^{\oplus}~ u_{3}^{\oplus}~ \ulinex}\otimes \rho_{\ulinex}.\nonumber
\end{eqnarray}
\end{theorem}

\begin{remark}
\label{Rem:afterstep3} 
The inner bound characterized above subsumes the current known largest inner bound for the $3-$CQIC achievable using unstructured IID codes. Indeed, choosing $U_{ji} = \phi$ for $(i,j) \in \dbrackthree$ we recover the latter inner bound.
\end{remark}

\begin{remark}
\label{Rem:ExtInnerBnd}
    The above inner bound is characterized  for a $3-$CQIC. We can extend this via the standard technique of regularization \cite[Sec.~20.1, Eq.~20.7]{BkWilde_2017} to characterize an inner bound to the bit communicating capacity region of a general $3-$user quantum interference channel. We omit a mathematical characterization of this inner bound in the interest of brevity. 
\end{remark}

\section*{Declaration}
Every word, phrase and sentence published in this article is the \textbf{sole creation of the authors}. Explicitly stated, \textbf{none} of the generative AI chatbots such as ChatGPT or Deepseek have been utilized in generating \textbf{any} of the words, phrases or sentences.

\appendices
\section{Inner bound for the 3-to-1 CQIC using unstructured IID codes}
\label{App:Innerbound3to1CQICIIDCodes}
The proofs of Propositions \ref{Prop:strictsuboptimalityiidcodesforadditiveinterference} and \ref{Prop:nonadditiveinterference} rely on the characterization of the best known inner bound using unstructured IID codes for the specific class of channels we refer to as a 3-to-1 CQIC. Specifically, as one notices in examples \ref{Ex:3CQICRoleOfCosetCds}, \ref{Ex:3CQICadditivenoncommutative}, and \ref{Ex:nonadditiveinterference}, users 2 and 3 do not see interference, only user 1 sees interference, so this is referred to as 3-to-1 CQIC, which is defined below.
\begin{definition}
    A $3-$CQIC $(\rho_{\ulinex}\in\mathcal{D}\left(\mathcal{H}_{\ulineY}\right) : \ulinex \in \mathcal{\ulineX}, \kappa_j : j \in [3])$ is a 3-to-1 CQIC if (i) for every $\Delta \in \mathcal{P}\left( \mathcal{H}_{Y_2}\right)$, $\tr\left( \left(I \otimes \Delta \otimes I \right) \rho_{x_1,x_2,x_3}\right)=\tr\left( \left(I \otimes \Delta \otimes I \right) \rho_{\tilde{x}_1,\tilde{x}_2,\tilde{x}_3}\right)$ for every $\ulinex, \uline{\tilde{x}} \in \mathcal{\ulineX}$ satisfying $x_2=\tilde{x}_2$, and (ii) for every $\mu \in \mathcal{P}\left( \mathcal{H}_{Y_3}\right)$, $\tr\left( \left(I \otimes I \otimes \mu \right) \rho_{x_1,x_2,x_3}\right)=\tr\left( \left(I \otimes I \otimes \mu \right) \rho_{\tilde{x}_1,\tilde{x}_2,\tilde{x}_3}\right)$ for every $\ulinex, \uline{\tilde{x}} \in \mathcal{\ulineX}$ satisfying $x_3=\tilde{x}_3$. 
\end{definition}
{
Since users 2 and 3 do not suffer interference, user 1 doesn't need to split its message, and it suffices only for users 2 and 3 to split their messages, so a Han-Kobayashi~\cite{198101TIT_HanKob} strategy will only employ message splitting via superposition coding for users 2 and 3. The following inner bound can be proven via simultaneous decoding. Since each user only needs to decode into two codebooks, one can use the technique developed in \cite[Sec.~V.B]{201206TIT_FawHaySavSenWil} for simultaneously decoding two codebooks.}
We use these inner bounds to prove the strict suboptimality of unstructured IID codes in Appendices \ref{App:additiveinterference} and \ref{App:Nonadditive3to1CQIC}.
\begin{theorem}
    A rate-cost triple $(R_1,R_2,R_3,\tau_1,\tau_2,\tau_3)$ is achievable if there exists (i) two finite sets $\mathcal{U}_2, \mathcal{U}_3$, and (ii) a PMF $p_{U_2U_3\ulineX}=p_{X_1} p_{U_2X_2} p_{U_3X_3}$ defined on $\mathcal{U}_2 \times \mathcal{U}_3 \times \mathcal{\ulineX}$ such that $\mathbb{E}[\kappa_j(X_j)]\leq \tau_j : j\in [3]$ and 
    \begin{eqnarray}
        R_1 &<& I(X_1;Y_1|U_2,U_3) \label{Bnd1:3to1CQICIIDCodes} \\
        R_j &<& I(U_jX_j;Y_j) : j=2,3 \label{Bnd2:3to1CQICIIDCodes} \\
        R_1 + R_j &<& I(U_jX_1;Y_1|{U_{\msout{j}}}) + I(X_j;Y_j|U_{j}) : j=2,3, \, \msout{j} \neq j \label{Bnd3:3to1CQICIIDCodes} \\
       R_1 + R_2 + R_3 &<& I(U_2,U_3,X_1;Y_1) + I(X_2;Y_2|U_2) + I(X_3;Y_3|U_3) \label{Bnd4:3to1CQICIIDCodes}
    \end{eqnarray}
    holds, where all the information quantities are computed with respect to the state 
    \begin{eqnarray}
       \rho^{U_2U_3\ulineX\ulineY} \define \sum_{u_2,u_3,\ulinex} p_{X_1}(x_1) p_{U_2X_2}(u_2,x_2) p_{U_3X_3}(u_3,x_3) \ketbra{u_2,u_3,\ulinex} \otimes \rho^{Y_1Y_2Y_3}_{x_1,x_2,x_3}. \nonumber 
    \end{eqnarray}
\end{theorem}

%%%%%%%%%%%%%%%%%%%%%%%%%%%%%%%%%%%%%%%%%%%%%%%
\section{Proof of Proposition \ref{Prop:strictsuboptimalityiidcodesforadditiveinterference} and \ref{Prop:nonadditiveinterference}}
\label{App:Additiveandnonadditiveinterference}

\subsection{Proof of Proposition \ref{Prop:strictsuboptimalityiidcodesforadditiveinterference}}
\label{App:additiveinterference}

Consider the $3-$CQIC defined in Ex.~\ref{Ex:3CQICadditivenoncommutative}. It can be verified that $\CalC_k=1-h_b(\delta_k)$ for $k=2,3$. From Fact \ref{fact:examples}, we have $\CalC_1=h_b(\frac{1+\sqrt{1-4\tau(1-\tau) \sin(\varphi)^2}}{2})$ and $C_1=h_b(\frac{1+\cos(\varphi)}{2})$. We prove that the rate triple $(\CalC_1,\CalC_2,\CalC_3)$ is not achievable using unstructured IID codes under the condition 
\begin{eqnarray}
    \CalC_1 + \CalC_2 + \CalC_3 > C_1 ,\label{Eqn:Conditionsofexamples}
\end{eqnarray}
We prove this by contradiction. Suppose there exists PMF $p_{X_1}p_{U_2X_2}p_{U_3X_3}$ on $\CalU_2 \times \CalU_3 \times \mathcal{\ulineX}$ such that $(\CalC_1,\CalC_2,\CalC_3)$ is achievable. From \eqref{Bnd4:3to1CQICIIDCodes}, we have 
\begin{eqnarray}
    \CalC_1 + \CalC_2 + \CalC_3 \leq I(U_2,U_3,X_1;Y_1) + I(X_2;Y_2|U_2) + I(X_3;Y_3|U_3). \nonumber 
\end{eqnarray}
If $H(X_2|U_2)=H(X_3|U_3)=0$, then we have 
\begin{eqnarray}
    \CalC_1 + \CalC_2 + \CalC_3 \leq I(X_2,X_3,X_1;Y_1) \leq C_1 \nonumber 
\end{eqnarray}
which violates the condition \eqref{Eqn:Conditionsofexamples}. Hence, we should have $H(X_2|U_2)>0$ or $H(X_3|U_3)>0$. Without loss of generality, we assume  $H(X_2|U_2)>0$. Using the independence of $(U_2,X_2)$ and $(U_3,X_3)$, we have 
\begin{eqnarray}
    0 < H(X_2|U_2)&=&H(X_2|X_3,U_3,U_2) \nonumber \\
    &=&H(X_2 \oplus X_3|X_3,U_3,U_2) \nonumber \\
    &\leq& H(X_2 \oplus X_3|U_3,U_2). \nonumber
\end{eqnarray}
Hence, $H(X_2 \oplus X_3 |U_2,U_3)>0$, which guarantees the existence of $(u_2^*,u_3^*) \in \CalU_2 \times \CalU_3$ such that $H(X_2 \oplus X_3|U_2=u_2^*,U_3=u_3^*)>0$.

\begin{lemma}
\label{lemma:expsuboptimality}
Let $\CalX=\mathcal{B}=\{0,1\}$, $\CalH_Y=\mathbb{C}^2$ and $\rho_x=\ketbra{0} \mathds{1}\{x=0\} + \ketbra{v_{\varphi}} \mathds{1}\{x=1\}$. Then $I(X;Y) \leq I(X;Y|B)$, where the mutual information $I(.,.)$ is with respect to the state 
\[\rho^{XBY}=\sum_x \sum_b p_X(x) p_B(b) \ketbra{x,b} \otimes \rho_{x \oplus b}.\]
Moreover, if $H(Y|B)>0$, and $H(Y|X)>0$, then the inequality is strict, i.e., $I(X;Y) < I(X;Y|B)$.
\end{lemma}
\begin{proof} From Fact \ref{fact:examples}, we have 
\begin{eqnarray}
    I(X;Y)= h_b\left(f\left(p_X(1)*p_B(1)\right)\right)-h_b\left(f\left(p_B(1)\right)\right),  \mbox{ and }  I(X;Y|B)=h_b\left(f\left(p_X(1)\right)\right), \nonumber 
\end{eqnarray}
where $f(t)=\frac{1+\sqrt{1-4t(1-t)\sin(\varphi)^2}}{2}$ for $t \in [0,\frac{1}{2}]$.
It is easy to verify that if (1) $H(Y|B)=0$, then $I(X;Y)=I(X;Y|B)=0$ and if (2) $H(Y|X)=0$, then $I(X;Y)=I(X;Y|B)=h_b\left(f(p_{X}(1))\right)$. Now, if $H(Y|B)>0$ and $H(Y|X)>0$, we define $g(t)=h_b\left(f\left(p_X(1)*t\right)\right) - h_b\left(f(t)\right)$ for $t \in [0,\frac{1}{2}]$. Observe that $g(p_{B}(1))=I(X;Y)$ and $g(0)=I(X;Y|B)$.
Therefore, we need to show that $g(p_B(1))< g(0)$. 

\med By Taylor series, $g(p_B(1))=g(0)+p_B(1)g^{\prime}(\tilde{t})$ for some $\tilde{t} \in (0,p_B(1))$. Hence, it suffices to prove $g^{\prime}(t) <0$ for all $t \in (0,\frac{1}{2})$.
We abbreviate $p_X(1)=p$ and we calculate $g^{\prime}(t)$. It can be verified that 
\begin{eqnarray}
    g^{\prime}(t)= \log \left(\Big(\frac{1-f(t*p)}{f(t*p)}\Big)^{(1-2p)f^{\prime}(t*p)}\right)  -\log \left(\Big(\frac{1-f(t)}{f(t)}\Big)^{f^{\prime}(t)}\right). \nonumber
\end{eqnarray}
Since $\log(x^{a+b})= \log(x^a)+ \log(x^b)$, we split the first term into two separate terms. We obtain 
\begin{eqnarray}
    g^{\prime}(t)= \log \left(\Big(\frac{1-f(t*p)}{f(t*p)}\Big)^{f^{\prime}(t*p)}\right) + \log \left(\Big(\frac{1-f(t*p)}{f(t*p)}\Big)^{-2pf^{\prime}(t*p)}\right) -\log \left(\Big(\frac{1-f(t)}{f(t)}\Big)^{f^{\prime}(t)}\right). \nonumber
\end{eqnarray}
\noindent Note that (i) for $t \in (0,\frac{1}{2})$, $ 0 \leq \frac{1-f(t*p)}{f(t*p)} \leq 1$ and $-2p f^{\prime}(t*p) > 0$. Therefore,
\begin{eqnarray}
\log \left(\Big(\frac{1-f(t*p)}{f(t*p)}\Big)^{-2pf^{\prime}(t*p)}\right) \leq 0. \nonumber     
\end{eqnarray}
(ii) Define $\gamma(t)= \log\left(\Big(\frac{1-f(t)}{f(t)}\Big)^{f^{\prime}(t)}\right)$. Since $f$ is convex ($f^{\prime \prime} \geq 0 $), $\frac{(f^{\prime}(t))^2}{f(t)(1-f(t))}>0$, and $ \log (\frac{1-f(t)}{f(t)}) \leq 0 $ for $t \in (0, \frac{1}{2})$, we have 
\begin{eqnarray}
\gamma^{\prime}(t)=f^{\prime \prime}(t) \log (\frac{1-f(t)}{f(t)}) - \frac{(f^{\prime}(t))^2}{f(t)(1-f(t))} \leq 0. \nonumber     
\end{eqnarray}
Therefore, $\gamma(t*p) < \gamma(t)$, which implies 
\begin{eqnarray}
    \log \left(\Big(\frac{1-f(t*p)}{f(t*p)}\Big)^{f^{\prime}(t*p)}\right) < \log \left(\Big(\frac{1-f(t)}{f(t)}\Big)^{f^{\prime}(t)}\right). \nonumber 
\end{eqnarray}
Combining (i) and (ii), we conclude that $g^{\prime}(t) < 0$ for
$t \in (0,\frac{1}{2})$.
\end{proof}

\noindent Next, observe that, 
\begin{eqnarray}
\CalC_1 < I(X_1;Y_1|U_2,U_3) &=& \sum_{u_2} \sum_{u_3} p_{U_2}(u_2) p_{U_3}(u_3) I(X_1;Y_1|U_2=u_2,U_3=u_3) \nonumber \\
&<& \sum_{u_2} \sum_{u_3} p_{U_2}(u_2) p_{U_3}(u_3) I(X_1;Y_1| X_2 \oplus X_3, U_2=u_2,U_3=u_3) \nonumber \\
&=& I(X_1;Y_1| X_2 \oplus X_3, U_2,U_3) \nonumber \\
&=&h_b(f(p_X(1))) \nonumber \\
&\leq&  \CalC_1, \nonumber 
\end{eqnarray}
where the first inequality follows from (i) the existence of $(u_2^*,u_3^*) \in \CalU_2 \times \CalU_3$ such that $H(X_2 \oplus X_3|U_2=u_2^*,U_3=u_3^*)>0$, which implies $H(Y_1|X_1,U_2=u_2^*,U_3=u_3^*)>0$ and (ii) by substituting $p_{X_1}$ for $p_{X}$ and $p_{X_2 \oplus X_3|U_2,U_3}$ for $p_{B}$ in lemma \ref{lemma:expsuboptimality}. The last inequality follows from the cost constraint on $X_1$.
Thus, the rate triple $(\CalC_1,\CalC_2,\CalC_3)$ is not achievable.

\subsection{Proof of Proposition \ref{Prop:nonadditiveinterference}}
\label{App:Nonadditive3to1CQIC}
Consider the $3-$CQIC defined in Ex.~\ref{Ex:nonadditiveinterference}. It can be verified that $\CalC_j= h_b(\tau_j * \delta_j) -h_b(\delta_j)$ for $j=2,3$. From Fact \ref{fact:examples}, we have $\CalC_1 = h_b(f(\tau_1))$ and $C_1=h_b(f(\tau_1 *\beta))$, where $\beta=\tau_2 + \tau_3 -\tau_2\tau_3$ and $f(t)=\frac{1+\sqrt{1-4t(1-t)\sin(\varphi)^2}}{2}$ for $t \in [0,\frac{1}{2}]$.

We prove that the rate triple $(\CalC_1,\CalC_2,\CalC_3)$ is not achievable using unstructured codes under the condition
\begin{eqnarray}
\CalC_1 + \CalC_2 + \CalC_3 < C_1.
 \label{Eq:Conditionforproofofpropnonadditiveinterference}
\end{eqnarray}
We prove this by contradiction. Suppose there exists a PMF $p_{X_1}p_{U_2X_2}p_{U_3X_3}$ defined on $\CalU_2 \times \CalU_3 \times \mathcal{\ulineX}$ such that $(\CalC_1,\CalC_2,\CalC_3)$ is achievable under the condition \eqref{Eq:Conditionforproofofpropnonadditiveinterference}.
We define a characterization of such PMF $p_{X_1}p_{U_2X_2}p_{U_3X_3}$ and use it to derive a contradiction. Observe that 

\noindent (i) from \eqref{Bnd2:3to1CQICIIDCodes} and for $j=2,3$, we have
\begin{eqnarray}
R_j &\leq& I\left(U_jX_j;Y_j\right) \nonumber \\
    &=& H\left(Y_j\right)-H\left(Y_j|U_jX_j\right) \nonumber \\
    &=&H\left(\sum_{x_j} p_{X_j}(x_j) \rho_{x_j}^{Y_j}\right)-\sum_{u_j}\sum_{x_j}p_{U_jX_j}(u_j,x_j) H\left(\rho_{x_j}^{Y_j}\right) \nonumber \\
    &=&h_b\left(p_{X_j}(1)*\delta_j\right)-h_b\left(\delta_j\right)  \nonumber \\
    &\leq& h_b\left(\tau_j*\delta_j\right)-h_b\left(\delta_j\right) \nonumber,
\end{eqnarray}
where the last inequality follows from the following observations $0 \leq p_{X_j}(1) \leq \tau_j < \frac{1}{2}$,
\begin{eqnarray}
    p_{X_j}(1)*\delta_j&=&p_{X_j}(1)(1-2\delta_j)+\delta_j \nonumber \\
    &\leq& \tau_j(1-2\delta_j)+\delta_j \nonumber \\
    &<& \frac{1}{2}(1-2\delta_j)+\delta_j \nonumber \\
    &=& \frac{1}{2}, \nonumber 
\end{eqnarray} 
and the fact that $h_b(.)$ is strictly increasing on $ [0,\frac{1}{2}]$, which implies $h_b(p_{X_j}(1)*\delta_j)$ is strictly increasing in $p_{X_j}(1)$. Note that $R_j=h_b\left(\tau_j*\delta_j\right)-h_b\left(\delta_j\right)$ if and only if $p_{X_j}(1)=\tau_j$.

\noindent (ii) From \eqref{Bnd1:3to1CQICIIDCodes}, we have 
\begin{eqnarray}
    R_1 \leq I\left(X_1;Y_1|U_2,U_3\right)= \sum_{u_2,u_3} p_{U_2}(u_2) p_{U_3}(u_3) I\left(X_1;Y_1|U_2=u_2,U_3=u_3\right), \nonumber 
    \end{eqnarray}
    substituting $p_{X_1}$ for $p_X$ and $p_{X_2 \lor X_3|U_2U_3}(.|u_2,u_3)$ for $p_B$ in lemma \ref{lemma:expsuboptimality}, we obtain 
    \begin{eqnarray}
     R_1 &\leq& \sum_{u_2,u_3} p_{U_2}(u_2) p_{U_3}(u_3) I\left(X_1;Y_1|X_2 \lor X_3, U_2=u_2,U_3=u_3\right) \nonumber \\ 
    &=&\sum_{\substack{u_2, u_3 \\ x_2 \lor x_3}} p_{U_2 U_3 X_2 \lor X_3}(u_2,u_3,x_2 \lor x_3) I\left(X_1;Y_1|X_2 \lor X_3=x_2 \lor x_3, U_2=u_2,U_3=u_3\right) \nonumber \\
    &=& \sum_{\substack{u_2, u_3 \\ x_2 \lor x_3}} p_{U_2 U_3 X_2 \lor X_3}(u_2,u_3,x_2 \lor x_3) \nonumber \\
    &&\left[ H\left(Y_1|X_2 \lor X_3=x_2 \lor x_3, U_2=u_2,U_3=u_3\right)-H\left(Y_1|X_1,X_2 \lor X_3=x_2 \lor x_3, U_2=u_2,U_3=u_3\right)\right] \nonumber \\
    &=& \sum_{ x_2 \lor x_3} p_{X_2 \lor X_3}(x_2 \lor x_3) \left[ H\left(\sum_{x_1}p_{X_1}(x_1) \rho^{Y_1}_{x_1 \oplus (x_2 \lor x_3)}\right)- \sum_{x_1} p_{X_1}(x_1) H\left(\rho^{Y_1}_{x_1 \oplus (x_2 \lor x_3)}\right)\right]. \nonumber 
    \end{eqnarray}
    since $H\left(\rho^{Y_1}_{x_1 \oplus (x_2 \lor x_3)}\right)=0$, and $H\left(\sum_{x_1}p_{X_1}(x_1) \rho^{Y_1}_{x_1 \oplus 1}\right)=H\left(\sum_{x_1}p_{X_1}(x_1) \rho^{Y_1}_{x_1}\right)=h_b\left(f\left(p_{X_1}(1)\right)\right)$, we obtain
    \begin{eqnarray}
    R_1 &\leq&  \sum_{x_2 \lor x_3} p_{X_2 \lor X_3}(x_2 \lor x_3) H\left(\sum_{x_1}p_{X_1}(x_1) \rho_{x_1 \oplus (x_2 \lor x_3)}^{Y_1}\right) \nonumber \\
&=& H\left(\sum_{x_1}p_{X_1}(x_1) \rho_{x_1}^{Y_1}\right) \nonumber \\
&=& h_b\left(f\left(p_{X_1}(1)\right)\right) \nonumber \\
&\leq& h_b\left(f\left(\tau_1\right)\right), \nonumber
\end{eqnarray}
\noindent where the last inequality follows from 
$0 \leq p_{X_1}(1) \leq \tau_1 < \frac{1}{2}$ 
and from
\[
h_b\left(f\left(t\right)\right)' = \frac{(2t-1)\sin^2(\varphi)}{\sqrt{1 - 4 t (1-t) \sin^2(\varphi)}} \, 
\log\left(\frac{1-f(t)}{f(t)}\right) > 0, \:\: t \in (0, 1/2),
\]
which implies that $h_b\left(f\left(p_{X_1}(1)\right)\right)$ is strictly increasing in \(p_{X_1}(1)\). Moreover, $R_1=h_b\left(f\left(\tau_1\right)\right)$ if and only if $p_{X_1}(1)=\tau_1$.

\noindent (iii) Suppose $H(X_2|U_2)= H(X_3|U_3)=0$. Therefore, from \eqref{Bnd4:3to1CQICIIDCodes}, we have 
\begin{eqnarray}
    R_1+ R_2+R_3 &\leq& I\left(X_1,X_2,X_3;Y_1\right) \nonumber \\
    &=&H\left(Y_1\right)-H\left(Y_1|X_1,X_2,X_3\right) \nonumber \\
    &=&H\left(\sum_{\ulinex}p_{\ulineX}(\ulinex) \rho_{x_1 \oplus (x_2 \lor x_3)}^{Y_1}\right) - \sum_{\ulinex} p_{\ulineX}(\ulinex) H\left(\rho_{x_1 \oplus (x_2 \lor x_3)}^{Y_1}\right) \nonumber \\
    &=&h_b\left(f\left(p_{X_1}(1) * p_{X_2 \lor X_3}(1)\right)\right) \nonumber \\
    &=&h_b\left(f\left(\tau_1*\beta\right)\right), \nonumber
 \end{eqnarray}
where the last inequality follows from substituting $p_{X_1}(1)= \tau_1$ and $p_{X_j}(1)=\tau_j : j=2,3$ as obtained in (i) and (ii). This violates the hypothesis \eqref{Eq:Conditionforproofofpropnonadditiveinterference}. Hence, we should have $H(X_2|U_2)>0$ or $H(X_3|U_3)>0$. 

\noindent (iv) Without loss of generality, we assume $H(X_2|U_2)>0$ . Therefore, there exists $u_2^* \in \CalU_2$ such that $p_{U_2}(u_2^*)>0$ and $H(X_2|U_2=u_2^*)>0$ which implies $p_{X_2|U_2}(x_2|u_2^*)>0$ for $x_2 \in \{0,1\}$. Additionally, from (i), we have $0< 1-\tau_3=p_{X_3}(0)=\sum_{u_3} p_{U_3}(u_3) p_{X_3|U_3}(0|u_3)$ which ensures the existence of $u_3^* \in \CalU_3$ such that $p_{U_3}(u_3^*)>0$ and $0<p_{X_3|U_3}(0|u_3^*) \leq 1$. Hence, there exists $(u_2^*,u_3^*) \in \CalU_2 \times \CalU_3$ such that 
\begin{eqnarray}
    \mbox{ (a) } p_{U_2}(u_2^*)>0, \: p_{U_3}(u_3^*)>0, \mbox{ (b) } 0<p_{X_2|U_2 U_3}(x_2|u_2^*,u_3^*)<1, \mbox{ for }  x_2 \in \{0,1\},\mbox{ and (c) }  0<p_{X_3|U_2 U_3}(0|u_2^*,u_3^*) \leq 1, \nonumber
\end{eqnarray} 
where (b) and (c) follow from the independence of the pair $(U_2,X_2)$ and $(U_3,X_3)$. Therefore, $0< p_{X_2 \lor X_3 |U_2 U_3}(x|u_2^*,u_3^*) < 1, \mbox{ for } x \in \{0,1\}$.
Hence 
\begin{eqnarray}
H\left(X_2 \lor X_3|U_2,U_3\right)=\sum_{u_2,u_3} p_{U_2}(u_2) p_{U_3}(u_3) H\left(X_2 \lor X_3| U_2=u_2, U_3=u_3\right) > 0. \nonumber    
\end{eqnarray}

\noindent (v) From \eqref{Bnd1:3to1CQICIIDCodes}, we have 
\begin{eqnarray}
     R_1 &\leq& I\left(X_1;Y_1|U_2,U_3\right) \nonumber \\
     &=& \sum_{u_2,u_3} p_{U_2}(u_2) p_{U_3}(u_3) I\left(X_1;Y_1|U_2=u_2,U_3=u_3\right). \nonumber 
\end{eqnarray}
From (iv), there exists $(u_2^*,u_3^*) \in \CalU_2 \times \CalU_3$ such that 
\begin{eqnarray}
    H\left( Y_1|X_1,U_2=u_2^*,U_3=u_3^*\right) &=& \sum_{x_1} p_{X_1}(x_1) H\left( \sum_{x_2 \lor x_3} p_{X_2 \lor X_3 | U_2U_3}(x_2 \lor x_3|u_2^*,u_3^*) \rho_{x_1 \oplus (x_2 \lor x_3)}^{Y_1}\right) \nonumber \\
    &=& H\left( \sum_{x_2 \lor x_3} p_{X_2 \lor X_3 | U_2U_3}(x_2 \lor x_3|u_2^*,u_3^*) \rho_{ x_2 \lor x_3}^{Y_1} \right) \nonumber \\
    &=&h_b\left( f\left(p_{X_2 \lor X_3 |U_2U_3}(1| u_2^*,u_3^*)\right) \right) >0.\nonumber 
\end{eqnarray}
And since $ 0< \tau_1 = p_{X_1}(1)$, we have 
\begin{eqnarray}
    H\left( Y_1 | X_2 \lor X_3 , U_2=u_2^*, U_3=u_3^*\right) &=& \sum_{x_2 \lor x_3} p_{X_2 \lor X_3 | U_2 U_3}(x_2 \lor x_3|u_2^*,u_3^*) H\left( \sum_{x_1} p_{X_1}(x_1) \rho_{x_1 \oplus (x_2 \lor x_3)}^{Y_1} \right) \nonumber \\
    &=& H\left( \sum_{x_1} p_{X_1}(x_1) \rho_{x_1}^{Y_1} \right) \nonumber \\
    \hspace{-0.25in}&=&h_b\left(f\left(p_{X_1}(1)\right) \right) >0. \nonumber 
\end{eqnarray}
Consequently, by substituting $p_{X_1}$ for $p_X$ and $p_{X_2 \lor X_3 |U_2 U_3}(.|u_2^*,u_3^*)$ for $p_B$ in lemma \ref{lemma:expsuboptimality}, we have
\begin{eqnarray}
R_1 &<& \sum_{u_2,u_3} p_{U_2}(u_2) p_{U_3}(u_3) I\left(X_1;Y_1|X_2 \lor X_3, U_2=u_2,U_3=u_3\right) \nonumber \\
&=& h_b\left(f\left(\tau_1\right)\right).    \nonumber
\end{eqnarray}
The last equality follows from the same steps derived in (ii). Hence $R_1<h_b\left(f(\tau_1)\right)=\CalC_1$. We have thus derived a contradiction $\CalC_1=R_1 < \CalC_1$.

\section{Proof of Proposition \ref{Prop:Dec1Cuffterm}}
\label{App:Dec1Cuffterm}
Observe that
\begin{eqnarray}
t_1 &\leq& \frac{1}{|\mathcal{\ulineM}|} \sum_{\ulinem} \sum_{a_2, a_3}  \left|\mathds{1}{\left\{a_2(m_2)=a_2\right\}} \mathds{1}{\left\{a_3(m_3)=a_3\right\}}-\frac{r_{U_2}^n(u_2^n(a_2,m_2))}{\Prime^{k_2}}\frac{r_{U_3}^n(u_3^n(a_3,m_3))}{\Prime^{k_3}}\right|, \nonumber \\
&&\left|\tr \left\{  \left( I - \mu^{Y_1}_{m_1,s} \right)  \left( \rho_{x_1^n(m_1),f_2^n(u_2^n(a_2,m_2)),f_3^n(u_3^n(a_3,m_3))}^{Y_1} \otimes \ketbra{0}\right)  \right\}\right| \nonumber \\
&\leq& \frac{1}{|\mathcal{\ulineM}|} \sum_{\ulinem} \sum_{a_2, a_3}  \left|\mathds{1}{\left\{a_2(m_2)=a_2\right\}} \mathds{1}{\left\{a_3(m_3)=a_3\right\}}-\frac{r_{U_2}^n(u_2^n(a_2,m_2))}{\Prime^{k_2}}\frac{r_{U_3}^n(u_3^n(a_3,m_3))}{\Prime^{k_3}}\right|, \nonumber
\end{eqnarray}
We evaluate the expectation over $C_2(m_2)$ and $C_3(m_3)$. We obtain 
\begin{eqnarray}
\mathbb{E}\left[t_1\right] &\leq& \frac{1}{|\mathcal{\ulineM}|} \sum_{\ulinem} \sum_{a_2, a_3}  \sum_{c_2(m_2)} \sum_{c_3(m_3)} P\left( C_2(m_2)=\left(u_2^n(a_2,m_2): a_2 \in \CalF_{\Prime}^{k_2}\right)\right) \nonumber \\ &&P\left(C_3(m_3)=\left(u_3^n(a_3,m_3): a_3 \in \CalF_{\Prime}^{k_3}\right)\right) \nonumber \\ && \left|\frac{r_{U_2}^n(u_2^n(a_2,m_2))}{\sum_{\Tilde{a}_2}  r_{U_2}^n(u_2^n(\Tilde{a}_2,m_2))}\frac{r_{U_3}^n(u_3^n(a_3,m_3))}{\sum_{\Tilde{a}_3}  r_{U_3}^n(u_3^n(\Tilde{a}_3,m_3))  } -\frac{r_{U_2}^n(u_2^n(a_2,m_2))}{\Prime^{k_2}}\frac{r_{U_3}^n(u_3^n(a_3,m_3))}{\Prime^{k_3}}\right| \nonumber \\
&=&\frac{1}{|\mathcal{\ulineM}|} \sum_{\ulinem} \sum_{a_2, a_3}  \sum_{c_2(m_2)} \sum_{c_3(m_3)} P\left( C_2(m_2)=\left(u_2^n(a_2,m_2): a_2 \in \CalF_{\Prime}^{k_2}\right)\right) \nonumber \\ && P\left( C_3(m_3)=\left(u_3^n(a_3,m_3): a_3 \in \CalF_{\Prime}^{k_3}\right) \right) \nonumber \\ &&\frac{r_{U_2}^n(u_2^n(a_2,m_2))}{\sum_{\Tilde{a}_2}  r_{U_2}^n(u_2^n(\Tilde{a}_2,m_2))}\frac{r_{U_3}^n(u_3^n(a_3,m_3))}{\sum_{\Tilde{a}_3}  r_{U_3}^n(u_3^n(\Tilde{a}_3,m_3))  } \left|1-\frac{1}{\Prime^{k_2 + k_3}} \sum_{\tilde{a}_2} \sum_{\tilde{a}_3} r_{U_2}^n(u_2^n(\tilde{a}_2,m_2)) r_{U_3}^n(u_3^n(\tilde{a}_3,m_3))\right| \nonumber \\
&=& \frac{1}{|\mathcal{\ulineM}|} \sum_{\ulinem}  \sum_{c_2(m_2)} \sum_{c_3(m_3)} P\left( C_2(m_2)=\left(u_2^n(a_2,m_2): a_2 \in \CalF_{\Prime}^{k_2}\right)\right) \nonumber \\ && P\left(C_3(m_3)=\left(u_3^n(a_3,m_3): a_3 \in \CalF_{\Prime}^{k_3}\right) \right)\left|\frac{1}{\Prime^{k_2 + k_3}} \sum_{\tilde{a}_2} \sum_{\tilde{a}_3} r_{U_2}^n(u_2^n(\tilde{a}_2,m_2)) r_{U_3}^n(u_3^n(\tilde{a}_3,m_3))-1\right| \nonumber \\
&=&  \mathbb{E}\left[ \left|\frac{1}{\Prime^{k_2 + k_3}} \sum_{\tilde{a}_2} \sum_{\tilde{a}_3} r_{U_2}^n(U_2^n(\tilde{a}_2,M_2)) r_{U_3}^n(U_3^n(\tilde{a}_3,M_3))-1\right| \right],\nonumber 
\end{eqnarray}
\begin{comment}
where the first equality follows from 
\begin{eqnarray}
 \sum_{a_2}\frac{r_{U_2}^n(u_2^n(a_2,m_2))}{\sum_{\Tilde{a}_2}  r_{U_2}^n(u_2^n(\Tilde{a}_2,m_2))
}=1 \quad \mbox{and} \quad     \sum_{a_3}\frac{r_{U_3}^n(u_3^n(a_3,m_3))}{\sum_{\Tilde{a}_3}  r_{U_3}^n(u_3^n(\Tilde{a}_3,m_3))  } =1. \nonumber 
\end{eqnarray}    
\end{comment}
Using the `cloud mixing' lemma \cite[Lemma~19]{CuffPhDThesis}, we have $\mathbb{E}[t_1] \leq \epsilon,$ if
\begin{eqnarray}
  \frac{k_j}{n} \log(\Prime) > \log(\Prime) - H(U_j), \mbox{ for } j=2,3. \nonumber  
\end{eqnarray}

\section{ Proof of Proposition \ref{Prop:3to1CQICclosnessofstates}: Closeness of the states}
\label{App:3to1CQICClosenessOfStates}
We mimic the steps in \cite{202103SAD_Sen} in this proof.
For any $\ket{h} \in \boldsymbol{\CalH_{Y_1}^G}$, we have 
\begin{eqnarray}
    \norm{\CalT_{d_1^n,d_2^n,\eta}(\ketbra{h})-\ketbra{h}}_1 \leq 2 \norm{\CalT_{d_1^n,d_2^n,\eta}(\ket{h})-\ket{h}}_2. \nonumber 
\end{eqnarray}
Observe that
\begin{eqnarray}
\norm{\CalT_{d_1^n,d_2^n,\eta}(\ket{h})-\ket{h}}_2^2 &=& \left( \CalT_{d_1^n,d_2^n,\eta}(\bra{h}) - \bra{h} \right) \left( \CalT_{d_1^n,d_2^n,\eta}(\ket{h}) - \ket{h} \right) \nonumber \\
    &=& <\CalT_{d_1^n,d_2^n,\eta}(h),\CalT_{d_1^n,d_2^n,\eta}(h)>-<\CalT_{d_1^n,d_2^n,\eta}(h),h>-<h,\CalT_{d_1^n,d_2^n,\eta}(h)>+<h,h> \nonumber \\
    &=& 2-\frac{2}{\sqrt{1 + 2 \eta^2}} \nonumber, 
\end{eqnarray}
    where the last equality follows from the fact that $\CalT_{d_1^n,d_2^n,\eta}$ is an isometry and 
    \begin{eqnarray}
        <\CalT_{d_1^n,d_2^n,\eta}(h),h>=<h,\CalT_{d_1^n,d_2^n,\eta}(h)>= \frac{1}{\sqrt{1 + 2 \eta^2}}. \nonumber 
    \end{eqnarray} 
    
\noindent Using this, we obtain 
\begin{eqnarray}
    \norm{\CalT_{d_1^n,d_2^n,\eta}(\ketbra{h})-\ketbra{h}}_1 \leq 2 \norm{\CalT_{d_1^n,d_2^n,\eta}(\ket{h})-\ket{h}}_2 \leq 2 \sqrt{2-2e^{-2\eta^2}} \leq 4 \eta, \nonumber 
\end{eqnarray}
where the first inequality follows from $\frac{1}{\sqrt{1 + 2 \eta^2}} > e^{-2 \eta^2}$ and the last inequality follows from using  $e^{-x} \geq 1-x$.

\section{Proof of Proposition \ref{Prop:Dec1firsttermHay}}
\label{App:Dec1FirsttermHay}
We first prove a property of the decoding POVM elements $G_{x_1^n,u^n}^{\mathcal{J}}$, for $\mathcal{J}\in  \{1,2,3\}$, that will be applied in this appendix.
For $\mathcal{J}=1$, we have 
\begin{eqnarray}
    \sum_{x_1^n} \sum_{u^n} p_{X_1}^n(x_1^n) p_{U_2 \oplus U_3}^n(u^n) \tr \left\{ G^{1}_{x_1^n,u^n} \rho^{Y_1}_{x_1^n,u^n}\right\}
    &=& \sum_{(x_1^n,u^n) \in T_{\delta}^n(p_{X_1} p_{U_2 \oplus U_3})} p_{X_1}^n(x_1^n) p_{U_2 \oplus U_3}^n(u^n) \tr \left\{ \Pi_{u^n}^{Y_1} \: \Pi^{Y_1}_{x_1^n,u^n} \: \Pi_{u^n}^{Y_1} \rho^{Y_1}_{x_1^n,u^n}  \right\} \nonumber \\
    &=& \sum_{(x_1^n,u^n) \in T_{\delta}^n(p_{X_1} p_{U_2 \oplus U_3})} p_{X_1}^n(x_1^n) p_{U_2 \oplus U_3}^n(u^n)  \tr \left\{  \Pi^{Y_1}_{x_1^n,u^n} \: \Pi_{u^n}^{Y_1} \: \rho^{Y_1}_{x_1^n,u^n} \: \Pi_{u^n}^{Y_1}    \right\} \nonumber \\
    &\geq&  \sum_{(x_1^n,u^n) \in T_{\delta}^n(p_{X_1} p_{U_2 \oplus U_3})} p_{X_1}^n(x_1^n) p_{U_2 \oplus U_3}^n(u^n)  \tr \left\{  \Pi^{Y_1}_{x_1^n,u^n}  \rho^{Y_1}_{x_1^n,u^n}  \right\}  \nonumber \\
    &-& \sum_{(x_1^n,u^n) \in T_{\delta}^n(p_{X_1} p_{U_2 \oplus U_3})} p_{X_1}^n(x_1^n) p_{U_2 \oplus U_3}^n(u^n)  \norm{\Pi_{u^n}^{Y_1} \: \rho^{Y_1}_{x_1^n,u^n} \: \Pi_{u^n}^{Y_1}-\rho^{Y_1}_{x_1^n,u^n}}_{1} \nonumber \\
    &\geq& 1-\epsilon - 2 \sqrt{\epsilon} \nonumber,
\end{eqnarray}
The first equality follows from the definition of $G^1_{x_1^n,u^n}$ in \eqref{Eqn:3to1CQICtheoriginalpovmelements}. The second equality follows from the cyclicity of the trace. The first inequality follows from the trace inequality $\tr\left( \Lambda \rho\right) \geq \tr\left( \Lambda \sigma \right) - \norm{\rho - \sigma}_1$, for $0 \leq \Lambda, \rho, \sigma \leq I$. The last inequality follows from (i) the Gentle Operator Lemma \cite{BkWilde_2017} and (ii) from the typicality property
\begin{eqnarray}
     \tr \left\{  \Pi^{Y_1}_{x_1^n,u^n}  \rho^{Y_1}_{x_1^n,u^n}  \right\}  \geq 1-\epsilon, \mbox{ for }(x_1^n,u^n) \in T_{\delta}^n(p_{X_1} p_{U_2 \oplus U_3}). \nonumber 
\end{eqnarray}
The same argument applies for $\mathcal{J} \in \{2, 3\}$.

Now, we evaluate the expectation of $t_{2.1.1}$ over the codebook generation distribution. We obtain
\begin{eqnarray}
\mathbb{E}\left[t_{2.1.1}\right]&=& \frac{1}{|\mathcal{M}_1|} \frac{1}{\Prime^{k_{\sfb} + l_{\sfb}}} \frac{1}{|\mathcal{D}_1|^n} \frac{1}{|\mathcal{D}_2|^n}\sum_{m_1} \sum_{s}  \sum_{u^n} \sum_{x_1^n} \sum_{d_1^n, d_2^n}  p_{U_2 \oplus U_3}^n(u^n) p_{X_1}^n(x_1^n) \nonumber \\
&& \tr \left\{  \left( I - \gamma^{*}_{x_1^n,d^n_1,u^n,d_2^n} \right)   \theta_{x_1^n,d^n_1,u^n,d^n_2} \right\} \nonumber  \\
&\overset{(a)}{=}& \frac{1}{|\mathcal{M}_1|} \frac{1}{\Prime^{k_{\sfb} + l_{\sfb}}} \frac{1}{|\mathcal{D}_1|^n} \frac{1}{|\mathcal{D}_2|^n}\sum_{m_1} \sum_{s} \sum_{u^n} \sum_{x_1^n} \sum_{d_1^n, d_2^n}  p_{U_2 \oplus U_3}^n(u^n) p_{X_1}^n(x_1^n) \nonumber \\
   &&\left[\tr \left\{ \theta_{x_1^n,d^n_1,u^n,d^n_2} \right\} - \tr \left\{    \Pi_{\boldsymbol{\CalH_{Y_1}^G} }\left(I_{\boldsymbol{\CalH_{Y_1}^{e}}}-  \beta^{*}_{x_1^n,d_1^n,u^n,d_2^n}\right) \theta_{x_1^n,d^n_1,u^n,d^n_2} \left(I_{\boldsymbol{\CalH_{Y_1}^{e}}}-  \beta^{*}_{x_1^n,d_1^n,u^n,d_2^n}\right) \Pi_{\boldsymbol{\CalH_{Y_1}^G} } \right\} \right]\nonumber  \\
&\overset{(b)}{\leq}& \frac{4}{|\mathcal{M}_1|} \frac{1}{\Prime^{k_{\sfb} + l_{\sfb}}} \frac{1}{|\mathcal{D}_1|^n} \frac{1}{|\mathcal{D}_2|^n} \sum_{m_1} \sum_{s}  \sum_{u^n} \sum_{x_1^n} \sum_{d_1^n, d_2^n}  p_{U_2 \oplus U_3}^n(u^n) p_{X_1}^n(x_1^n) \nonumber\\ 
&&\tr \left\{ \left(I_{\boldsymbol{\CalH_{Y_1}^{e}}}-\Pi_{\boldsymbol{\CalH_{Y_1}^G} }+\beta^{*}_{x_1^n,d_1^n,u^n,d_2^n} \right)\theta_{x_1^n,d^n_1,u^n,d^n_2} \right\} \nonumber  \\
&\overset{(c)}{\leq}& \frac{4}{|\mathcal{M}_1|} \frac{1}{\Prime^{k_{\sfb} + l_{\sfb}}} \frac{1}{|\mathcal{D}_1|^n} \frac{1}{|\mathcal{D}_2|^n} \sum_{m_1} \sum_{s}  \sum_{u^n} \sum_{x_1^n} \sum_{d_1^n, d_2^n}  p_{U_2 \oplus U_3}^n(u^n) p_{X_1}^n(x_1^n) \nonumber \\ 
    &&\left[\tr \left\{ \beta^{*}_{x_1^n,d_1^n,u^n,d_2^n} \left(\rho^{Y_1}_{x_1^n,u^n}\otimes \ketbra{0}\right) \right\} + \norm{\theta_{x_1^n,d^n_1,u^n,d^n_2}- \left(\rho^{Y_1}_{x_1^n,u^n}\otimes \ketbra{0}\right)}_1 \right] \nonumber  
\end{eqnarray}
where (a) follows from (i) the definition $\gamma^{*}_{x_1^n,d^n_1,u^n,d_2^n}$ in \eqref{Eqn:3to1CQICPovmelementfordecoder1}, (ii) the cyclicity of the trace and (iii) the property of projectors, (b) follows from applying the non-commutative union bound \cite[Fact.3]{202103SAD_Sen}, and (c) follows from (i) substituting $\theta_{x_1^n,d^n_1,u^n,d^n_2}$ with  $\left(\rho^{Y_1}_{x_1^n,u^n}\otimes \ketbra{0}\right)$ and (ii) because $\tr\left\{\left(I_{\boldsymbol{\CalH_{Y_1}^{e}}}-\Pi_{\boldsymbol{\CalH_{Y_1}^G} }  \right) \left(\rho^{Y_1}_{x_1^n,u^n}\otimes \ketbra{0}\right) \right\}=0$, since $\left( \rho^{Y_1}_{x_1^n,u^n}\otimes \ketbra{0} \right)$ lives in $\boldsymbol{\CalH_{Y_1}^G}$ and $I_{\boldsymbol{\CalH_{Y_1}^{e}}}-\Pi_{\boldsymbol{\CalH_{Y_1}^G}}$ is a projector onto the complement of $\boldsymbol{\CalH_{Y_1}^G}$.

\noindent Now applying Prop.~\ref{Prop:3to1CQICclosnessofstates} and \cite[Corollary 1]{202103SAD_Sen}, we obtain  
\begin{eqnarray}
\mathbb{E}\left[t_{2.1.1}\right] &\leq& 16\eta + \frac{6}{\eta^2} \frac{4}{|\mathcal{M}_1|} \frac{1}{\Prime^{k_{\sfb} + l_{\sfb}}} \frac{1}{|\mathcal{D}_1|^n} \frac{1}{|\mathcal{D}_2|^n} \sum_{m_1} \sum_{s}  \sum_{u^n} \sum_{x_1^n} \sum_{d_1^n, d_2^n}  p_{U_2 \oplus U_3}^n(u^n) p_{X_1}^n(x_1^n) \nonumber \\ 
    &&\left[\tr \left\{ \olineB^1_{x_1^n,u^n} \left(\rho^{Y_1}_{x_1^n,u^n}\otimes \ketbra{0}\right) \right\} + \tr \left\{ \olineB^2_{x_1^n,u^n} \left(\rho^{Y_1}_{x_1^n,u^n}\otimes \ketbra{0}\right) \right\} + \tr \left\{ \olineB^3_{x_1^n,u^n} \left(\rho^{Y_1}_{x_1^n,u^n}\otimes \ketbra{0}\right) \right\} \right] \nonumber  \\
&\overset{(a)}{=}& 16 \eta + \frac{6}{\eta^2} \frac{4}{|\mathcal{M}_1|} \frac{1}{\Prime^{k_{\sfb} + l_{\sfb}}} \frac{1}{|\mathcal{D}_1|^n} \frac{1}{|\mathcal{D}_2|^n} \sum_{m_1} \sum_{s}  \sum_{u^n} \sum_{x_1^n} \sum_{d_1^n, d_2^n}  p_{U_2 \oplus U_3}^n(u^n) p_{X_1}^n(x_1^n) \nonumber \\ 
    &&\left[1-\tr \left\{ G^1_{x_1^n,u^n} \rho^{Y_1}_{x_1^n,u^n} \right\} + 1-\tr \left\{ G^2_{x_1^n,u^n} \rho^{Y_1}_{x_1^n,u^n}\right\} + 1-\tr \left\{ G^3_{x_1^n,u^n} \rho^{Y_1}_{x_1^n,u^n} \right\} \right] \nonumber  \\
&\overset{(b)}{\leq}& 16 \eta +  \frac{72}{\eta^2} (\epsilon + 2\sqrt{\epsilon})  \nonumber,
\end{eqnarray}
where (a) follows from (i) the definition of $\olineB_{x_1^n,u^n}^\mathcal{J}$, for $\mathcal{J} \in \{1,2,3\}$ in \eqref{Eqn:3to1CQICcomplementprojector} and (ii) from Gelfand–Naimark’s Thm.~\cite[Thm.~3.7]{BkHolevo_2019}, and (b) follows from the property stated at the beginning of this appendix, for all $\mathcal{J} \in \{1,2,3\}$.
Chossing $\eta=\epsilon^{\frac{1}{5}}$, we have 
\begin{eqnarray}
    \mathbb{E}[t_{2.1.1}] \leq 16 \eta^{\frac{1}{5}} + 72 \epsilon^{\frac{3}{5}} + 144 \epsilon^{\frac{1}{10}}. \nonumber
\end{eqnarray}
\section{Proof of Proposition \ref{Prop:Dec1SecondtermHay}}
\label{App:Dec1SecondtermHay}

\subsubsection*{Analysis of $t_{2.1.2}$} We evaluate the expectation of $t_{2.1.2}$ over the codebook generation distribution. We obtain 
\begin{eqnarray}
\mathbb{E}\left[t_{2.1.2}\right] &=& \frac{1}{|\mathcal{M}_1|} \frac{1}{|\mathcal{D}_1|^{2n}} \frac{1}{|\CalD_2|^n} \frac{1}{\Prime^{k_{\sfb} + l_{\sfb}}} \sum_{m_1} \sum_{\Tilde{m}_1 \neq m_1} \sum_{s}  \sum_{u^n} \sum_{\Tilde{x}_1^n,x_1^n} \sum_{\Tilde{d}_1^n}  
      \sum_{d_1^n, d_2^n}  p_{X_1}^n(\Tilde{x}_1^n) p_{X_1}^n(x_1^n) 
     p_{U_2 \oplus U_3}^n(u^n) \nonumber \\
     &&\tr \left\{  \gamma^{*}_{\Tilde{x}_1^n,\Tilde{d}^n_1,u^n,d_2^n}   \theta_{x_1^n,d^n_1,u^n,d^n_2} \right\}\nonumber \\
     &=&  \frac{1}{|\mathcal{M}_1|} \frac{1}{|\mathcal{D}_1|^n} \frac{1}{|\mathcal{D}_2|^n}  \frac{1}{\Prime^{k_{\sfb} + l_{\sfb}}} \sum_{m_1} \sum_{\Tilde{m}_1 \neq m_1}\sum_{s}  \sum_{u^n} \sum_{\Tilde{x}_1^n} \sum_{\Tilde{d}_1^n, d_2^n}  p_{X_1}^n(\Tilde{x}_1^n)  p_{U_2 \oplus U_3}^n(u^n) \tr \left\{  \gamma^{*}_{\Tilde{x}_1^n,\Tilde{d}^n_1,u^n,d_2^n}   \theta_{u^n,d^n_2} \right\}\nonumber \\
      &\leq& \frac{1}{|\mathcal{M}_1|} \frac{1}{|\mathcal{D}_1|^n} \frac{1}{|\mathcal{D}_2|^n}  \frac{1}{\Prime^{k_{\sfb} + l_{\sfb}}} \sum_{m_1} \sum_{\Tilde{m}_1}\sum_{s}  \sum_{u^n} \sum_{\Tilde{x}_1^n} \sum_{\Tilde{d}_1^n, d_2^n} p_{X_1}^n(\Tilde{x}_1^n)  p_{U_2 \oplus U_3}^n(u^n) \tr \left\{  \gamma^{*}_{\Tilde{x}_1^n,\Tilde{d}^n_1,u^n,d_2^n}  \CalT_{d_2^n,\eta}\left( \rho_{u^n}^{Y_1} \otimes \ketbra{0}\right) \right\}\nonumber \\
      &+& \frac{1}{|\mathcal{M}_1|} \frac{1}{|\mathcal{D}_1|^n} \frac{1}{|\mathcal{D}_2|^n}  \frac{1}{\Prime^{k_{\sfb} + l_{\sfb}}} \sum_{m_1} \sum_{\Tilde{m}_1}\sum_{s}  \sum_{u^n} \sum_{\Tilde{x}_1^n} \sum_{\Tilde{d}_1^n, d_2^n}  p_{X_1}^n(\Tilde{x}_1^n)  p_{U_2 \oplus U_3}^n(u^n) \nonumber \\
      &&\tr \left\{  \gamma^{*}_{\Tilde{x}_1^n,\Tilde{d}^n_1,u^n,d_2^n}  \CalN_{d_2^n,\eta}\left( \rho_{u^n}^{Y_1} \otimes \ketbra{0} \right) \right\}, \label{Eqn:3to1CQICsecondterminhayashinagaokainequality}
\end{eqnarray}
where the first equality follows from 
\begin{eqnarray}
     \theta_{u^n,d_2^n} \define  \frac{1}{|\CalD_1|^n} \sum_{d_1^n} \sum_{x_1^n} p_{X_1}^n(x_1^n) \theta_{x_1^n,d_1^n,u^n,d_2^n}, \nonumber 
\end{eqnarray}
and the first inequality is obtained from 
\begin{eqnarray}
  \theta_{u^n,d_2^n} &=& \frac{1+\eta^2}{1+2\eta^2} \CalT_{d_2^n,\eta}\left( \rho_{u^n}^{Y_1} \otimes \ketbra{0}\right) + \CalN_{d_2^n,\eta}\left( \rho_{u^n}^{Y_1} \otimes \ketbra{0} \right), \mbox{ where }\nonumber \\
     \CalN_{d_2^n,\eta}\left( \rho_{u^n}^{Y_1} \otimes \ketbra{0} \right) &\define& \theta_{u^n,d_2^n} - \frac{1+\eta^2}{1+2\eta^2} \CalT_{d_2^n,\eta}\left( \rho_{u^n}^{Y_1} \otimes \ketbra{0}\right), \nonumber 
\end{eqnarray}
\noindent as stated in \cite{202103SAD_Sen}. We upper-bound the first term in \eqref{Eqn:3to1CQICsecondterminhayashinagaokainequality} using
\begin{eqnarray}
\gamma^*_{\Tilde{x}_1^n,\Tilde{d}^n_1,u^n,d_2^n} \leq \left( I_{\boldsymbol{\CalH_{Y_1}^{e}}}-  \beta^{*}_{\Tilde{x}_1^n,\Tilde{d}^n_1,u^n,d_2^n}\right) \leq \left( I_{\boldsymbol{\CalH_{Y_1}^{e}}}-  \beta^{1}_{\Tilde{x}_1^n,\Tilde{d}^n_1,u^n,d_2^n}\right), \nonumber     
\end{eqnarray}
we obtain 
 \begin{eqnarray}
 &&\frac{1}{|\mathcal{M}_1|} \frac{1}{|\mathcal{D}_1|^n} \frac{1}{|\mathcal{D}_2|^n}  \frac{1}{\Prime^{k_{\sfb} + l_{\sfb}}} \sum_{m_1} \sum_{\Tilde{m}_1}\sum_{s}  \sum_{u^n} \sum_{\Tilde{x}_1^n} \sum_{\Tilde{d}_1^n, d_2^n}  p_{X_1}^n(\Tilde{x}_1^n)  p_{U_2 \oplus U_3}^n(u^n) \tr \left\{  \gamma^{*}_{\Tilde{x}_1^n,\Tilde{d}^n_1,u^n,d_2^n}  \CalT_{d_2^n,\eta}\left( \rho_{u^n}^{Y_1} \otimes \ketbra{0}\right) \right\}\nonumber \\
   &\leq& \frac{1}{|\mathcal{M}_1|} \frac{1}{|\mathcal{D}_1|^n} \frac{1}{|\mathcal{D}_2|^n}  \frac{1}{\Prime^{k_{\sfb} + l_{\sfb}}} \sum_{m_1} \sum_{\Tilde{m}_1} \sum_{s}  \sum_{u^n} \sum_{\Tilde{x}_1^n} \sum_{\Tilde{d}_1^n, d_2^n}  p_{X_1}^n(\Tilde{x}_1^n)  p_{U_2 \oplus U_3}^n(u^n) \nonumber \\
   &&\tr \left\{  \left( I_{\boldsymbol{\CalH_{Y_1}^{e}}}-  \beta^{1}_{\Tilde{x}_1^n,\Tilde{d}^n_1,u^n,d_2^n}\right)  \CalT_{d_2^n,\eta}\left( \rho_{u^n}^{Y_1} \otimes \ketbra{0}\right) \right\}\nonumber \\
    &=& \frac{1}{|\mathcal{M}_1|} \frac{1}{|\mathcal{D}_1|^n} \frac{1}{|\mathcal{D}_2|^n}  \frac{1}{\Prime^{k_{\sfb} + l_{\sfb}}} \sum_{m_1} \sum_{\Tilde{m}_1} \sum_{s}  \sum_{u^n} \sum_{\Tilde{x}_1^n} \sum_{\Tilde{d}_1^n, d_2^n}  p_{X_1}^n(\Tilde{x}_1^n)  p_{U_2 \oplus U_3}^n(u^n) \nonumber \\
    &&\tr \left\{  \left( I_{\boldsymbol{\CalH_{Y_1}^{e}}}-  \beta^{1}_{\Tilde{x}_1^n,\Tilde{d}^n_1,u^n,d_2^n}\right) \Pi_{\CalT_{d_2^n,\eta}\left(\boldsymbol{\CalH_{Y_1}^G}\right)}  \CalT_{d_2^n,\eta}\left( \rho_{u^n}^{Y_1} \otimes \ketbra{0}\right) \right\}\nonumber \\
         &\overset{(a)}{=}& \frac{1}{|\mathcal{M}_1|} \frac{1}{|\mathcal{D}_1|^n} \frac{1}{|\mathcal{D}_2|^n}  \frac{1}{\Prime^{k_{\sfb} + l_{\sfb}}} \sum_{m_1} \sum_{\Tilde{m}_1} \sum_{s}  \sum_{u^n} \sum_{\Tilde{x}_1^n} \sum_{\Tilde{d}_1^n, d_2^n}  p_{X_1}^n(\Tilde{x}_1^n)  p_{U_2 \oplus U_3}^n(u^n)  \nonumber \\
         &&\tr \left\{  \left( I_{\CalT_{d_2^n,\eta}\left(\boldsymbol{\CalH_{Y_1}^G}\right)}-  \beta^{1}_{\Tilde{x}_1^n,\Tilde{d}^n_1,u^n,d_2^n}\right)   \CalT_{d_2^n,\eta}\left( \rho_{u^n}^{Y_1} \otimes \ketbra{0}\right) \right\}\nonumber \\
      &\overset{(b)}{=}& \frac{1}{|\mathcal{M}_1|} \frac{1}{|\mathcal{D}_1|^n} \frac{1}{|\mathcal{D}_2|^n}  \frac{1}{\Prime^{k_{\sfb} + l_{\sfb}}} \sum_{m_1} \sum_{\Tilde{m}_1} \sum_{s}  \sum_{u^n} \sum_{\Tilde{x}_1^n} \sum_{\Tilde{d}_1^n, d_2^n}  p_{X_1}^n(\Tilde{x}_1^n)  p_{U_2 \oplus U_3}^n(u^n)  \tr \left\{  \left( I-  \olineB^{1}_{\Tilde{x}_1^n,u^n}\right)   \left( \rho_{u^n}^{Y_1} \otimes \ketbra{0}\right) \right\}\nonumber \\
        &\overset{(c)}{=}& \frac{1}{|\mathcal{M}_1|} \frac{1}{|\mathcal{D}_1|^n} \frac{1}{|\mathcal{D}_2|^n}  \frac{1}{\Prime^{k_{\sfb} + l_{\sfb}}} \sum_{m_1} \sum_{\Tilde{m}_1}\sum_{s}  \sum_{u^n} \sum_{\Tilde{x}_1^n} \sum_{\Tilde{d}_1^n, d_2^n}  p_{X_1}^n(\Tilde{x}_1^n)  p_{U_2 \oplus U_3}^n(u^n)  \tr \left\{ G^{1}_{\Tilde{x}_1^n,u^n} \rho_{u^n}^{Y_1} \right\}\nonumber \\
        &\overset{(d)}{=}& \frac{1}{|\mathcal{M}_1|} \frac{1}{|\mathcal{D}_1|^n} \frac{1}{|\mathcal{D}_2|^n}  \frac{1}{\Prime^{k_{\sfb} + l_{\sfb}}} \sum_{m_1} \sum_{\Tilde{m}_1}\sum_{s}  \sum_{(\Tilde{x}_1^n,u^n) \in T_{\delta}^n(p_{X_1} p_{U_2 \oplus U_3})} \sum_{\Tilde{d}_1^n, d_2^n}  p_{X_1}^n(\Tilde{x}_1^n)  p_{U_2 \oplus U_3}^n(u^n)  \tr \left\{ 
 \Pi_{u^n}^{Y_1} \: \Pi^{Y_1}_{\Tilde{x}_1^n,u^n} \: \Pi_{u^n}^{Y_1} \rho_{u^n}^{Y_1} \right\}\nonumber \\
       &\overset{(e)}{=}& \frac{1}{|\mathcal{M}_1|} \frac{1}{|\mathcal{D}_1|^n} \frac{1}{|\mathcal{D}_2|^n}  \frac{1}{\Prime^{k_{\sfb} + l_{\sfb}}} \sum_{m_1} \sum_{\Tilde{m}_1}\sum_{s}  \sum_{(\Tilde{x}_1^n,u^n) \in T_{\delta}^n(p_{X_1} p_{U_2 \oplus U_3})}
       \sum_{\Tilde{d}_1^n, d_2^n}  p_{X_1}^n(\Tilde{x}_1^n)  p_{U_2 \oplus U_3}^n(u^n)  \tr \left\{ 
 \Pi^{Y_1}_{\Tilde{x}_1^n,u^n} \: \Pi_{u^n}^{Y_1} \rho_{u^n}^{Y_1}\: \Pi_{u^n}^{Y_1} \right\}\nonumber \\ &\overset{(f)}{\leq}& 2^{-n \left( I(X_1;Y_1|U_2 \oplus U_3) -2\delta \right)} 2^{nR_1}, \nonumber 
 \end{eqnarray}
 where (a) follows because the support of $ I_{\boldsymbol{\CalH_{Y_1}^{e}}}-  \beta^{1}_{\Tilde{x}_1^n,\Tilde{d}^n_1,u^n,d_2^n}$ is contained in $ \CalT_{d_2^n,\eta}\left(\boldsymbol{\CalH_{Y_1}^G}\right)$, (b) follows because $\CalT_{d_2^n,\eta}$ is an isometry, (c) follows from (i) the definition of $\olineB^1_{\Tilde{x}_1^n,u^n}$ in \eqref{Eqn:3to1CQICcomplementprojector} and (ii) from Gelfand–Naimark’s Thm.~\cite[Thm.~3.7]{BkHolevo_2019}, (d) follows from the definition of $G^1_{\tilde{x}_1^n,u^n}$ in \eqref{Eqn:3to1CQICtheoriginalpovmelements}, (e) follows from the cyclicity of the trace and finally (f) follows from the typicality properties
 \begin{eqnarray}
     \Pi_{u^n}^{Y_1} \: \rho^{Y_1}_{u^n} \: \Pi^{Y_1}_{u^n} &\leq& 2^{-n(H(Y_1|U_2 \oplus U_3) - \delta)}, \mbox{ and } \nonumber \\
     \tr\{\Pi^{Y_1}_{\tilde{x}_1^n,u^n} \: \Pi_{u^n}^{Y_1}\} &\leq&  \tr\{\Pi^{Y_1}_{\tilde{x}_1^n,u^n} \}\nonumber \\ 
     &\leq& 2^{n(H(Y_1|X_1,U_2 \oplus U_3) + \delta )}, \mbox{ for }  (\tilde{x}_1^n,u^n) \in T_{\delta}^n(p_{X_1} p_{U_2 \oplus U_3}). \nonumber
 \end{eqnarray}
 
\noindent Next, we upper-bound the second term in \eqref{Eqn:3to1CQICsecondterminhayashinagaokainequality} using the following inequality
 \begin{eqnarray}
     &&|\tr\left( AB\right)| \leq \norm{AB}_1 \leq \min\{\norm{A}_1 \norm{B}_{\infty}, \norm{A}_{\infty} \norm{B}_{1}\}. \label{Eqn:Traceinequality} 
 \end{eqnarray}
We obtain
\begin{eqnarray}
&&\frac{1}{|\mathcal{M}_1|} \frac{1}{|\mathcal{D}_1|^n} \frac{1}{|\mathcal{D}_2|^n}  \frac{1}{\Prime^{k_{\sfb} + l_{\sfb}}} \sum_{m_1} \sum_{\Tilde{m}_1} \sum_{s}  \sum_{u^n} \sum_{\Tilde{x}_1^n} \sum_{\Tilde{d}_1^n, d_2^n}  p_{X_1}^n(\Tilde{x}_1^n)  p_{U_2 \oplus U_3}^n(u^n) \tr \left\{  \gamma^{*}_{\Tilde{x}_1^n,\Tilde{d}^n_1,u^n,d_2^n}  \CalN_{d_2^n,\eta}\left( \rho_{u^n}^{Y_1} \otimes \ketbra{0} \right) \right\} \nonumber \\ 
&\leq&\frac{1}{|\mathcal{M}_1|} \frac{1}{|\mathcal{D}_1|^n} \frac{1}{|\mathcal{D}_2|^n}  \frac{1}{\Prime^{k_{\sfb} + l_{\sfb}}} \sum_{m_1} \sum_{\Tilde{m}_1} \sum_{s}  \sum_{u^n} \sum_{\Tilde{x}_1^n} \sum_{\Tilde{d}_1^n, d_2^n}  p_{X_1}^n(\Tilde{x}_1^n)  p_{U_2 \oplus U_3}^n(u^n) \norm{\gamma^{*}_{\Tilde{x}_1^n,\Tilde{d}^n_1,u^n,d_2^n}}_1  \norm{\CalN_{d_2^n,\eta}\left( \rho_{u^n}^{Y_1} \otimes \ketbra{0} \right)}_{\infty} \nonumber \\ 
&\leq&  2^n |\CalH_{Y_1}|^n \: \frac{3 \eta }{\sqrt{|\CalD_1|^n}} 2^{nR_1}\nonumber,
\end{eqnarray}
where the last inequality is obtained from (i)
\begin{eqnarray}
\norm{\gamma^{*}_{\Tilde{x}_1^n,\Tilde{d}^n_1,u^n,d_2^n}}_1 \leq \norm{I_{\boldsymbol{\CalH_{Y_1}^{e}}}- \beta^{*}_{\Tilde{x}_1^n,\Tilde{d}^n_1,u^n,d_2^n}}_{\infty}^2 \norm{\Pi_{\boldsymbol{\CalH_{Y_1}^G}}}_1 \leq 2^n |\CalH_{Y_1}|^n, \label{Eqn:normgamma*}  
\end{eqnarray}
and from (ii) the inequality stated in \cite{202103SAD_Sen}
\begin{eqnarray}
     \norm{\CalN_{d_2^n,\eta}\left( \rho_{u^n}^{Y_1} \otimes \ketbra{0} \right)}_{\infty} \leq \frac{3 \eta }{\sqrt{|\CalD_2|^n}} \nonumber.
\end{eqnarray}

\noindent If we choose $|\CalD_1|\geq \left( 3 \eta 2^n |\CalH_{Y_1}|^n 2^{n \left( I(X_1;Y_1|U_2 \oplus U_3) - 2 \delta \right)}\right)^{\frac{2}{n}}$, then we have 
\begin{eqnarray}
    \mathbb{E}\left[t_{2.1.2}\right] &\leq& 2^{nR_1} \left( 2^{-n \left( I(X_1;Y_1|U_2 \oplus U_3) -2\delta \right)}+ 2^n |\CalH_{Y_1}|^n \: \frac{3 \eta}{\sqrt{|\CalD_1|^n}}\right) \nonumber \\
    &\leq& 2 \: 2^{-n \left( I(X_1;Y_1|U_2 \oplus U_3) -2\delta-R_1 \right)}. \nonumber 
\end{eqnarray}
Hence, for $n$ sufficiently large, we have $\mathbb{E}[t_{2.1.2}] \leq \epsilon,$ if
\begin{eqnarray}
R_1 < I(X_1;Y_1|U_2 \oplus U_3). \nonumber
\end{eqnarray}

\subsubsection*{Analysis of $t_{2.1.3}$}
We evaluate the expectation of $t_{2.1.3}$ over the codebook generation distribution. We get

\begin{eqnarray}
\mathbb{E}\left[t_{2.1.3}\right] &=& \frac{1}{|\mathcal{M}_1|} \frac{1}{|\mathcal{D}_1|^n} \frac{1}{|\CalD_2|^{2n}}\frac{1}{\Prime^{k_{\sfb} + l_{\sfb}}} \sum_{m_1} \sum_{s}\sum_{\Tilde{s} \neq s}   \sum_{x_1^n} \sum_{d_1^n, d_2^n} \sum_{u^n,\Tilde{u}^n} \sum_{\Tilde{d}_2^n} 
     p_{U_2 \oplus U_3}^n(u^n) p_{X_1}^n(x_1^n) q_{U_2 \oplus U_3}^n(\Tilde{u}^n) \nonumber \\
     &&\tr \left\{  \gamma^{*}_{x_1^n,d^n_1,\Tilde{u}^n,\Tilde{d}_2^n}   \theta_{x_1^n,d^n_1,u^n,d^n_2} \right\}\nonumber \\
&=& \frac{1}{|\mathcal{M}_1|} \frac{1}{|\mathcal{D}_1|^n} \frac{1}{|\mathcal{D}_2|^n} \frac{1}{\Prime^{k_{\sfb} + l_{\sfb}}} \sum_{m_1} \sum_{s}\sum_{\Tilde{s} \neq s}   \sum_{x_1^n} \sum_{d_1^n, \Tilde{d}_2^n} \sum_{\Tilde{u}^n}  
      p_{X_1}^n(x_1^n) q_{U_2 \oplus U_3}^n(\Tilde{u}^n) \tr \left\{\gamma^{*}_{x_1^n,d^n_1,\Tilde{u}^n,\Tilde{d}_2^n}    \theta_{x_1^n,d^n_1} \right\}\nonumber \\
&\leq& \frac{1}{|\mathcal{M}_1|} \frac{1}{|\mathcal{D}_1|^n} \frac{1}{|\mathcal{D}_2|^n} \frac{1}{\Prime^{k_{\sfb} + l_{\sfb}}} \sum_{m_1} \sum_{s}\sum_{\Tilde{s}}   \sum_{x_1^n} \sum_{d_1^n, \Tilde{d}_2^n} \sum_{\Tilde{u}^n}  
      p_{X_1}^n(x_1^n) q_{U_2 \oplus U_3}^n(\Tilde{u}^n) \tr \left\{  \gamma^{*}_{x_1^n,d^n_1,\Tilde{u}^n,\Tilde{d}_2^n}    \CalT_{d_1^n,\eta}\left(\rho^{Y_1}_{x_1^n} \otimes \ketbra{0} \right) \right\}\nonumber \\
&+& \frac{1}{|\mathcal{M}_1|} \frac{1}{|\mathcal{D}_1|^n} \frac{1}{|\mathcal{D}_2|^n} \frac{1}{\Prime^{k_{\sfb} + l_{\sfb}}} \sum_{m_1} \sum_{s}\sum_{\Tilde{s}}   \sum_{x_1^n} \sum_{d_1^n, \Tilde{d}_2^n} \sum_{\Tilde{u}^n}  
      p_{X_1}^n(x_1^n) q_{U_2 \oplus U_3}^n(\Tilde{u}^n) \tr \left\{  \gamma^{*}_{x_1^n,d^n_1,\Tilde{u}^n,\Tilde{d}_2^n}    \CalN_{d_1^n,\eta}\left(\rho^{Y_1}_{x_1^n} \otimes \ketbra{0} \right) \right\}\!, \label{Eqn:3to1CQICthirdterminhayashinagaokainequality}
\end{eqnarray}
where the second equality follows from 
\begin{eqnarray}
    \theta_{x_1^n,d_1^n} \define \frac{1}{|\CalD_2|^n} \sum_{d_2^n} \sum_{u^n} p_{U_2 \oplus U_3}^n(u^n)  \theta_{x_1^n,d_1^n,u^n,d_2^n}, \nonumber 
\end{eqnarray}
and the first inequality follows from 
\begin{eqnarray}
    \theta_{x_1^n,d_1^n} &=& \frac{1 + \eta^2}{1+2\eta^2} \CalT_{d_1^n,\eta} \left( \rho_{x_1^n}^{Y_1} \otimes \ketbra{0} \right) + \CalN_{d_1^n,\eta}\left( \rho_{x_1^n}^{Y_1} \otimes \ketbra{0} \right), \mbox{ where } \nonumber \\
      \CalN_{d_1^n,\eta}\left( \rho_{x_1^n}^{Y_1} \otimes \ketbra{0} \right) &\define& \theta_{x_1^n,d_1^n} - \frac{1 + \eta^2}{1+2\eta^2} \CalT_{d_1^n,\eta} \left( \rho_{x_1^n}^{Y_1} \otimes \ketbra{0} \right),\nonumber 
\end{eqnarray}
as stated in \cite{202103SAD_Sen}. Consider the first term in \eqref{Eqn:3to1CQICthirdterminhayashinagaokainequality}, we have 
\begin{eqnarray}
    &&\frac{1}{|\mathcal{M}_1|} \frac{1}{|\mathcal{D}_1|^n} \frac{1}{|\mathcal{D}_2|^n} \frac{1}{\Prime^{k_{\sfb} + l_{\sfb}}} \sum_{m_1} \sum_{s,\Tilde{s}}   \sum_{x_1^n} \sum_{d_1^n, \Tilde{d}_2^n} \sum_{\Tilde{u}^n}  
      p_{X_1}^n(x_1^n) q_{U_2 \oplus U_3}^n(\Tilde{u}^n) \tr \left\{  \gamma^{*}_{x_1^n,d^n_1,\Tilde{u}^n,\Tilde{d}_2^n}    \CalT_{d_1^n,\eta}\left(\rho^{Y_1}_{x_1^n} \otimes \ketbra{0} \right) \right\}\nonumber \\
      &\overset{(a)}{\leq}& \frac{1}{|\mathcal{M}_1|} \frac{1}{|\mathcal{D}_1|^n} \frac{1}{|\mathcal{D}_2|^n} \frac{1}{\Prime^{k_{\sfb} + l_{\sfb}}} \sum_{m_1} \sum_{s,\Tilde{s}}   \sum_{x_1^n} \sum_{d_1^n, \Tilde{d}_2^n} \sum_{\Tilde{u}^n}  
      p_{X_1}^n(x_1^n) q_{U_2 \oplus U_3}^n(\Tilde{u}^n)  \tr \left\{  
\left( I_{\boldsymbol{\CalH_{Y_1}^{e}}}-  \beta^{2}_{x_1^n,d^n_1,\Tilde{u}^n,\Tilde{d}_2^n}\right)    \CalT_{d_1^n,\eta}\left(\rho^{Y_1}_{x_1^n} \otimes \ketbra{0} \right) \right\}\nonumber \\
&=& \frac{1}{|\mathcal{M}_1|} \frac{1}{|\mathcal{D}_1|^n} \frac{1}{|\mathcal{D}_2|^n} \frac{1}{\Prime^{k_{\sfb} + l_{\sfb}}} \sum_{m_1} \sum_{s,\Tilde{s}}  \sum_{x_1^n} \sum_{d_1^n, \Tilde{d}_2^n} \sum_{\Tilde{u}^n} 
p_{X_1}^n(x_1^n) q_{U_2 \oplus U_3}^n(\Tilde{u}^n) \nonumber \\ 
&&\tr \left\{  
\left( I_{\boldsymbol{\CalH_{Y_1}^{e}}}-  \beta^{2}_{x_1^n,d^n_1,\Tilde{u}^n,\Tilde{d}_2^n}\right) 
\Pi_{\CalT_{d_1^n,\eta}\left(\boldsymbol{\CalH_{Y_1}^G}\right)}    \CalT_{d_1^n,\eta}\left(\rho^{Y_1}_{x_1^n} \otimes \ketbra{0} \right) \right\}\nonumber \\
&\overset{(b)}{=}&\frac{1}{|\mathcal{M}_1|} \frac{1}{|\mathcal{D}_1|^n} \frac{1}{|\mathcal{D}_2|^n} \frac{1}{\Prime^{k_{\sfb} + l_{\sfb}}} \sum_{m_1} \sum_{s,\Tilde{s}}   \sum_{x_1^n} \sum_{d_1^n, \Tilde{d}_2^n} \sum_{\Tilde{u}^n}  
      p_{X_1}^n(x_1^n) q_{U_2 \oplus U_3}^n(\Tilde{u}^n) \nonumber \\ 
      && \tr \left\{\left(I_{\CalT_{d_1^n,\eta}\left(\boldsymbol{\CalH_{Y_1}^G}\right)}-  \beta^{2}_{x_1^n,d^n_1,\Tilde{u}^n,\Tilde{d}_2^n}\right) 
    \CalT_{d_1^n,\eta}\left(\rho^{Y_1}_{x_1^n} \otimes \ketbra{0} \right) \right\}\nonumber \\
    &\overset{(c)}{=}& \frac{1}{|\mathcal{M}_1|} \frac{1}{|\mathcal{D}_1|^n} \frac{1}{|\mathcal{D}_2|^n} \frac{1}{\Prime^{k_{\sfb} + l_{\sfb}}} \sum_{m_1} \sum_{s,\Tilde{s}}   \sum_{x_1^n} \sum_{d_1^n, \Tilde{d}_2^n} \sum_{\Tilde{u}^n}  
      p_{X_1}^n(x_1^n) q_{U_2 \oplus U_3}^n(\Tilde{u}^n) \tr \left\{  \left( I-  \olineB^{2}_{x_1^n,\Tilde{u}^n}\right)
   \left(\rho^{Y_1}_{x_1^n} \otimes \ketbra{0} \right) \right\}\nonumber \\
    &\overset{(d)}{=}& \frac{1}{|\mathcal{M}_1|} \frac{1}{|\mathcal{D}_1|^n} \frac{1}{|\mathcal{D}_2|^n} \frac{1}{\Prime^{k_{\sfb} + l_{\sfb}}} \sum_{m_1} \sum_{s,\Tilde{s}}  \sum_{x_1^n} \sum_{d_1^n, \Tilde{d}_2^n} \sum_{\Tilde{u}^n} 
      p_{X_1}^n(x_1^n) q_{U_2 \oplus U_3}^n(\Tilde{u}^n) \tr \left\{  G^{2}_{x_1^n,\Tilde{u}^n} \rho_{x_1^n}^{Y_1} \right\}\nonumber \\
      &\overset{(e)}{=}& \frac{1}{|\mathcal{M}_1|} \frac{1}{|\mathcal{D}_1|^n} \frac{1}{|\mathcal{D}_2|^n} \frac{1}{\Prime^{k_{\sfb} + l_{\sfb}}} \sum_{m_1} \sum_{s,\Tilde{s}} \sum_{(x_1^n,\Tilde{u}^n) \in T_{\delta}^n(p_{X_1} p_{U_2 \oplus U_3}) } \sum_{d_1^n, \Tilde{d}_2^n}   
      p_{X_1}^n(x_1^n) q_{U_2 \oplus U_3}^n(\Tilde{u}^n) \tr \left\{   \Pi^{Y_1}_{x_1^n} \: \Pi^{Y_1}_{x_1^n,\Tilde{u}^n} \:  \Pi^{Y_1}_{x_1^n} \: \rho_{x_1^n}^{Y_1} \right\}\nonumber \\
      &\overset{(f)}{\leq}& \frac{1}{|\mathcal{M}_1|} \frac{1}{|\mathcal{D}_1|^n} \frac{1}{|\mathcal{D}_2|^n} \frac{1}{\Prime^{k_{\sfb} + l_{\sfb}}} \sum_{m_1} \sum_{s,\Tilde{s}}  \sum_{(x_1^n,\Tilde{u}^n) \in T_{\delta}^n(p_{X_1} p_{U_2 \oplus U_3})} \sum_{d_1^n, \Tilde{d}_2^n}   
      p_{X_1}^n(x_1^n) 2^{-n \left( D(p_{U_2 \oplus U_3} ||q_{U_2 \oplus U_3}) - \tilde{\delta} \right)}p_{U_2 \oplus U_3}^n(\Tilde{u}^n) \nonumber \\ 
      &&\tr \left\{  \Pi^{Y_1}_{x_1^n,\Tilde{u}^n} \: \Pi^{Y_1}_{x_1^n} \:\rho_{x_1^n}^{Y_1} \: \Pi^{Y_1}_{x_1^n} \right\}\nonumber \\
      &\overset{(g)}{\leq}&  2^{-n\left( I(U_2 \oplus U_3;Y_1|X_1)- 2 \delta \right)}  \: 2^{-n \left( D(p_{U_2 \oplus U_3} ||q_{U_2 \oplus U_3}) - \tilde{\delta} - \frac{(k_{\sfb}+l_{\sfb})}{n}\log(\Prime)\right)} \nonumber, 
\end{eqnarray}
where (a) follows from applying the following bound 
\begin{eqnarray}
    && \gamma^*_{\Tilde{x}_1^n,\Tilde{d}^n_1,u^n,d_2^n} \leq \left( I_{\boldsymbol{\CalH_{Y_1}^{e}}}-  \beta^{*}_{\Tilde{x}_1^n,\Tilde{d}^n_1,u^n,d_2^n}\right) \leq \left( I_{\boldsymbol{\CalH_{Y_1}^{e}}}-  \beta^{2}_{\Tilde{x}_1^n,\Tilde{d}^n_1,u^n,d_2^n}\right)\nonumber, 
\end{eqnarray}
(b) follows because the support of $I_{\boldsymbol{\CalH_{Y_1}^{e}}}-  \beta^{2}_{\Tilde{x}_1^n,\Tilde{d}^n_1,u^n,d_2^n}$ is contained in  $\CalT_{d_1^n,\eta}\left(\boldsymbol{\CalH_{Y_1}^G}\right)$, (c) follows because $\CalT_{d_1^n,\eta}$ is an isometry, (d) follows from (i) the definition of $\olineB^2_{x_1^n,\Tilde{u}^n}$ in \eqref{Eqn:3to1CQICcomplementprojector} and (ii) from Gelfand–Naimark’s Thm.~\cite[Thm.~3.7]{BkHolevo_2019}, (e) follows from (i) the definition of $G^2_{x_1^n, \tilde{u}^n}$ in \eqref{Eqn:3to1CQICtheoriginalpovmelements} and from (ii) $G^2_{x_1^n, \tilde{u}^n} = 0$ if $(x_1^n, \tilde{u}^n) \notin T_{\delta}^n(p_{X_1} p_{U_2 \oplus U_3})$, 
(f) follows from (i) the cyclicity of the trace and from (ii) 
\begin{eqnarray}
    q_{U_2 \oplus U_3}^n(\Tilde{u}^n) \leq 2^{-n\left( D(p_{U_2 \oplus U_3} || q_{U_2 \oplus U_3}) - \tilde{\delta} \right)}p_{U_2 \oplus U_3}^n(\Tilde{u}^n), \mbox{ for } \Tilde{u}^n \in T_{\delta}^n(p_{U_2 \oplus U_3}), \label{Eqn:3to1CQICdivergenceproperty}
\end{eqnarray}
and finally (g) follows from the typicality properties 
\begin{eqnarray}
    \Pi^{Y_1}_{x_1^n} \:\rho_{x_1^n}^{Y_1} \: \Pi^{Y_1}_{x_1^n}  &\leq& 2^{-n(H(Y_1|X_1) - \delta)}, \mbox{ and } \nonumber \\
    \tr \left\{  \Pi^{Y_1}_{x_1^n,\Tilde{u}^n} \: \Pi^{Y_1}_{x_1^n} \right\} &\leq&  \tr \left\{  \Pi^{Y_1}_{x_1^n,\Tilde{u}^n}\right\} \nonumber \\
    &\leq& 2^{n(H(Y_1|X_1,U_2 \oplus U_3) + \delta )},\mbox{ for } (x_1^n,\tilde{u}^n) \in T_{\delta}^n(p_{X_1} p_{U_2 \oplus U_3}^n). \nonumber 
\end{eqnarray}

Next, consider the second term in \eqref{Eqn:3to1CQICthirdterminhayashinagaokainequality}, 
\begin{eqnarray}
    &&\frac{1}{|\mathcal{M}_1|} \frac{1}{|\mathcal{D}_1|^n} \frac{1}{|\mathcal{D}_2|^n} \frac{1}{\Prime^{k_{\sfb} + l_{\sfb}}} \sum_{m_1} \sum_{s,\Tilde{s}}   \sum_{x_1^n} \sum_{d_1^n, \Tilde{d}_2^n} \sum_{\Tilde{u}^n} 
      p_{X_1}^n(x_1^n) q_{U_2 \oplus U_3}^n(\Tilde{u}^n) \tr \left\{  \gamma^{*}_{x_1^n,d^n_1,\Tilde{u}^n,\Tilde{d}_2^n}    \CalN_{d_1^n,\eta}\left(\rho^{Y_1}_{x_1^n} \otimes \ketbra{0} \right) \right\}\nonumber \\
      &\overset{(a)}{=}& \frac{1}{|\mathcal{M}_1|} \frac{1}{|\mathcal{D}_1|^n} \frac{1}{|\mathcal{D}_2|^n} \frac{1}{\Prime^{k_{\sfb} + l_{\sfb}}} \sum_{m_1} \sum_{s,\Tilde{s}}   \sum_{x_1^n} \sum_{d_1^n, \Tilde{d}_2^n} \sum_{\Tilde{u}^n \in T_{\delta}^n(p_{U_2 \oplus U_3})}  
      p_{X_1}^n(x_1^n) q_{U_2 \oplus U_3}^n(\Tilde{u}^n) \tr \left\{  \gamma^{*}_{x_1^n,d^n_1,\Tilde{u}^n,\Tilde{d}_2^n}    \CalN_{d_1^n,\eta}\left(\rho^{Y_1}_{x_1^n} \otimes \ketbra{0} \right) \right\}\nonumber \\
      &\overset{(b)}{\leq}& \frac{1}{|\mathcal{M}_1|} \frac{1}{|\mathcal{D}_1|^n} \frac{1}{|\mathcal{D}_2|^n} \frac{1}{\Prime^{k_{\sfb} + l_{\sfb}}} \sum_{m_1} \sum_{s,\Tilde{s}}  \sum_{x_1^n} \sum_{d_1^n, \Tilde{d}_2^n} \sum_{\Tilde{u}^n \in T_{\delta}^n(p_{U_2 \oplus U_3})}  
      p_{X_1}^n(x_1^n) 2^{-n \left( D(p_{U_2 \oplus U_3}||q_{U_2 \oplus U_3}) - \tilde{\delta}\right)}p_{U_2 \oplus U_3}^n(\Tilde{u}^n) \nonumber \\
      &&\tr \left\{  \gamma^{*}_{x_1^n,d^n_1,\Tilde{u}^n,\Tilde{d}_2^n}    \CalN_{d_1^n,\eta}\left(\rho^{Y_1}_{x_1^n} \otimes \ketbra{0} \right) \right\} \nonumber \\
      &\overset{(c)}{\leq}&  \frac{1}{|\mathcal{M}_1|} \frac{1}{|\mathcal{D}_1|^n} \frac{1}{|\mathcal{D}_2|^n} \frac{1}{\Prime^{k_{\sfb} + l_{\sfb}}} \sum_{m_1} \sum_{s,\Tilde{s}}  \sum_{x_1^n} \sum_{d_1^n, \Tilde{d}_2^n} \sum_{\Tilde{u}^n \in T_{\delta}^n(p_{U_2 \oplus U_3})}  
      p_{X_1}^n(x_1^n) 2^{-n \left( D(p_{U_2 \oplus U_3}||q_{U_2 \oplus U_3}) - \tilde{\delta}\right)} p_{U_2 \oplus U_3}^n(\Tilde{u}^n) \nonumber \\
      && \norm{ \gamma^{*}_{x_1^n,d^n_1,\Tilde{u}^n,\Tilde{d}_2^n}   }_1 \norm{\CalN_{d_1^n,\eta}\left(\rho^{Y_1}_{x_1^n} \otimes \ketbra{0} \right)}_{\infty} \nonumber \\
      &\overset{(d)}{\leq}&  2^n |\CalH_{Y_1}|^n \: \frac{3 \eta }{\sqrt{|\CalD_2|^n}}
 \: 2^{-n \left( D(p_{U_2 \oplus U_3} ||q_{U_2 \oplus U_3}) - \tilde{\delta} - \frac{(k_{\sfb}+l_{\sfb})}{n}\log(\Prime)\right)},
 \nonumber 
\end{eqnarray}
where (a) follows because $\gamma^*_{x_1^n,d_1^n,u^n,d_2^n}=0$, if $u^n \notin T_{\delta}^n(p_{U_2 \oplus U_3})$,
(b) follows from the inequality in \eqref{Eqn:3to1CQICdivergenceproperty}, (c) follows from the trace inequality in \eqref{Eqn:Traceinequality} and (d) follows from \eqref{Eqn:normgamma*} and from 
\begin{eqnarray}
    \norm{\CalN_{d_1^n,\eta}\left( \rho_{x_1^n}^{Y_1} \otimes \ketbra{0} \right)}_{\infty} \leq \frac{3 \eta }{\sqrt{|\CalD_1|^n}}, \nonumber
\end{eqnarray}
as stated in \cite{202103SAD_Sen}. Hence, if we choose $|\CalD_2| \geq \left( 3 \eta 2^n |\CalH_{Y_1}|^n 2^{-n \left( I(U_2 \oplus U_3;Y_1|X_1) - 2 \delta \right)}\right)$, we have 
\begin{eqnarray}
\mathbb{E}\left[ t_{2.1.3}\right] &\leq& 2^{-n \left( D(p_{U_2 \oplus U_3}||q_{U_2 \oplus U_3}) - \tilde{\delta} - \frac{(k_{\sfb}+l_{\sfb})}{n} \log(\Prime)\right) } \left( 2^{-n\left( I(U_2 \oplus U_3;Y_1|X_1)- 2 \delta \right)} + 2^n |\CalH_{Y_1}|^n \: \frac{3 \eta}{\sqrt{|\CalD_2|^n}}\right) \nonumber \\ 
&\leq& 2 \: 2^{-n \left( I(U_2 \oplus U_3;Y_1|X_1) + D(p_{U_2 \oplus U_3}||q_{U_2 \oplus U_3}) - \tilde{\delta} - 2 \delta  - \frac{(k_{\sfb}+l_{\sfb})}{n} \log(\Prime)\right) } \nonumber.
\end{eqnarray}
Therefore, we conclude that, for $n$ sufficiently large, we have $\mathbb{E}[t_{2.1.3}] \leq \epsilon,$ if
\begin{eqnarray}
\frac{(k_{\sfb}+l_{\sfb})}{n} \log(\Prime) &<& I(U_2 \oplus U_3;Y_1|X_1) + D(p_{U_2 \oplus U_3}||q_{U_2 \oplus U_3}) \nonumber \\
&=& I(U_2 \oplus U_3;Y_1|X_1) + \log(\Prime) - H(U_2 \oplus U_3). \nonumber     
\end{eqnarray}

%The analysis of $t_{2.1.4}$ follows the same steps as $t_{2.1.3}$. Therefore,
%\begin{eqnarray}
 %\mathbb{E}[t_{2.1.4}] \leq \epsilon \quad \mbox{if} \quad \frac{(k_{\sfb}+l_{\sfb})}{\Prime} < I(U;Y_1|X_1) + D(p_U||q_U) = I(U;Y_1|X_1) + \log(\Prime) -H(U). \nonumber   
%\end{eqnarray}

%The analysis of $t_{2.1.5}$ and $t_{2.1.6}$ are identical. Hence, we describe only one of them.

\subsubsection*{Analysis of $t_{2.1.4}$} We evaluate the expectation of $t_{2.1.4}$ over the codebook generation distribution. We obtain 
\begin{eqnarray}
\mathbb{E}\left[t_{2.1.4}\right] &=& \frac{1}{|\mathcal{M}_1|} \frac{1}{\Prime^{k_{\sfb} + l_{\sfb}}} \frac{1}{|\mathcal{D}_1|^{2n}} \frac{1}{|\mathcal{D}_2|^{2n}} \sum_{m_1} \sum_{\Tilde{m}_1 \neq m_1} \sum_{s} \sum_{\Tilde{s} \neq s}    \sum_{\Tilde{x}_1^n} \sum_{\Tilde{d}_1^n, \Tilde{d}_2^n} \sum_{\Tilde{u}^n} \sum_{x_1^n} \sum_{d_1^n,d_2^n} \sum_{u^n}  p_{X_1}^n(\Tilde{x}_1^n) q_{U_2 \oplus U_3}^n(\Tilde{u}^n) \nonumber \\ 
&&p_{X_1}^n(x_1^n) p_{U_2 \oplus U_3}^n(u^n) \tr \left\{ \gamma^{*}_{\Tilde{x}_1^n,\Tilde{d}_1^n,\Tilde{u}^n,\Tilde{d}_2^n} \theta_{x_1^n,d_1^n,u^n,d_2^n}\right\} \nonumber \\
&=&  \frac{1}{|\mathcal{M}_1|} \frac{1}{\Prime^{k_{\sfb} + l_{\sfb}}} \frac{1}{|\mathcal{D}_1|^n} \frac{1}{|\mathcal{D}_2|^n} \sum_{m_1} \sum_{\Tilde{m}_1 \neq m_1} \sum_{s} \sum_{\Tilde{s} \neq s} \sum_{\Tilde{x}_1^n} \sum_{\Tilde{d}_1^n, \Tilde{d}_2^n} \sum_{\Tilde{u}^n}   p_{X_1}^n(\Tilde{x}_1^n) q_{U_2 \oplus U_3}^n(\Tilde{u}^n)  \tr\left\{ \gamma^*_{\Tilde{x}_1^n,\Tilde{d}_1^n,\Tilde{u}^n,\Tilde{d}_2^n} \theta^{\otimes n }\right\} \nonumber \\
&\leq&\frac{1}{|\mathcal{M}_1|} \frac{1}{\Prime^{k_{\sfb} + l_{\sfb}}} \frac{1}{|\mathcal{D}_1|^n} \frac{1}{|\mathcal{D}_2|^n} \sum_{m_1} \sum_{\Tilde{m}_1} \sum_{s} \sum_{\Tilde{s}}
    \sum_{\Tilde{x}_1^n} \sum_{\Tilde{d}_1^n, \Tilde{d}_2^n} \sum_{\Tilde{u}^n}   p_{X_1}^n(\Tilde{x}_1^n) q_{U_2 \oplus U_3}^n(\Tilde{u}^n)  \nonumber \\
    &&\tr\left\{ \gamma^*_{\Tilde{x}_1^n,\Tilde{d}_1^n,\Tilde{u}^n,\Tilde{d}_2^n} \left( \left( \rho^{Y_1}\right)^{\otimes n } \otimes \ketbra{0} \right)\right\} + \frac{1}{|\mathcal{M}_1|} \frac{1}{\Prime^{k_{\sfb} + l_{\sfb}}} \frac{1}{|\mathcal{D}_1|^n} \frac{1}{|\mathcal{D}_2|^n} \sum_{m_1} \sum_{\Tilde{m}_1} \sum_{s} \sum_{\Tilde{s}}
    \sum_{\Tilde{x}_1^n} \sum_{\Tilde{d}_1^n, \Tilde{d}_2^n} \sum_{\Tilde{u}^n}     \nonumber \\ 
    && p_{X_1}^n(\Tilde{x}_1^n) q_{U_2 \oplus U_3}^n(\Tilde{u}^n) \tr\left\{ \gamma^*_{\Tilde{x}_1^n,\Tilde{d}_1^n,\Tilde{u}^n,\Tilde{d}_2^n} \left( \CalN_{\eta}\left( \rho^{Y_1}\right)^{\otimes n } \otimes \ketbra{0} \right)\right\}, \label{Eqn:3to1CQIClastterminhayashinagaokainequality}
\end{eqnarray}
where the second equality follows from
 \begin{eqnarray}
     \theta^{\otimes n} \define \frac{1}{|\mathcal{D}_1|^n}\frac{1}{|\mathcal{D}_2|^n} \sum_{d_1^n}\sum_{d_2^n} \sum_{x_1^n} \sum_{u^n} p_{X_1}^n(x_1^n) p_{U_2 \oplus U_3}^n(u^n)  \theta_{x_1^n,d_1^n,u^n,d_2^n},\nonumber 
 \end{eqnarray}
and the first inequality follows from
 \begin{eqnarray}
    \theta^{\otimes n} &=& \frac{1}{1+2\eta^2} \left( \left( \rho^{Y_1}\right)^{\otimes n} \otimes \ketbra{0}\right) + \CalN_{\eta}\left( \left( \rho^{Y_1}\right)^{\otimes n} \otimes \ketbra{0}\right), \mbox{ where }\nonumber \\
      \CalN_{\eta}\left( \left( \rho^{Y_1}\right)^{\otimes n} \otimes \ketbra{0}\right) &\define& \theta^{\otimes n} - \frac{1}{1+2\eta^2} \left( \left( \rho^{Y_1}\right)^{\otimes n} \otimes \ketbra{0}\right),\nonumber 
 \end{eqnarray}
as stated in \cite{202103SAD_Sen}. Using 
\begin{eqnarray}
\gamma^*_{\Tilde{x}_1^n,\Tilde{d}_1^n,\Tilde{u}^n,\Tilde{d}_2^n} \leq \left( I_{\boldsymbol{\CalH_{Y_1}^{e}}}-  \beta^{*}_{\Tilde{x}_1^n,\Tilde{d}_1^n,\Tilde{u}^n,\Tilde{d}_2^n}\right) \leq \left( I_{\boldsymbol{\CalH_{Y_1}^{e}}}-  \beta^3_{\Tilde{x}_1^n,\Tilde{d}_1^n,\Tilde{u}^n,\Tilde{d}_2^n}\right), \nonumber 
\end{eqnarray}
we bound the first term in \eqref{Eqn:3to1CQIClastterminhayashinagaokainequality}, as follows
\begin{eqnarray}
    &&\frac{1}{|\mathcal{M}_1|} \frac{1}{\Prime^{k_{\sfb} + l_{\sfb}}} \frac{1}{|\mathcal{D}_1|^n} \frac{1}{|\mathcal{D}_2|^n} \sum_{m_1} \sum_{\Tilde{m}_1} \sum_{s,\Tilde{s}}   \sum_{\Tilde{x}_1^n} \sum_{\Tilde{d}_1^n, \Tilde{d}_2^n} \sum_{\Tilde{u}^n}   p_{X_1}^n(\Tilde{x}_1^n) q_{U_2 \oplus U_3}^n(\Tilde{u}^n)  \tr\left\{ \gamma^*_{\Tilde{x}_1^n,\Tilde{d}_1^n,\Tilde{u}^n,\Tilde{d}_2^n} \left( \left( \rho^{Y_1}\right)^{\otimes n } \otimes \ketbra{0} \right)\right\} \nonumber \\
    &\leq&  \frac{1}{|\mathcal{M}_1|} \frac{1}{\Prime^{k_{\sfb} + l_{\sfb}}} \frac{1}{|\mathcal{D}_1|^n} \frac{1}{|\mathcal{D}_2|^n} \sum_{m_1} \sum_{\Tilde{m}_1} \sum_{s,\Tilde{s}}   \sum_{\Tilde{x}_1^n} \sum_{\Tilde{d}_1^n, \Tilde{d}_2^n} \sum_{\Tilde{u}^n}  p_{X_1}^n(\Tilde{x}_1^n) q_{U_2 \oplus U_3}^n(\Tilde{u}^n)  \nonumber \\ 
    &&\tr\left\{\left( I_{\boldsymbol{\CalH_{Y_1}^{e}}}-  \beta^3_{\Tilde{x}_1^n,\Tilde{d}_1^n,\Tilde{u}^n,\Tilde{d}_2^n}\right)  \left( \left( \rho^{Y_1}\right)^{\otimes n } \otimes \ketbra{0} \right)\right\} \nonumber \\
    &=& \frac{1}{|\mathcal{M}_1|} \frac{1}{\Prime^{k_{\sfb} + l_{\sfb}}} \frac{1}{|\mathcal{D}_1|^n} \frac{1}{|\mathcal{D}_2|^n} \sum_{m_1} \sum_{\Tilde{m}_1} \sum_{s,\Tilde{s}}   \sum_{\Tilde{x}_1^n} \sum_{\Tilde{d}_1^n, \Tilde{d}_2^n} \sum_{\Tilde{u}^n}  p_{X_1}^n(\Tilde{x}_1^n) q_{U_2 \oplus U_3}^n(\Tilde{u}^n)  \nonumber \\ 
    &&\tr\left\{\left( I_{\boldsymbol{\CalH_{Y_1}^{e}}}-  \beta^3_{\Tilde{x}_1^n,\Tilde{d}_1^n,\Tilde{u}^n,\Tilde{d}_2^n}\right)  \Pi_{\boldsymbol{\CalH_{Y_1}^G}}\left( \left( \rho^{Y_1}\right)^{\otimes n } \otimes \ketbra{0} \right)\right\} \nonumber \\
&\overset{(a)}{=}& \frac{1}{|\mathcal{M}_1|} \frac{1}{\Prime^{k_{\sfb} + l_{\sfb}}} \frac{1}{|\mathcal{D}_1|^n} \frac{1}{|\mathcal{D}_2|^n} \sum_{m_1} \sum_{\Tilde{m}_1} \sum_{s,\Tilde{s}}   \sum_{\Tilde{x}_1^n} \sum_{\Tilde{d}_1^n, \Tilde{d}_2^n} \sum_{\Tilde{u}^n} p_{X_1}^n(\Tilde{x}_1^n) q_{U_2 \oplus U_3}^n(\Tilde{u}^n) \tr\left\{\left( I_{\boldsymbol{\CalH_{Y_1}^{G}}}-  \olineB^3_{\Tilde{x}_1^n,\Tilde{u}^n}\right)  \left( \left( \rho^{Y_1}\right)^{\otimes n } \otimes \ketbra{0} \right)\right\} \nonumber \\
&\overset{(b)}{=}& \frac{1}{|\mathcal{M}_1|} \frac{1}{\Prime^{k_{\sfb} + l_{\sfb}}} \frac{1}{|\mathcal{D}_1|^n} \frac{1}{|\mathcal{D}_2|^n} \sum_{m_1} \sum_{\Tilde{m}_1} \sum_{s,\Tilde{s}}   \sum_{\Tilde{x}_1^n} \sum_{\Tilde{d}_1^n, \Tilde{d}_2^n} \sum_{\Tilde{u}^n} p_{X_1}^n(\Tilde{x}_1^n) q_{U_2 \oplus U_3}^n(\Tilde{u}^n)  \tr\left\{ G^3_{\Tilde{x}_1^n,\Tilde{u}^n} \left( \rho^{Y_1}\right)^{\otimes n } \right\} \nonumber \\
&\overset{(c)}=& \frac{1}{|\mathcal{M}_1|} \frac{1}{\Prime^{k_{\sfb} + l_{\sfb}}} \frac{1}{|\mathcal{D}_1|^n} \frac{1}{|\mathcal{D}_2|^n} \sum_{m_1} \sum_{\Tilde{m}_1} \sum_{s,\Tilde{s}}   \sum_{\Tilde{d}_1^n, \Tilde{d}_2^n} \sum_{(\Tilde{x}_1^n,\Tilde{u}^n) \in T_{\delta}^n(p_{X_1} p_{U_2 \oplus U_3})} \hspace{-0.3in} p_{X_1}^n(\Tilde{x}_1^n) q_{U_2 \oplus U_3}^n(\Tilde{u}^n)  \tr\left\{ \Pi^{Y_1} \: \Pi^{Y_1}_{\Tilde{x}_1^n,\Tilde{u}^n} \: \Pi^{Y_1} \: \left( \rho^{Y_1}\right)^{\otimes n } \right\} \nonumber \\
&\overset{(d)}{\leq}& \frac{1}{|\mathcal{M}_1|} \frac{1}{\Prime^{k_{\sfb} + l_{\sfb}}} \frac{1}{|\mathcal{D}_1|^n} \frac{1}{|\mathcal{D}_2|^n} \sum_{m_1} \sum_{\Tilde{m}_1} \sum_{s,\Tilde{s}}  \sum_{\Tilde{d}_1^n, \Tilde{d}_2^n} \sum_{(\Tilde{x}_1^n, \Tilde{u}^n) \in T_{\delta}^n(p_{X_1} p_{U_2 \oplus U_3})}p_{X_1}^n(\Tilde{x}_1^n) 2^{-n\left( D(p_{U_2 \oplus U_3}||q_{U_2 \oplus U_3}) - \tilde{\delta} \right)}p_{U_2 \oplus U_3}^n(\Tilde{u}^n)  \nonumber \\
&&\tr\left\{ \Pi^{Y_1}_{\Tilde{x}_1^n,\Tilde{u}^n} \: \Pi^{Y_1} \: \left( \rho^{Y_1}  \right)^{\otimes n } \: \Pi^{Y_1} \right\} \nonumber \\
&\overset{(e)}{\leq}& 2^{-n \left( I(X_1,U_2 \oplus U_3;Y_1) - 2 \delta \right)} 2^{-n\left(D(p_{U_2 \oplus U_3}||q_{U_2 \oplus U_3}) - \tilde{\delta} -R_1 -\frac{(k_{\sfb}+l_{\sfb})}{n} \log(\Prime)\right)} \nonumber,
\end{eqnarray}
where (a) follows because (i) 
$\beta^3_{\Tilde{x}_1^n,\Tilde{d}_1^n,\Tilde{u}^n,\Tilde{d}_2^n}=\olineB^3_{\Tilde{x}_1^n,\Tilde{u}^n}$ and because (ii) the support of $I_{\boldsymbol{\CalH_{Y_1}^{e}}}-  \olineB^3_{\Tilde{x}_1^n,\Tilde{u}^n}$ is contained in $\boldsymbol{\CalH_{Y_1}^G}$, (b) follows from (i) the definition of $\olineB^3_{\Tilde{x}_1^n,\Tilde{u}^n}$ in \eqref{Eqn:3to1CQICcomplementprojector}
and (ii) from Gelfand–Naimark’s Thm.~\cite[Thm.~3.7]{BkHolevo_2019}, (c) follows from (i) the definition of $G^3_{\tilde{x}_1^n,\tilde{u}^n}$ in \eqref{Eqn:3to1CQICtheoriginalpovmelements}
and from (ii) $G^3_{\tilde{x}_1^n,\tilde{u}^n}=0$ if $(\tilde{x}_1^n, \tilde{u}^n) \notin T_{\delta}^n(p_{X_1} p_{U_2 \oplus U_3})$, (d) follows from 
the property defined in \eqref{Eqn:3to1CQICdivergenceproperty} and from (ii) the cyclicity of the trace, and finally (e) follows from the typical projetor properties 
\begin{eqnarray}
\Pi^{Y_1} \: \left( \rho^{Y_1}  \right)^{\otimes n } \: \Pi^{Y_1} &\leq& 2^{-n(H(Y_1) - \delta)} \: \Pi^{Y_1}, \mbox{ and } \nonumber \\  \tr\left\{\Pi^{Y_1}_{\tilde{x}_1^n, \tilde{u}^n}  \: \Pi^{Y_1}\right\} &\leq& \tr\left\{\Pi^{Y_1}_{\tilde{x}_1^n, \tilde{u}^n} \right\} \nonumber \\ 
&\leq&2^{n(H(Y_1|X_1,U_2 \oplus U_3) + \delta)}, \mbox{ for } (\tilde{x}_1^n, \tilde{u}^n) \in T_{\delta}^n(p_{X_1} p_{U_2 \oplus U_3}) . \nonumber 
\end{eqnarray}

Next, using the inequality in \eqref{Eqn:normgamma*}, the property defined in \eqref{Eqn:3to1CQICdivergenceproperty} and 
the inequality stated in \cite{202103SAD_Sen}
\begin{eqnarray}
      \norm{\CalN_{\eta}\left( \left( \rho^{Y_1}\right)^{\otimes n} \otimes \ketbra{0}\right)}_{\infty} \leq \frac{3 \eta }{\sqrt{\max\{|\CalD_1|^n, |\CalD_2|^n \}}}, \nonumber 
\end{eqnarray}
we upper-bound the second term in \eqref{Eqn:3to1CQIClastterminhayashinagaokainequality}, as follows 
\begin{eqnarray}
&&\frac{1}{|\mathcal{M}_1|} \frac{1}{\Prime^{k_{\sfb} + l_{\sfb}}} \frac{1}{|\mathcal{D}_1|^n} \frac{1}{|\mathcal{D}_2|^n} \sum_{m_1} \sum_{\Tilde{m}_1} \sum_{s,\Tilde{s}}   \sum_{\Tilde{x}_1^n} \sum_{\Tilde{d}_1^n} \sum_{\Tilde{u}^n} \sum_{\Tilde{d}_2^n}  p_{X_1}^n(\Tilde{x}_1^n) q_{U_2 \oplus U_3}^n(\Tilde{u}^n) \tr\left\{ \gamma^*_{\Tilde{x}_1^n,\Tilde{d}_1^n,\Tilde{u}^n,\Tilde{d}_2^n} \left( \CalN_{\eta}\left( \rho^{Y_1}\right)^{\otimes n } \otimes \ketbra{0} \right)\right\}  \nonumber \\
&\leq& 2^n |\CalH_{Y_1}|^n \: \frac{3 \eta}{\sqrt{\max\{|\CalD_1|^n,|\CalD_2|^n\}}}  \: 2^{-n\left(D(p_{U_2 \oplus U_3}||q_{U_2 \oplus U_3}) - \tilde{\delta} - R_1 - \frac{(k_{\sfb}+l_{\sfb})}{n} \log(\Prime)\right)}. \nonumber
\end{eqnarray}
\noindent Hence, if we choose $\max\{|\CalD_1|^n,|\CalD_2|^n\} \geq \left( 3 \eta 2^n |\CalH_{Y_1}|^n 2^{-n \left( I(X_1,U_2 \oplus U_3;Y_1) - 2 \delta \right)}\right)$, we have 
\begin{eqnarray}
\mathbb{E} \left[ t_{2.1.4}\right] &\leq&  2^{-n\left(D(p_{U_2 \oplus U_3}||q_{U_2 \oplus U_3}) - \tilde{\delta} - R_1 - \frac{(k_{\sfb}+l_{\sfb})}{n} \log(\Prime)\right)} \left( 2^{-n \left( I(X_1,U_2 \oplus U_3;Y_1) - 2 \delta \right)} + 2^n |\CalH_{Y_1}|^n \frac{3 \eta}{\sqrt{\max\{|\CalD_1|^n,|\CalD_2|^n\}}}\right) \nonumber \\
&\leq& 2 \: 2^{-n \left( I(X_1,U_2 \oplus U_3;Y_1) + D(p_{U_2 \oplus U_3}||q_{U_2 \oplus U_3}) - \tilde{\delta} - 2 \delta - R_1 - \frac{(k_{\sfb}+l_{\sfb})}{n} \log(\Prime)\right)}. \nonumber
\end{eqnarray}
\noindent Therefore, for $n$ sufficiently large, we have $\mathbb{E}[t_{2.1.4}] \leq \epsilon,$ if
\begin{eqnarray}
 R_1 + \frac{(k_{\sfb}+l_{\sfb})}{n} \log(\Prime)  &<& I(X_1,U_2 \oplus U_3;Y_1) + D(p_{U_2 \oplus U_3}||q_{U_2 \oplus U_3}) \nonumber \\
 &=&  I(X_1,U_2 \oplus U_3;Y_1) + \log(\Prime) - H(U_2 \oplus U_3). \nonumber 
\end{eqnarray} 
\begin{comment}
%\med Similarly, we have 
%\begin{eqnarray}
 %   \mathbb{E}[t_{2.1.5}] \leq \epsilon \quad \mbox{if} \quad R_1 + \frac{k}{n} \log(\Prime) < I(X_1,U;Y_1) + D(p_U||q_U) = I(X_1,U;Y_1) + \log(\Prime) - H(U). \nonumber 
%\end{eqnarray}    
\end{comment}

\section{Proof of Proposition \ref{Prop:Dec2FirsttermHay}}
\label{App:Dec2FirsttermHay}
We evaluate the expectation of $t_{2.1}$ over the codebook generation distribution. We obtain
\begin{eqnarray}
\mathbb{E}\left[t_{2.1}\right] &=& \frac{1}{|\mathcal{\ulineM}|} \frac{1}{\Prime^{k_2 + k_3}} \sum_{\ulinem} \sum_{a_2, a_3}  \sum_{u_2^n} \sum_{u_3^n} \sum_{x_1^n} q_{U_2}^n(u_2^n)  q_{U_3}^n(u_3^n) p_{X_1}^n(x_1^n) r_{U_2}^n(u_2^n) r_{U_3}^n(u_3^n) \nonumber \\
    &&\tr \left\{ \left( I - \Pi^{Y_2} \Pi^{Y_2}_{u_2^n} \Pi^{Y_2} \right)  \rho_{x_1^n,f_2^n(u_2^n),f_3^n(u_3^n)}^{Y_2}   \right\} \nonumber \\
&\overset{(a)}{=}&\frac{1}{|\mathcal{\ulineM}|} \frac{1}{\Prime^{k_2 + k_3}} \sum_{\ulinem} \sum_{a_2, a_3}  \sum_{u_2^n} \sum_{u_3^n} \sum_{x_1^n} p_{U_2}^n(u_2^n)  p_{U_3}^n(u_3^n) p_{X_1}^n(x_1^n)  \tr \left\{ \left( I - \Pi^{Y_2} \Pi^{Y_2}_{u_2^n} \Pi^{Y_2} \right)  \rho_{x_1^n,f_2^n(u_2^n),f_3^n(u_3^n)}^{Y_2}   \right\} \nonumber \\
&\overset{(b)}{=}&\frac{1}{|\mathcal{M}_2|} \frac{1}{\Prime^{k_2}} \sum_{m_2} \sum_{a_2}  \sum_{u_2^n} p_{U_2}^n(u_2^n) \tr \left\{ \left( I - \Pi^{Y_2} \Pi^{Y_2}_{u_2^n} \Pi^{Y_2} \right)  \rho_{u_2^n}^{Y_2}   \right\} \nonumber \\
&=&\frac{1}{|\mathcal{M}_2|} \frac{1}{\Prime^{k_2}} \sum_{m_2} \sum_{a_2} \left[\sum_{u_2^n \in T_{\delta}^n(p_{U_2})} p_{U_2}^n(u_2^n) \tr \left\{ \left( I - \Pi^{Y_2} \Pi^{Y_2}_{u_2^n} \Pi^{Y_2} \right)  \rho_{u_2^n}^{Y_2}   \right\} + \sum_{u_2^n\notin T_{\delta}^n(p_{U_2})} p_{U_2}^n(u_2^n)\right] \nonumber, 
\end{eqnarray}
where (a) follows from using $q_{U_2}^n(u_2^n)r_{U_2}^n(u_2^n)=p_{U_2}^n(u_2^n)$ and $q_{U_3}^n(u_3^n)r_{U_3}^n(u_3^n)=p_{U_3}^n(u_3^n)$, and (b) follows from the definition of $\rho_{u_2^n}^{Y_2}$ in \eqref{Eqn:3to1CQICassociateddensityoperatorsforuser2.1}.
Observe that,  
\begin{eqnarray}
    \tr \left\{ \Pi^{Y_2} \Pi^{Y_2}_{u_2^n} \Pi^{Y_2} \rho^{Y_2}_{u_2^n} \right\} &=& \tr \left\{ \Pi^{Y_2}_{u_2^n} \Pi^{Y_2} \rho^{Y_2}_{u_2^n} \Pi^{Y_2} \right\} \nonumber \\
    &\geq& \tr \left\{ \Pi^{Y_2}_{u_2^n} \rho^{Y_2}_{u_2^n}\right\} - \norm{\Pi^{Y_2} \rho^{Y_2}_{u_2^n} \Pi^{Y_2} - \rho^{Y_2}_{u_2^n}}_1 \nonumber \\
    &\geq& 1- \epsilon -2 \sqrt{\epsilon}, \nonumber 
\end{eqnarray}
where the first equality follows from the cyclicity of the trace, the first inequality follows from using $\tr\left(\Lambda \rho \right) \geq \tr\left(\Lambda \sigma \right) - \norm{\rho - \sigma }_1$, for $0 \leq \Lambda, \rho, \sigma \leq I$ and the second inequality follows from (i) the Gentle Operator Lemma \cite{BkWilde_2017} and from (ii) the conditional typical property 
\begin{eqnarray}
    \tr \left\{ \Pi^{Y_2}_{u_2^n} \rho^{Y_2}_{u_2^n} \right\} \geq 1 - \epsilon, \mbox{ for } u_2^n \in T_{\delta}^n(p_{U_2}).  \nonumber 
\end{eqnarray}
Hence, we have 
\begin{eqnarray}
    \mathbb{E}[t_{2.1}] &\leq& \left( 1- \left( 1- \epsilon - 2 \sqrt{\epsilon} \right) \right) + \epsilon \nonumber \\
    &=& 2 \epsilon + 2\sqrt{\epsilon}. \nonumber 
\end{eqnarray}
\section{Proof of Proposition \ref{Prop:Dec2SecondtermHay}}
\label{App:Dec2SecondtermHay}
We evaluate the expectation of $t_{2.3}$ over the codebook generation distribution. We obtain
\begin{eqnarray}
\mathbb{E}\left[t_{2.3}\right] &=&\frac{1}{|\mathcal{\ulineM}|} \frac{1}{\Prime^{k_2 + k_3}} \sum_{\ulinem} \sum_{a_2, a_3}  \sum_{\Tilde{m}_2 \neq m_2}\sum_{\Tilde{a}_2 } \sum_{\Tilde{u}_2^n}  \sum_{u_2^n} \sum_{u_3^n} \sum_{x_1^n} q_{U_2}^n(\Tilde{u}_2^n) q_{U_2}^n(u_2^n) q_{U_3}^n(u_3^n) p_{X_1}^n(x_1^n) \nonumber \\ 
&&r_{U_2}^n(u_2^n) r_{U_3}^n(u_3^n) \tr \left\{ \Pi^{Y_2} \Pi^{Y_2}_{\Tilde{u}_2^n} \Pi^{Y_2}  \rho_{x_1^n,f_2^n(u_2^n),f_3^n(u_3^n)}^{Y_2}   \right\} \nonumber \\
&\overset{(a)}{=}&\frac{1}{|\mathcal{\ulineM}|} \frac{1}{\Prime^{k_2 + k_3}} \sum_{\ulinem} \sum_{a_2, a_3} \sum_{\Tilde{m}_2 \neq m_2}\sum_{\Tilde{a}_2 }\sum_{\Tilde{u}_2^n}  \sum_{u_2^n} \sum_{u_3^n} \sum_{x_1^n} q_{U_2}^n(\Tilde{u}_2^n) p_{U_2}^n(u_2^n) p_{U_3}^n(u_3^n) p_{X_1}^n(x_1^n) \nonumber \\
&&\tr \left\{ \Pi^{Y_2} \Pi^{Y_2}_{\Tilde{u}_2^n} \Pi^{Y_2}  \rho_{x_1^n,f_2^n(u_2^n),f_3^n(u_3^n)}^{Y_2}   \right\} \nonumber \\
&\overset{(b)}{=}&\frac{1}{|\mathcal{\ulineM}|} \sum_{\ulinem} \sum_{\Tilde{m}_2 \neq m_2}\sum_{\Tilde{a}_2 } \sum_{\Tilde{u}_2^n\in T_{\delta}^n(p_{U_2})}  q_{U_2}^n(\Tilde{u}_2^n) \tr \left\{ \Pi^{Y_2} \Pi^{Y_2}_{\Tilde{u}_2^n} \Pi^{Y_2}  \left(\rho^{Y_2}\right)^{\otimes n}   \right\} \nonumber \\
&\overset{(c)}{\leq}& \frac{1}{|\mathcal{\ulineM}|}  \sum_{\ulinem}  \sum_{\Tilde{m}_2} \sum_{\Tilde{a}_2 } \sum_{\Tilde{u}_2^n\in T_{\delta}^n(p_{U_2})}  2^{-n\left(D(p_{U_2}||q_{U_2})    - \delta_2\right)}p_{U_2}^n(\Tilde{u}_2^n) \tr \left\{  \Pi^{Y_2}_{\Tilde{u}_2^n} \Pi^{Y_2}  \left(\rho^{Y_2}\right)^{\otimes n}   \Pi^{Y_2}\right\} \nonumber \\
&\overset{(d)}{\leq}&\frac{1}{|\mathcal{\ulineM}|} \sum_{\ulinem}   \sum_{\Tilde{m}_2} \sum_{\Tilde{a}_2 } \sum_{\Tilde{u}_2^n\in T_{\delta}^n(p_{U_2})}  2^{-n\left(D(p_{U_2}||q_{U_2}) - \delta_2\right)}p_{U_2}^n(\Tilde{u}_2^n) \tr \left\{  \Pi^{Y_2}_{\Tilde{u}_2^n} \: \Pi^{Y_2}\right\} 2^{-n\left(H(Y_2)-\delta\right)} \nonumber \\
&\overset{(e)}{\leq} &\frac{1}{|\mathcal{\ulineM}|}  \sum_{\ulinem}  \sum_{\Tilde{m}_2} \sum_{\Tilde{a}_2 } \sum_{\Tilde{u}_2^n\in T_{\delta}^n(p_{U_2})}  2^{-n\left(D(p_{U_2}||q_{U_2}) - \delta_2\right)}p_{U_2}^n(\Tilde{u}_2^n)  2^{-n\left(I(Y_2;U_2)-2\delta\right)} \nonumber \\
&\leq& 2^{-n\left( I(U_2;Y_2) + D(p_{U_2}||q_{U_2}) - 2\delta - \delta_2 - \frac{k_2+l_2}{n} \log(\Prime)\right)} \nonumber,
\end{eqnarray}
where (a) follows from $q_{U_2}^n(u_2^n) r_{U_2}^n(u_2^n)=p_{U_2}^n(u_2^n)$ and $q_{U_3}^n(u_3^n) r_{U_3}^n(u_3^n)=p_{U_3}^n(u_3^n)$, (b) follows from the definition of $\left(\rho^{Y_2}\right)^{\otimes n}$ in \eqref{Eqn:3to1CQICassociateddensityoperatorsforuser2}, (c) follows from (i) the cyclicity of the trace and (ii) from 
\begin{eqnarray}
    q_{U_2}^n(u_2^n) \leq 2^{-n\left(D(p_{U_2}||q_{U_2}) - \delta_2\right)}p_{U_2}^n(u_2^n), \mbox{ for } u_2^n \in T_{\delta}^n(p_{U_2}), \nonumber
\end{eqnarray}
(d) and (e) follow from the typicality properties, i.e., for
 $u_2^n \in T_{\delta}^n(p_{U_2})$, we have  
\begin{eqnarray}
     \Pi^{Y_2}  \left(\rho^{Y_2}\right)^{\otimes n}   \Pi^{Y_2} &\leq& 2^{-n\left(H(Y_2)-\delta \right)} \: \Pi^{Y_2}, \mbox{ and } \nonumber \\
     \tr \left\{  \Pi^{Y_2}_{u_2^n} \: \Pi^{Y_2} \right\} &\leq& \tr \left\{  \Pi^{Y_2}_{u_2^n}\right\} \nonumber \\
     &\leq& 2^{n \left( H(Y_2|U_2) + \delta \right)}.  \nonumber
\end{eqnarray}
Hence, for $n$ sufficiently large, we have $   \mathbb{E}[t_{2.3}] \leq \epsilon,$ if
\begin{eqnarray}
 \frac{k_2+l_2}{n}\log(\Prime) &<& I(U_2;Y_2) + D(p_{U_2}||q_{U_2}) \nonumber \\
 &=& I(U_2;Y_2) + \log(\Prime) -H(U_2). \nonumber 
\end{eqnarray}

Following the same steps as above, for $n$ sufficiently large, we have $\mathbb{E}[t_{2.2}] \leq \epsilon,$ if
\begin{eqnarray}
\frac{k_2}{n}\log(\Prime) &<& I(U_2;Y_2) + D(p_{U_2}||q_{U_2}) \nonumber \\
&=& I(U_2;Y_2) + \log(\Prime) - H(U_2). \nonumber
\end{eqnarray}

\section{Proof of Proposition \ref{Prop:4CQMACFirsttermHay}}
\label{App:4CQMACFirsttermHay}

We evaluate the expectation of $t_{1.1}$ over the codebook generation distribution. We obtain
\begin{eqnarray}
\mathbb{E}[t_{1.1}]&=&\frac{1}{|\mathcal{\ulineM}|} \frac{1}{|\mathcal{\ulineD}|^n} \sum_{\ulinem} \sum_{\ulined^n} \sum_{\ulinez^n} p^n_{\ulineZ}(\ulinez^n) \tr \left\{ \left(I-\gamma^*_{\ulinez^n, \ulined^n}\right) \theta_{\ulinez^n, \ulined^n} \right\} \nonumber \\
&=&\frac{1}{|\mathcal{\ulineM}|} \frac{1}{|\mathcal{\ulineD}|^n} \sum_{\ulinem} \sum_{\ulined^n} \sum_{\ulinez^n} p^n_{\ulineZ}(\ulinez^n) \left[\tr \left\{\theta_{\ulinez^n, \ulined^n} \right) - \tr \left\{ \Pi_{\boldsymbol{\CalH_{Y_G}}} \left(I_{\boldsymbol{\CalH_{Y}^{e}}}-  \beta^{*}_{\ulinez^{n},\ulined^{n}}\right) \theta_{\ulinez^n, \ulined^n} \left(I_{\boldsymbol{\CalH_{Y}^{e}}}-  \beta^{*}_{\ulinez^{n},\ulined^{n}}\right) \Pi_{\boldsymbol{\CalH_{Y_G}}} \right\} \right],\nonumber 
    \end{eqnarray}
    where the last equality follows from (i) the definition of $\gamma^*_{\ulinez^n, \ulined^n}$ in \eqref{Eqn:3CQICgammadef}, (ii) the cyclicity of the trace and (iii) the property of the projector i.e., $\Pi_{\boldsymbol{\CalH_{Y_G}}}^2=\Pi_{\boldsymbol{\CalH_{Y_G}}}$. Subsequently, we apply the non-commutative union bound \cite[Fact.3]{202103SAD_Sen}. We obtain
    \begin{eqnarray}
\mathbb{E}[t_{1.1}] &\leq& \frac{4}{|\mathcal{\ulineM}|} \frac{1}{|\mathcal{\ulineD}|^n} \sum_{\ulinem} \sum_{\ulined^n} \sum_{\ulinez^n} p^n_{\ulineZ}(\ulinez^n) \tr \left\{ \left(I-\Pi_{\boldsymbol{\CalH_{Y_G}}} + \beta^*_{\ulinez^n, \ulined^n}\right) \theta_{\ulinez^n, \ulined^n} \right\} \nonumber \\
    &\leq& \frac{4}{|\mathcal{\ulineM}|} \frac{1}{|\mathcal{\ulineD}|^n} \sum_{\ulinem} \sum_{\ulined^n} \sum_{\ulinez^n} p^n_{\ulineZ}(\ulinez^n) \tr \left\{ \left(I-\Pi_{\boldsymbol{\CalH_{Y_G}}} + \beta^*_{\ulinez^n, \ulined^n}\right) \left(\xi_{\ulinez^n} \otimes \ketbra{0}\right) \right\}  \nonumber \\
&+& \frac{4}{|\mathcal{\ulineM}|} \frac{1}{|\mathcal{\ulineD}|^n} \sum_{\ulinem} \sum_{\ulined^n} \sum_{\ulinez^n} p^n_{\ulineZ}(\ulinez^n) \norm{\theta_{\ulinez^n, \ulined^n}-\left(\xi_{\ulinez^n} \otimes \ketbra{0}\right)}_1  \label{Eqn:goingbacktooriginalstate},
\end{eqnarray}
where the last inequality follows from the trace inequality $\tr(\Lambda \rho) \leq \tr(\Lambda \sigma) + \norm{\rho - \sigma}_1$, with $0 \leq \Lambda, \rho, \sigma \leq I$.
From Prop.~\ref{Prop:3CQICClosnessOFstates}, we bound the second term in \eqref{Eqn:goingbacktooriginalstate} by $12 \eta$ and since the state $\left( \xi_{\ulinez^n} \otimes \ketbra{0} \right)$ lives in $\boldsymbol{\CalH_{Y_G}}$ and $I-\Pi_{\boldsymbol{\CalH_{Y_G}}}$ is a projector onto its complement, we are left with
\begin{eqnarray}
   \mathbb{E}[t_{1.1}] &\leq&  \frac{4}{|\mathcal{\ulineM}|} \frac{1}{|\mathcal{\ulineD}|^n} \sum_{\ulinem} \sum_{\ulined^n} \sum_{\ulinez^n} p^n_{\ulineZ}(\ulinez^n) \tr \left\{ \beta^*_{\ulinez^n, \ulined^n} \left(\xi_{\ulinez^n} \otimes \ketbra{0}\right) \right\}
+  12 \eta \nonumber \\
&\leq& \frac{42}{\eta^2}\frac{4}{|\mathcal{\ulineM}|} \frac{1}{|\mathcal{\ulineD}|^n} \sum_{\ulinem} \sum_{\ulined^n} \sum_{\ulinez^n} p^n_{\ulineZ}(\ulinez^n) \left[ \sum_{S \subseteq [4]} \tr \left\{ \olineB^S_{\ulinez^n} \left(\xi_{\ulinez^n} \otimes \ketbra{0}\right) \right\} \right]
+  12 \eta \nonumber \\
&=&\frac{42}{\eta^2}\frac{4}{|\mathcal{\ulineM}|} \frac{1}{|\mathcal{\ulineD}|^n} \sum_{\ulinem} \sum_{\ulined^n} \left[ 15- \sum_{S \subseteq [4]} \sum_{\ulinez^n} p^n_{\ulineZ}(\ulinez^n) \tr \left\{ G^S_{\ulinez^n} \xi_{\ulinez^n}  \right\} \right]
+  12 \eta \nonumber,
\end{eqnarray}
where the second inequality follows from \cite[Corollary.1]{202103SAD_Sen} and the equality follows from (i) the definition of $\olineB_{\ulinez^n}^S$ in \eqref{Eqn:3CQICComplementprojector} and (ii) from Gelfand–Naimark’s Thm.~\cite[Thm.~3.7]{BkHolevo_2019}. For $S \subseteq [4]$, observe that,
\begin{eqnarray}
    \sum_{\ulinez^n} p_{\ulineZ}^n(\ulinez^n) \tr\left\{ G^S_{\ulinez^n} \xi_{\ulinez^n} \right\} &=& \sum_{\ulinez^n \in T_{\delta}^n(p_{\ulineZ})} p_{\ulineZ}^n(\ulinez^n) \tr\left\{ \Pi_{z_{S^{c}}^{n}} \Pi_{\ulinez^{n}}\Pi_{z_{S^{c}}^{n}} \xi_{\ulinez^n} \right\} \nonumber \\
    &\overset{(a)}{=}& \sum_{\ulinez^n \in T_{\delta}^n(p_{\ulineZ})} p_{\ulineZ}^n(\ulinez^n) \tr\left\{  \Pi_{\ulinez^{n}}\Pi_{z_{S^{c}}^{n}} \xi_{\ulinez^n} \Pi_{z_{S^{c}}^{n}} \right\} \nonumber \\
&\overset{(b)}{\geq}& \sum_{\ulinez^n \in T_{\delta}^n(p_{\ulineZ})} p_{\ulineZ}^n(\ulinez^n) \tr\left\{  \Pi_{\ulinez^{n}} \xi_{\ulinez^n}  \right\} - \sum_{\ulinez^n \in T_{\delta}^n(p_{\ulineZ})} p_{\ulineZ}^n(\ulinez^n) \norm{ \Pi_{z_{S^{c}}^{n}} \xi_{\ulinez^n} \Pi_{z_{S^{c}}^{n}} - \xi_{\ulinez^n}}_1 \nonumber  \\
&\overset{(c)}{\geq}& 1- \epsilon - 2 \sqrt{\epsilon}, \nonumber
\end{eqnarray}
where (a) follows from the cyclicity of the trace, (b) follows from the trace inequality $\tr\left( \Lambda \rho\right) \geq \tr\left( \Lambda \sigma \right) - \norm{\rho - \sigma}_1$, for $0 \leq \Lambda, \rho, \sigma \leq I$, and (c) follows from  (i) the Gentle Operator Lemma \cite{BkWilde_2017} and (ii) from 
\begin{eqnarray}
     \tr\left\{  \Pi_{\ulinez^{n}} \xi_{\ulinez^n}  \right\}  \geq 1-\epsilon, \mbox{ for }  \ulinez^n \in T_{\delta}^n(p_{\ulineZ}). \nonumber 
\end{eqnarray}

\noindent Hence, using the above inequality, we obtain 
\begin{eqnarray}
    && \mathbb{E}[t_{1.1}] \leq \frac{168}{\eta^2} (15\epsilon + 30 \sqrt{\epsilon}) + 12 \eta.\nonumber 
\end{eqnarray}
With $\eta=\epsilon^{\frac{1}{5}}$, we have
\begin{eqnarray}
    && \mathbb{E}[t_{1.1}] \leq 168 (15\epsilon^{\frac{3}{5}} + 30 \epsilon^{\frac{1}{10}}) + 12 \epsilon^{\frac{1}{5}}. \nonumber 
\end{eqnarray}

\section{Proof of Proposition \ref{Prop:4CQMACSectermHay}}
\label{App:4CQMACSectermHay}

 We elaborate on the analysis of $\sum_{S \subseteq [4]} t_{1S}$, and in doing so we leverage the \textit{smoothing} and \textit{augmentation} \cite{202103SAD_Sen} properties of the tilting maps we have defined. As is standard, this is broken into fifteen terms—four singleton errors, six pair errors, four triple errors, and one quadruple error. We begin with the quadruple error.
 \begin{comment}
 %The analysis of the quadruple error is slightly different from the analysis of the remaining fourteen terms. To analyze the quadruple error, we first prove that the difference between the average of the original state and the average of the tilted state over $\mathcal{\ulineZ}^n$ is small. This is done analogously to the steps used in \cite{202103SAD_Sen}. Having done this, the rest of the analysis is straightforward, since the corresponding projector has not been tilted, and we are left with it operating on the average of the original state. 
\end{comment}
\med Consider $t_{1.S}$, for $S=4$ and evaluate the expectation over the codebook generation distribution. We obtain
 \begin{eqnarray}
\mathbb{E}[t_{1.4}]&=& \frac{1}{|\mathcal{\ulineM}|}  \frac{1}{|\mathcal{\ulineD}|^{2n}}  \sum_{\ulinem}  \sum_{\tilde{\ulinem} \neq \ulinem}  \sum_{\uline{\tilde{d}}^n}  \sum_{\uline{\tilde{z}}^n}  \sum_{\ulined^n}  \sum_{\ulinez^n}  p_{\ulineZ}^n(\uline{\tilde{z}}^n)  p_{\ulineZ}^n(\ulinez^n) \tr \left\{  \gamma^*_{\uline{\tilde{z}}^n, \uline{\tilde{d}}^n}  \theta_{\ulinez^n, \ulined^n}  \right\} \nonumber \\
&\overset{(a)}{=}& \frac{1}{|\mathcal{\ulineM}|}  \frac{1}{|\mathcal{\ulineD}|^{n}}  \sum_{\ulinem}  \sum_{\tilde{\ulinem} \neq \ulinem}  \sum_{\uline{\tilde{d}}^n}  \sum_{\uline{\tilde{z}}^n}  p_{\ulineZ}^n(\uline{\tilde{z}}^n) \tr \left\{  \gamma^*_{\uline{\tilde{z}}^n, \uline{\tilde{d}}^n}  \theta^{\otimes n}  \right\} \nonumber \\
&\overset{(b)}{\leq}& \frac{1}{\Omega(\eta)}  \frac{1}{|\mathcal{\ulineM}|}  \frac{1}{|\mathcal{\ulineD}|^{n}}  \sum_{\ulinem}  \sum_{\tilde{\ulinem}}  \sum_{\uline{\tilde{d}}^n}  \sum_{\uline{\tilde{z}}^n}  p_{\ulineZ}^n(\uline{\tilde{z}}^n) \tr \left\{  \gamma^*_{\uline{\tilde{z}}^n, \uline{\tilde{d}}^n}  \left(\xi^{\otimes n} \otimes \ketbra{0}\right)  \right\} \nonumber \\ 
&+& \frac{1}{|\mathcal{\ulineM}|}  \frac{1}{|\mathcal{\ulineD}|^{n}}  \sum_{\ulinem}  \sum_{\tilde{\ulinem}}  \sum_{\uline{\tilde{d}}^n}  \sum_{\uline{\tilde{z}}^n}  p_{\ulineZ}^n(\uline{\tilde{z}}^n) \tr \left\{  \gamma^*_{\uline{\tilde{z}}^n, \uline{\tilde{d}}^n}  \CalN_{\eta}\left(\xi^{\otimes n} \otimes \ketbra{0}\right)  \right\}, \label{Eqn:lasttermofhayashinagaokainequality}
 \end{eqnarray} 
 where (a) follows from 
 \begin{eqnarray}
     \theta^{\otimes n} \define \frac{1}{|\mathcal{\ulineD}|^n} \sum_{\ulined^n} \sum_{\ulinez^n} p_{\ulineZ}^n(\ulinez^n) \theta_{\ulinez^n, \ulined^n}, \nonumber 
 \end{eqnarray}
and (b) follows from 
\begin{eqnarray}
    \theta^{\otimes n} = \frac{1}{\Omega(\eta)} \left( \xi^{\otimes n } \otimes \ketbra{0} \right) + \mathcal{N}_{\eta}\left( \xi^{\otimes n } \otimes \ketbra{0} \right), \mbox{ where } \mathcal{N}_{\eta}\left( \xi^{\otimes n } \otimes \ketbra{0} \right) \define \theta^{\otimes n} - \frac{1}{\Omega(\eta)} \left( \xi^{\otimes n } \otimes \ketbra{0} \right), \nonumber 
\end{eqnarray}
as stated in \cite{202103SAD_Sen}. Consider the first term in \eqref{Eqn:lasttermofhayashinagaokainequality}, we have 
 \begin{eqnarray}
&&\frac{1}{\Omega(\eta)}  \frac{1}{|\mathcal{\ulineM}|}  \frac{1}{|\mathcal{\ulineD}|^{n}}  \sum_{\ulinem}  \sum_{\tilde{\ulinem} }  \sum_{\uline{\tilde{d}}^n}  \sum_{\uline{\tilde{z}}^n}  p_{\ulineZ}^n(\uline{\tilde{z}}^n) \tr \left\{  \gamma^*_{\uline{\tilde{z}}^n, \uline{\tilde{d}}^n}  \left(\xi^{\otimes n} \otimes \ketbra{0}\right)  \right\} \nonumber \\
&\overset{(a)}{\leq}& \frac{1}{|\mathcal{\ulineM}|}  \frac{1}{|\mathcal{\ulineD}|^{n}}  \sum_{\ulinem}  \sum_{\tilde{\ulinem} }  \sum_{\uline{\tilde{d}}^n}  \sum_{\uline{\tilde{z}}^n}  p_{\ulineZ}^n(\uline{\tilde{z}}^n) \tr \left\{  \left( I - \beta^{[4]}_{\uline{\tildez}^n,\uline{\tilde{d}}^n} \right) \left(\xi^{\otimes n} \otimes \ketbra{0}\right)  \right\} \nonumber \\
&\overset{(b)}{=}& \frac{1}{|\mathcal{\ulineM}|}  \frac{1}{|\mathcal{\ulineD}|^{n}}  \sum_{\ulinem}  \sum_{\tilde{\ulinem} }  \sum_{\uline{\tilde{d}}^n}  \sum_{\uline{\tilde{z}}^n}  p_{\ulineZ}^n(\uline{\tilde{z}}^n) \tr \left\{  \left( I - \olineB^{[4]}_{\uline{\tildez}^n} \right) \left(\xi^{\otimes n} \otimes \ketbra{0}\right)  \right\} \nonumber \\
&\overset{(c)}{=}& \frac{1}{|\mathcal{\ulineM}|}  \frac{1}{|\mathcal{\ulineD}|^{n}}  \sum_{\ulinem}  \sum_{\tilde{\ulinem} }  \sum_{\uline{\tilde{d}}^n}  \sum_{\uline{\tilde{z}}^n}  p_{\ulineZ}^n(\uline{\tilde{z}}^n) \tr \left\{G^{[4]}_{\uline{\tildez}^n} \xi^{\otimes n}   \right\} \nonumber \\
&\overset{(e)}{=}&  \frac{1}{|\mathcal{\ulineM}|}  \frac{1}{|\mathcal{\ulineD}|^{n}}  \sum_{\ulinem}  \sum_{\tilde{\ulinem} }  \sum_{\uline{\tilde{d}}^n}  \sum_{\uline{\tilde{z}}^n \in T_{\delta}^n(p_{\ulineZ})}  p_{\ulineZ}^n(\uline{\tilde{z}}^n) \tr \left\{   \Pi_{\uline{\tildez}^n}\Pi \xi^{\otimes n} \Pi   \right\} \nonumber \\
&\overset{(f)}{\leq}&  \frac{1}{|\mathcal{\ulineM}|}  \frac{1}{|\mathcal{\ulineD}|^{n}}  \sum_{\ulinem}  \sum_{\tilde{\ulinem} }  \sum_{\uline{\tilde{d}}^n}  \sum_{\uline{\tilde{z}}^n \in T_{\delta}^n(p_{\ulineZ})}  p_{\ulineZ}^n(\uline{\tilde{z}}^n) 2 ^{-n(H(Y)- \delta)} \tr \left\{   \Pi_{\uline{\tildez}^n}\Pi\right\} \nonumber \\
&\overset{(g)}{\leq}&2^{n(\tilde{R}_1+\tilde{R}_2 + \tilde{R}_3 + \tilde{R}_4)}  2^{-n(I(Y; \ulineZ)-2\delta)},\nonumber 
 \end{eqnarray}
where (a) follows from 
\begin{eqnarray}
\gamma^*_{\uline{\tilde{z}}^n, \uline{\tilde{d}}^n} \leq I-\beta^*_{\uline{\tilde{z}}^n, \uline{\tilde{d}}^n} \leq I-\beta^{[4]}_{\uline{\tilde{z}}^n, \uline{\tilde{d}}^n}, \nonumber    
\end{eqnarray}
(b) follows from $\beta^{[4]}_{\uline{\tilde{z}}^n, \uline{\tilde{d}}^n} \define \olineB_{\uline{\tilde{z}}^n}^{[4]}$, (c) follows from (i) the definition of  $\olineB_{\uline{\tilde{z}}^n}^{[4]}$ in \eqref{Eqn:3CQICComplementprojector} and from (ii) the Gelfand–Naimark’s Thm.~\cite[Thm.~3.7]{BkHolevo_2019}, (e) follows from (i) the definition of $G^{[4]}_{\uline{\tildez}^n}$ in \eqref{Eqn:3CQICOrigPOVM}
and from (ii) the cyclicity of the trace, and finally (f) and (g) follow from
\begin{eqnarray}
    \tr\left\{ \Pi_{\uline{\tildez}^n} \Pi \xi^{\otimes n} \Pi \right\} &\leq& 2^{-n(H(Y)- \delta)} \tr\left\{ \Pi_{\uline{\tildez}^n}  \Pi \right\} \nonumber \\
    &\leq& 2^{-n(H(Y)- \delta)} \tr\left\{ \Pi_{\uline{\tildez}^n}\right\} \nonumber \\
    &\leq&   2^{-n(I(Y;\ulineZ) - 2\delta )}, \mbox{ for } \uline{\tildez}^n \in T_{\delta}^n(p_{\ulineZ}). \nonumber 
\end{eqnarray}
Now consider the second term in \eqref{Eqn:lasttermofhayashinagaokainequality}, 
 \begin{eqnarray}
&&\frac{1}{|\mathcal{\ulineM}|}  \frac{1}{|\mathcal{\ulineD}|^{n}}  \sum_{\ulinem, \tilde{\ulinem} }  \sum_{\uline{\tilde{d}}^n}  \sum_{\uline{\tilde{z}}^n}  p_{\ulineZ}^n(\uline{\tilde{z}}^n) \tr \left\{  \gamma^*_{\uline{\tilde{z}}^n, \uline{\tilde{d}}^n}  \CalN_{\eta}\left(\xi^{\otimes n} \otimes \ketbra{0}\right)  \right\} \nonumber \\
&\leq&  \frac{1}{|\mathcal{\ulineM}|}  \frac{1}{|\mathcal{\ulineD}|^{n}}  \sum_{\ulinem, \tilde{\ulinem}}  \sum_{\uline{\tilde{d}}^n}  \sum_{\uline{\tilde{z}}^n}  p_{\ulineZ}^n(\uline{\tilde{z}}^n)  \norm{ \gamma^*_{\uline{\tilde{z}}^n, \uline{\tilde{d}}^n}}_1 \norm{ \CalN_{\eta}\left(\xi^{\otimes n} \otimes \ketbra{0}\right) }_{\infty} \nonumber \\ 
&\leq&  2^{n(\tilde{R}_1+\tilde{R}_2 + \tilde{R}_3 + \tilde{R}_4)} \: 2 |\CalH_Y|^n \: \frac{21 \eta}{\sqrt{\max_{S \subseteq [4]}\{|D_S|^n\}}}. \nonumber 
 \end{eqnarray}
The first inequality follows from the standard trace-norm inequality
\begin{eqnarray}
    \label{Eqn:3CQICTraceNormineq}
|\tr(AB)| \leq \norm{AB}_{1} \leq \norm{A}_{1} \norm{B}_{\infty}.
\end{eqnarray}
The last inequality follows from (i)
 \begin{eqnarray}
 &&  \norm{\gamma^*_{\uline{\tilde{z}}^n, \uline{\tilde{d}}^n}}_1 \leq \norm{I-\beta^*_{\uline{\tilde{z}}^n, \uline{\tilde{d}}^n}}_{\infty} \norm{\Pi_{\boldsymbol{\CalH_{Y_G}} }\left(I_{\boldsymbol{\CalH_{Y}^{e}}}-  \beta^{*}_{\ulinez^{n},\ulined^{n}}\right)}_1 \leq \norm{I-\beta^*_{\uline{\tilde{z}}^n, \uline{\tilde{d}}^n}}_{\infty}^2 \norm{\Pi_{\boldsymbol{\CalH_{Y_G}} }}_1 \leq 2|\CalH_{Y}|^n, \nonumber  
 \end{eqnarray}
 and from (ii) the inequality stated in \cite{202103SAD_Sen}
 \begin{eqnarray}
\norm{ \CalN_{\eta}\left(\xi^{\otimes n} \otimes \ketbra{0}\right) }_{\infty}  \leq   \frac{21 \eta}{\sqrt{\max_{S \subseteq [4]}\{|D_S|^n\}}}.\nonumber 
 \end{eqnarray}

\noindent The upper-bound on the $\mathbb{L}_{\infty}$-norm of $\CalN_{\eta}$ 
shows that the difference between $\theta^{\otimes n}$ and $\left(\xi^{\otimes n} \otimes \ketbra{0}\right)$ is small. This establishes the smoothing property. Choosing 
\begin{eqnarray}
    \max_{S \subseteq [4]}\{|D_S|\} \geq (42 \eta^2)^{\frac{1}{n}} |\CalH_Y|^2 2^{2(I(Y;\ulineZ)+2\delta)}, \nonumber 
\end{eqnarray}
we have 
\begin{eqnarray}
      \mathbb{E}[t_{1.4}] &\leq&  2^{n(\tilde{R}_1 + \tilde{R}_2 + \tilde{R}_3 + \tilde{R}_4)} \left( 2^{-n\left(I\left(Y; \ulineZ \right)-2\delta\right)} + 2 |\CalH_Y|^n \:  \frac{21 \eta}{\sqrt{\max_{S \subseteq [4]}\{|D_S|^n\}}} \right) \nonumber \\
      &\leq&    2 \: 2^{-n\left(I\left(Y; \ulineZ \right)-2\delta-(\tilde{R}_1 + \tilde{R}_2 + \tilde{R}_3 + \tilde{R}_4)\right)}. \nonumber 
\end{eqnarray}
Therefore, 
\begin{eqnarray}
    \mbox{if }  \tilde{R}_1 + \tilde{R}_2 + \tilde{R}_3 + \tilde{R}_4 < I\left(Y; \ulineZ \right), \mbox{ then } \mathbb{E}[t_{1.4}] \leq \epsilon. \nonumber 
\end{eqnarray}

\noindent Returning to the original notation (see Fig.~\ref{TabNotataion4CQMAC}), we have $    \mathbb{E}[t_{1.4}] \leq \epsilon,$ if
\begin{eqnarray}
 S_{12} + S_{13} + K_1 + L_1 + \max\{S_{21},S_{31}\} < I(Y_1;U_{12},U_{13},X_1,U_{21}\oplus U_{31}). \nonumber
\end{eqnarray}
The above bound is obtained when binning is ignored. However, since our analysis includes binning and, moreover, each of the codebooks is picked according to the corresponding marginal distribution, the right-hand side of the bound we obtain will contain an additional divergence term. Specifically, we obtain
\begin{eqnarray}
S_{12} + S_{13} + K_1 + L_1 + \max\{S_{21}, S_{31} \} &<& I(Y_1;U_{12},U_{13},X_1,U_{21}\oplus U_{31}) + D\left( p_{U_{12}U_{13}X_1} p_{U_{21}\oplus U_{31}} \Big|\Big| \frac{p_{X_1}}{|\mathcal{U}_{12}| |\mathcal{U}_{13}| \Prime_1}\right) \nonumber \\
&=& \log \left(|\mathcal{U}_{12}| \right) + \log \left(|\mathcal{U}_{13}| \right) + \log \left(\Prime_1 \right) + H(X_1) - H(U_{12}, U_{13}, X_1,U_{21}\oplus U_{23} |Y_1).\nonumber 
\end{eqnarray}
\noindent Note that the above bound is identical to the one obtained in  \eqref{Eqn3CQIC:3CQICStep1ChnlBnd5}, \eqref{Eqn3CQIC:3CQICStep1ChnlBnd6}
of the Thm.~\ref{Thm:3CQICStageIRateRegion}, with $A_1 = \{12,13\}$.

Having described the analysis of the quadruple error term, we are left with fourteen terms. The analysis of each of these fourteen terms is essentially identical. Therefore, we describe the analysis for a general $S \subsetneq [4]$. 
Consider
\begin{eqnarray}
    t_{1.S}=\frac{1}{|\mathcal{\ulineM}|} \sum_{\ulinem} \sum_{\tilde{m}_S \neq m_S}\tr \left\{ \gamma^*_{Z^n_S(\tilde{m}_S),D^n_S(\tilde{m}_S),Z^n_{S^c}(m_{S^c}), D^n_{S^c}(m_{S^c})} \theta_{\ulineZ^n(\ulinem),\ulineD^n(\ulinem)}\right\}. \nonumber
\end{eqnarray}
We evaluate the expectation over the codebook generation distribution. We obtain
\begin{eqnarray}
\mathbb{E}[t_{1.S}] &=& \frac{1}{|\mathcal{\ulineM}|} \frac{1}{|\mathcal{\ulineD}|^n} \frac{1}{|\mathcal{D}_S|^n} \sum_{\ulinem} \sum_{\tilde{m}_S \neq m_S} \sum_{\tilde{d}^n_S} \sum_{\tilde{z}^n_S} \sum_{\ulined^n} \sum_{\ulinez^n} p_{\ulineZ}^n(\ulinez^n) p_{Z_S}^n(\tilde{z}_S^n)\tr \left\{ \gamma^*_{\tilde{z}^n_S, \tilde{d}^n_S,z^n_{S^c}, d^n_{S^c}} \theta_{\ulinez^n,\ulined^n} \right\} \nonumber\\
&=&\frac{1}{|\mathcal{\ulineM}|} \frac{1}{|\mathcal{\ulineD}|^n} \sum_{\ulinem} \sum_{\tilde{m}_S \neq m_S} \sum_{\tilde{d}^n_S} \sum_{\tilde{z}^n_S}  \sum_{z_{S^c}^n} p_{Z_{S^c}}^n(z_{S^c}^n) p_{Z_S}^n(\tilde{z}_S^n)\tr \left\{ \gamma^*_{\tilde{z}^n_S, \tilde{d}^n_S,z^n_{S^c}, d^n_{S^c}} \theta_{z^n_{S^c},d^n_{S^c}} \right\} \nonumber \\
&\leq&  \frac{1}{|\mathcal{\ulineM}|} \frac{1}{|\mathcal{\ulineD}|^n} \sum_{\ulinem} \sum_{\tilde{m}_S} \sum_{\tilde{d}^n_S} \sum_{\tilde{z}^n_S}  \sum_{z_{S^c}^n} p_{Z_{S^c}}^n(z_{S^c}^n) p_{Z_S}^n(\tilde{z}_S^n)\tr \left\{ \gamma^*_{\tilde{z}^n_S, \tilde{d}^n_S,z^n_{S^c}, d^n_{S^c}} \CalT^{S^c}_{d^n_{S^c}, \eta}\left(\xi_{z^n_{S^c}} \otimes \ketbra{0}\right) \right\} \nonumber \\
&+& \frac{1}{|\mathcal{\ulineM}|} \frac{1}{|\mathcal{\ulineD}|^n} \sum_{\ulinem} \sum_{\tilde{m}_S} \sum_{\tilde{d}^n_S} \sum_{\tilde{z}^n_S}  \sum_{z_{S^c}^n} p_{Z_{S^c}}^n(z_{S^c}^n) p_{Z_S}^n(\tilde{z}_S^n)\tr \left\{ \gamma^*_{\tilde{z}^n_S, \tilde{d}^n_S,z^n_{S^c}, d^n_{S^c}} \CalN_{d^n_{S^c}}\left(\xi_{z^n_{S^c}} \otimes \ketbra{0}\right) \right\}, \label{Eqn:Oneofthefourteentermaftersplitting}
\end{eqnarray}
where the first equality follows from 
\begin{eqnarray}
    \theta_{z^n_S,d^n_S} \define \frac{1}{|\mathcal{D}_{S^c}|^n} \sum_{d^n_{S^c}} \sum_{z^n_{S^c}} p_{Z_{S^c}|Z_S}^n(z^n_{S^c}|z^n_S) \theta_{\ulinez^n, \ulined^n}, \nonumber 
\end{eqnarray}
and the inequality follows 
\begin{eqnarray}
    \theta{z^n_S, d^n_S}&=& \frac{\Omega(S, \eta)}{\Omega(\eta)} \mathcal{T}_{d^n_S, \eta} \left( \xi_{z^n_S} \otimes \ketbra{0}\right) + \mathcal{N}_{d^n_S, \eta}\left( \xi_{z^n_S} \otimes \ketbra{0}\right), \mbox{ where }\nonumber \\
      \mathcal{N}_{d^n_S, \eta}\left( \xi_{z^n_S} \otimes \ketbra{0}\right) &\define& \theta{z^n_S, d^n_S} - \frac{\Omega(S, \eta)}{\Omega(\eta)} \mathcal{T}_{d^n_S, \eta} \left( \xi_{z^n_S} \otimes \ketbra{0}\right), \nonumber 
\end{eqnarray}
as stated in \cite{202103SAD_Sen}. Consider the first term in \eqref{Eqn:Oneofthefourteentermaftersplitting}
\begin{eqnarray}
&&\frac{1}{|\mathcal{\ulineM}|} \frac{1}{|\mathcal{\ulineD}|^n} \sum_{\ulinem} \sum_{\tilde{m}_S} \sum_{\tilde{d}^n_S} \sum_{\tilde{z}^n_S}  \sum_{z_{S^c}^n} p_{Z_{S^c}}^n(z_{S^c}^n) p_{Z_S}^n(\tilde{z}_S^n)\tr \left\{ \gamma^*_{\tilde{z}^n_S, \tilde{d}^n_S,z^n_{S^c}, d^n_{S^c}} \CalT^{S^c}_{d^n_{S^c}, \eta}\left(\xi_{z^n_{S^c}} \otimes \ketbra{0}\right) \right\} \nonumber \\
&\overset{(a)}{\leq}& \frac{1}{|\mathcal{\ulineM}|} \frac{1}{|\mathcal{\ulineD}|^n} \sum_{\ulinem} \sum_{\tilde{m}_S} \sum_{\tilde{d}^n_S} \sum_{\tilde{z}^n_S}  \sum_{z_{S^c}^n} p_{Z_{S^c}}^n(z_{S^c}^n) p_{Z_S}^n(\tilde{z}_S^n)\tr \left\{ \left(I-\beta^S_{\tilde{z}^n_S, \tilde{d}^n_S,z^n_{S^c}, d^n_{S^c}}\right) \CalT^{S^c}_{d^n_{S^c}, \eta}\left(\xi_{z^n_{S^c}} \otimes \ketbra{0}\right) \right\} \nonumber \\
&=& \frac{1}{|\mathcal{\ulineM}|} \frac{1}{|\mathcal{\ulineD}|^n} \sum_{\ulinem} \sum_{\tilde{m}_S} \sum_{\tilde{d}^n_S} \sum_{\tilde{z}^n_S}  \sum_{z_{S^c}^n} p_{Z_{S^c}}^n(z_{S^c}^n) p_{Z_S}^n(\tilde{z}_S^n) \tr \left\{ \left(I-\beta^S_{\tilde{z}^n_S, \tilde{d}^n_S,z^n_{S^c}, d^n_{S^c}}\right) \Pi_{\CalT^{S^c}_{d^n_{S^c},\eta}(\boldsymbol{\CalH_{Y_G}})} \CalT^{S^c}_{d^n_{S^c}, \eta}\left(\xi_{z^n_{S^c}} \otimes \ketbra{0}\right) \right\} \nonumber \\
&\overset{(b)}{=}&\frac{1}{|\mathcal{\ulineM}|} \frac{1}{|\mathcal{\ulineD}|^n} \sum_{\ulinem} \sum_{\tilde{m}_S} \sum_{\tilde{d}^n_S} \sum_{\tilde{z}^n_S}  \sum_{z_{S^c}^n} p_{Z_{S^c}}^n(z_{S^c}^n) p_{Z_S}^n(\tilde{z}_S^n) \tr \left\{ \left(I_{\CalT_{d^n_{S^c},\eta}^{S^c}(\boldsymbol{\CalH_{Y_G}})}-\beta^S_{\tilde{z}^n_S, \tilde{d}^n_S,z^n_{S^c}, d^n_{S^c}}\right)  \CalT^{S^c}_{d^n_{S^c}, \eta}\left(\xi_{z^n_{S^c}} \otimes \ketbra{0}\right) \right\} \nonumber \\
&\overset{(c)}{=}& \frac{1}{|\mathcal{\ulineM}|} \frac{1}{|\mathcal{\ulineD}|^n} \sum_{\ulinem} \sum_{\tilde{m}_S} \sum_{\tilde{d}^n_S} \sum_{\tilde{z}^n_S}  \sum_{z_{S^c}^n} p_{Z_{S^c}}^n(z_{S^c}^n) p_{Z_S}^n(\tilde{z}_S^n)\tr \left\{ \left(I_{\boldsymbol{\CalH_{Y_G}}}-\olineB^S_{\tilde{z}^n_S,z^n_{S^c}}\right) \left(\xi_{z^n_{S^c}} \otimes \ketbra{0}\right) \right\} \nonumber \\
&\overset{(d)}{=}& \frac{1}{|\mathcal{\ulineM}|} \frac{1}{|\mathcal{\ulineD}|^n} \sum_{\ulinem} \sum_{\tilde{m}_S} \sum_{\tilde{d}^n_S} \sum_{\tilde{z}^n_S}  \sum_{z_{S^c}^n} p_{Z_{S^c}}^n(z_{S^c}^n) p_{Z_S}^n(\tilde{z}_S^n)\tr \left\{ G^S_{\tilde{z}^n_S,z^n_{S^c}} \xi_{z^n_{S^c}} \right\} \nonumber \\
&\overset{(e)}{=}& \frac{1}{|\mathcal{\ulineM}|} \frac{1}{|\mathcal{\ulineD}|^n} \sum_{\ulinem} \sum_{\tilde{m}_S} \sum_{\tilde{d}^n_S} \sum_{(\tilde{z}^n_S,z^n_{S^c}) \in T_{\delta}^n(p_{\ulineZ})} p_{Z_{S^c}}^n(z_{S^c}^n) p_{Z_S}^n(\tilde{z}_S^n)\tr \left\{  \Pi_{\tilde{z}^n_S,z^n_{S^c}}\Pi_{z_{S^{c}}^{n}}
\xi_{z^n_{S^c}} \Pi_{z_{S^{c}}^{n}}\right\} \nonumber \\
&\overset{(f)}{\leq}&
\frac{1}{|\mathcal{\ulineM}|} \frac{1}{|\mathcal{\ulineD}|^n} \sum_{\ulinem} \sum_{\tilde{m}_S} \sum_{\tilde{d}^n_S} \sum_{(\tilde{z}^n_S,z^n_{S^c}) \in T_{\delta}^n(p_{\ulineZ})}  p_{Z_{S^c}}^n(z_{S^c}^n) p_{Z_S}^n(\tilde{z}_S^n) 2^{-n(H(Y|Z_{S^c})-\delta)} \tr \left\{  \Pi_{\tilde{z}^n_S,z^n_{S^c}}\Pi_{z_{S^{c}}^{n}}\right\} \nonumber \\
&\overset{(g)}{\leq}& 2^{-n(I(Y; Z_S|Z_{S^c})-2\delta)} 2^{n \left(\sum_{s \in S} \tilde{R}_s\right)}, \nonumber
\end{eqnarray}
where (a) follows from 
\begin{eqnarray}
    \gamma^*_{\tilde{z}^n_S, \tilde{d}^n_S,z^n_{S^c}, d^n_{S^c}} \leq I-\beta^*_{\tilde{z}^n_S, \tilde{d}^n_S,z^n_{S^c}, d^n_{S^c}} \leq I- \beta^S_{\tilde{z}^n_S, \tilde{d}^n_S,z^n_{S^c}, d^n_{S^c}}, \nonumber
\end{eqnarray} 
(b) follows from $\beta^S_{\tilde{z}^n_S, \tilde{d}^n_S,z^n_{S^c}, d^n_{S^c}}$ having support contained in $\CalT^{S^c}_{d^n_{S^c},\eta}(\boldsymbol{\CalH_{Y_G}})$, (c) follows from extracting the isometry $\mathcal{T}_{d^n_{S^c}, \eta}$, (d) follows from (i) the definition of $\olineB^S_{\tilde{z}_S^n,z^n_{S^c}}$ in \eqref{Eqn:3CQICComplementprojector} and (ii) from Gelfand–Naimark’s Thm.~\cite[Thm.~3.7]{BkHolevo_2019}, (e) follows from (i) the definition of $G^S_{\tilde{z}^n_S,z^n_{S^c}}$ in \eqref{Eqn:3CQICOrigPOVM} and from (ii) the cyclicity of the trace, and finally (f) and (g)  follow from 
\begin{eqnarray}
\tr\left\{ \Pi_{\tilde{z}^n_S,z^n_{S^c}} \Pi_{z_{S^{c}}^{n}} \xi_{z^n_{S^c}} \Pi_{z_{S^{c}}^{n}} \right\} &\leq&  2^{-n(H(Y|Z_{S^c})-\delta)} \tr\left\{ \Pi_{\tilde{z}^n_S,z^n_{S^c}} \Pi_{z_{S^{c}}^{n}} \right\} \nonumber \\
&\leq&  2^{-n(H(Y|Z_{S^c})-\delta)} \tr\left\{ \Pi_{\tilde{z}^n_S,z^n_{S^c}}\right\} \nonumber \\
&\leq& 2^{-n \left(I\left(Y;Z_S|Z_{S^c}\right)-2\delta \right)}, \mbox{ for }(\tilde{z}^n_S,z^n_{S^c})\in T_{\delta}^n(p_{\ulineZ}). \nonumber 
\end{eqnarray}
Now consider the second term in \eqref{Eqn:Oneofthefourteentermaftersplitting}. Using the trace-norm inequality \eqref{Eqn:3CQICTraceNormineq} and the inequality 
\begin{eqnarray}
\norm{\CalN_{d^n_{S^c}}\left(\xi_{z^n_{S^c}} \otimes \ketbra{0}\right)}_{\infty}  \leq \frac{21  \eta}{\sqrt{|\CalD_S|^n}}. \nonumber 
\end{eqnarray}
as stated in \cite{202103SAD_Sen}, we obtain 
\begin{eqnarray}
&&\frac{1}{|\mathcal{\ulineM}|} \frac{1}{|\mathcal{\ulineD}|^n} \sum_{\ulinem} \sum_{\tilde{m}_S} \sum_{\tilde{d}^n_S} \sum_{\tilde{z}^n_S}  \sum_{z_{S^c}^n} p_{Z_{S^c}}^n(z_{S^c}^n) p_{Z_S}^n(\tilde{z}_S^n)\tr \left\{ \gamma^*_{\tilde{z}^n_S, \tilde{d}^n_S,z^n_{S^c}, d^n_{S^c}} \CalN_{d^n_{S^c}}\left(\xi_{z^n_{S^c}} \otimes \ketbra{0}\right) \right\} \nonumber \\
&\leq& \frac{1}{|\mathcal{\ulineM}|} \frac{1}{|\mathcal{\ulineD}|^n} \sum_{\ulinem} \sum_{\tilde{m}_S} \sum_{\tilde{d}^n_S} \sum_{\tilde{z}^n_S}  \sum_{z_{S^c}^n} p_{Z_{S^c}}^n(z_{S^c}^n) p_{Z_S}^n(\tilde{z}_S^n) \norm{ \gamma^*_{\tilde{z}^n_S, \tilde{d}^n_S,z^n_{S^c}, d^n_{S^c}}}_{1} \norm{\CalN_{d^n_{S^c}}\left(\xi_{z^n_{S^c}} \otimes \ketbra{0}\right)}_{\infty} \nonumber \\
&\leq&  2 |\CalH_Y|^n \: \frac{21  \eta}{\sqrt{|\CalD_S|^n}}2^{n\left(\sum_{s \in S} \tilde{R}_s\right)}. \nonumber 
\end{eqnarray}
Choosing $|\CalD_S| \geq (42 \eta)^{\frac{2}{n}}|\CalH_Y|^2 2^{2\left(I(Y;Z_S|Z_{S^c})-2 \delta\right)}$, 
\begin{comment}
%\begin{eqnarray}
 %   |\CalD_S| \geq (42 \eta)^{\frac{2}{n}}|\CalH_Y|^2 2^{2\left(I(Y;Z_S|Z_{S^c})-2 \delta\right)}, \nonumber 
%\end{eqnarray}    
\end{comment}
we have 
\begin{eqnarray}
    \mathbb{E}[t_{1.S}] &\leq& 2^{n\left(\sum_{s \in S} \tilde{R}_s\right)} \left( 2^{-n(I(Y;Z_S|Z_{S^c})-2\delta)}+ 2 |\CalH_Y|^n \: \frac{21  \eta}{\sqrt{|\CalD_S|^n}}\right)  \nonumber \\
    &\leq& 2 2^{-n\left(I(Y;Z_S|Z_{S^c})-2\delta -\left(\sum_{s \in S} \tilde{R}_s\right)\right)} \nonumber 
    \end{eqnarray}
Hence, $   \mathbb{E}[t_{1.S}]\leq \epsilon,$ if
\begin{eqnarray}
 \sum_{s \in S} \tilde{R}_s < I(Y;Z_S|Z_{S^c}). \nonumber 
\end{eqnarray}

\noindent As an example, we take $S=\{1,2,4\}$, i.e., $\mathbb{E}[t_{1.S}]\leq \epsilon,$ if
\begin{eqnarray}
 \tilde{R}_1+\tilde{R}_2+\tilde{R}_4  < I(Y;Z_1,Z_2,Z_4|Z_3). \nonumber      
\end{eqnarray}
Returning to the original notation (see Fig.~\ref{TabNotataion4CQMAC}), we have $   \mathbb{E}[t_{1.S}]\leq \epsilon,$ if
\begin{eqnarray}
 S_{12}+S_{13}+\max\{S_{21},S_{31}\} < I(Y_1;U_{12},U_{13},U_{21}\oplus U_{31}|X_1). \nonumber 
\end{eqnarray}
\noindent Now if the analysis is carried out with binning, we obtain the following bound 
\begin{eqnarray}
S_{12} + S_{13} + \max\{S_{21}, S_{31}\}  &<& I(Y_1;U_{12},U_{13}, U_{21} \oplus U_{31}|X_1)  + D\left( p_{U_{12}U_{13}} p_{U_{21} \oplus U_{31}} \Big|\Big| \frac{1}{|\mathcal{U}_{12}||\mathcal{U}_{13}| \log(\Prime_1)} \right) \nonumber \\
&=& \log \left( |\mathcal{U}_{12}|\right) + \log \left( |\mathcal{U}_{13}|\right) + \log \left( \Prime_1 \right) - H\left(U_{12}, U_{13}, U_{21} \oplus U_{31} | X_1, Y_1 \right), \nonumber
\end{eqnarray}
which corresponds to the bound \eqref{Eqn3CQIC:3CQICStep1ChnlBnd2}, \eqref{Eqn3CQIC:3CQICStep1ChnlBnd3} in Thm.~\ref{Thm:3CQICStageIRateRegion}, with $\mathcal{A}_1=\{12,13\}$.

\bibliographystyle{IEEEtran}
{
\bibliography{CosetCdsFor3CQChnls}

\end{document}